\documentclass[12pt]{iopart} 
\expandafter\let\csname equation*\endcsname=\relax 
\expandafter\let\csname endequation*\endcsname=\relax 
\usepackage{amsmath,amssymb,amsthm} 
\usepackage{braket} 
\renewcommand{\Re}{\operatorname{Re}}
\renewcommand{\Im}{\operatorname{Im}}
\usepackage{graphicx} 
\graphicspath{{Materials/}} 
\usepackage{xfrac}
\usepackage{cleveref}
\usepackage{dsfont}
\usepackage{listings}
\usepackage{xcolor}
\usepackage[section]{placeins}
\usepackage{appendix}
\begin{document}
\title[Quantum-jump vs stochastic Schr\"odinger dynamics]{Quantum-jump vs stochastic Schr\"odinger dynamics for Gaussian states with quadratic Hamiltonians and linear Lindbladians}
\author{Robson Christie$^*$, Jessica Eastman$^*$, Roman Schubert$^{**}$ and Eva-Maria Graefe$^*$}
\address{$^*$Department of Mathematics, Imperial College London, London SW7 2AZ,
United Kingdom}
\address{$^{**}$School of Mathematics, University of Bristol, Bristol, BS8 1TW, United Kingdom}
\ead{robson.christie13@imperial.ac.uk, j.eastman@imperial.ac.uk, Roman.Schubert@bristol.ac.uk and e.m.graefe@imperial.ac.uk}
\vspace{10pt}
\begin{indented}
\item[]\today
\end{indented}

\begin{abstract}
The dynamics of Gaussian states for open quantum systems described by Lindblad equations can be solved analytically for systems with quadratic Hamiltonians and linear Lindbladians, showing the familiar phenomena of dissipation and decoherence. It is well known that the Lindblad dynamics can be expressed as an ensemble average over stochastic pure-state dynamics, which can be interpreted as individual experimental implementations, where the form of the stochastic dynamics depends on the measurement setup. Here we consider quantum-jump and stochastic Schr\"odinger dynamics for initially Gaussian states. While both unravellings converge to the same Lindblad dynamics when averaged, the individual dynamics can differ qualitatively. For the stochastic Schr\"odinger equation, Gaussian states remain Gaussian during the evolution, with stochastic differential equations governing the evolution of the phase-space centre and a deterministic evolution of the covariance matrix. In contrast to this, individual pure-state dynamics arising from the quantum-jump evolution do not remain Gaussian in general. Applying results developed in the non-Hermitian context for Hagedorn wavepackets, we formulate a method to generate quantum-jump trajectories that is described entirely in terms of the evolution of an underlying Gaussian state. To illustrate the behaviours of the different unravellings in comparison to the Lindblad dynamics, we consider two examples in detail, which can be largely treated analytically, a harmonic oscillator subject to position measurement and a damped harmonic oscillator. In both cases, we highlight the differences as well as the similarities of the stochastic Schr\"odinger and the quantum-jump dynamics.
\end{abstract}

\section{Introduction}

A large class of open quantum systems can be described by Lindblad equations for the dynamics of density operators. In many atomic physics and quantum optics applications, this is an accurate description of physical phenomena. Lindblad dynamics can be viewed as an ensemble average over stochastic pure-state trajectories. The stochastic trajectories may be interpreted as the state of an individual experimental quantum system conditioned on a measurement record. The nature of the measurement record determines the types of trajectories in the ensemble \cite{wisemanBook,belavkin1989nondemolition,carmichael2009open}, and intriguingly can lead to strikingly different microscopic dynamics and behaviours, while averaging to the same Lindblad dynamics.
An example of this has been reported in \cite{bartolo2017} where it has been demonstrated that depending on the specific unravelling a dissipative Kerr dynamics can lead to switches between either odd and even Schr\"{o}dinger cat states or between coherent states of opposite phase. It has been observed in examples that individual stochastic trajectories can even display chaotic dynamics that are not seen in the ensemble average \cite{eastman2017tuning}.

Two of the most important unravellings are given by \textit{stochastic Schr\"{o}dinger equations} (SSEs) and \textit{quantum-jump} trajectories, respectively. The SSE dynamics is a good description of certain homodyne and heterodyne detection schemes in quantum optics arising from constant weak continuous measurements \cite{wiseman1996quantum,verstraelen2018gaussian}. Quantum-jump dynamics, on the other hand, may arise in photodetection experiments and is theoretically described by periods of deterministic evolution under an effective non-Hermitian Hamiltonian, stochastically interrupted by discrete measurements \cite{plenio1998quantum}. In both SSE and quantum-jump scenarios,  if the measurement channel is not recorded, then the best estimate for the state is obtained by averaging over all possibilities, resulting in Lindblad dynamics \cite{wiseman1996quantum}. There are other interpretations of the unravellings of the Lindblad equation, viewing them as candidate laws of nature in an effort to explain the wavefunction collapse in quantum measurement (see e.g. \cite{ghirardi1986unified,bassi2013models,brody2002efficient,gisin1992quantum} for more details).

In the present paper, we analyse in detail the structural differences of these unravellings as compared to the full Lindblad dynamics for the case of quadratic Hamiltonians and linear Lindbladians for initial Gaussian states. In these cases, the full quantum dynamics in the Lindblad case can be reduced to simple phase-space dynamics, as has been discussed in \cite{Brodier2004,graefebradley}. This follows the spirit of Heller's and Littlejohn's approach to closed system dynamics \cite{hellersemi,littlejohn1986semiclassical} where for a quadratic Hamiltonian initially Gaussian states remain Gaussian under time evolution. This idea carries through to the SSE \cite{strunz1998classical,halliwell1995quantum,verstraelen2018gaussian}. Since Gaussian states may be parameterised entirely by their centres and covariances, the dynamics of these systems may be transformed from state vector dynamics in infinite-dimensional Hilbert space to a handful of differential equations for classical phase-space observables. This simple approach does not carry through to the quantum-jump dynamics, which in general do not preserve Gaussian states for arbitrary linear Lindbladians. Here we present a different approach that builds on an extension of Hagedorn's wavepacket dynamics \cite{hagedorn1998raising}  that has been adapted to non-Hermitian systems in \cite{lasser2018non}. 

In detail, the paper is organised as follows. In \cref{sec:unravell} we introduce the Lindblad equation along with the quantum-jump and SSE unravellings. In \cref{sec:GaussDyn} we insert an initially Gaussian state into the SSE and derive a set of stochastic differential equations which is exact in the case of quadratic Hamiltonians and linear Lindbladians. The resulting parameter dynamics are less computationally expensive to simulate than the full state vector dynamics. In \cref{sec:HagJumps} we briefly review the non-Hermitian dynamics of Gaussian states and then apply the non-Hermitian Hagedorn wavepacket dynamics developed in \cite{lasser2018non} to formulate several semi-analytical algorithms for simulating quantum-jump trajectories. In \cref{sec:examples} we illustrate our findings for two examples which we consider in some detail: A harmonic oscillator Hamiltonian with either a position measurement, modelled by a Lindbladian $\hat L=\hat x$, or a radiative decay, modelled by a Lindbladian $\hat L=\hat a$.  We conclude with a summary and outlook in \cref{sec:conclusion}.

\section{Unravellings of the Lindblad Equation} \label{sec:unravell}
 The Lindblad equation was initially derived as the most general Markovian dynamical equation that preserves the trace, Hermiticity and positivity of the density matrix \cite{gorini1976,lindblad1976generators}. Physically, it can be used to describe certain quantum systems that are weakly coupled to a memoryless environment. In this spirit, dynamics of Lindblad form can be obtained for the reduced density matrix by averaging over the effect of a bath of quantum harmonic oscillators \cite{petruccione}. Any linear and Markovian (local in time) master equation that preserves the Hermiticity and trace of the density matrix may be expressed in Lindblad form,
 \begin{equation} \label{eq:lindblad}
		i \hbar \frac{d}{d t}\hat{\rho}(t)=[\hat{H},\hat{\rho}(t)]  + i \sum_{k}[\hat{L}_k \hat{\rho}(t) \hat{L}_{k }^{\dagger} -\frac{1}{2} \hat{L}_k^{\dagger} \hat{L}_k\hat{\rho}(t)-\frac{1}{2} \hat{\rho}(t)\hat{L}_k^{\dagger} \hat{L}_k].
\end{equation}
Here $\hat H$ is a Hamiltonian and ${\hat L_k}$ are general Lindblad operators, the properties of which are system-specific. For simplicity in what follows we shall confine the discussion to a single Lindblad operator.

Much like the deterministic Fokker-Planck equation for the dynamics of probability distributions in classical physics admits \textit{unravellings} in terms of single trajectories of the stochastic Langevin equation, the deterministic Lindblad equation for the dynamics of the density operator may be unravelled in terms of stochastic pure-state trajectories. There are infinitely many such unravellings that differ from each other in the stochastic driving processes. The two types of unravellings most commonly considered in the literature are SSE trajectories \cite{gisin1992quantum} driven with continuous Gaussian distributed noise, and quantum-jump \cite{plenio1998quantum} trajectories driven by discrete Poissonian distributed noise.

The SSE we consider here is given by \cite{gisin1992quantum,strunz1998classical}
\begin{equation} \label{eq:SSE}
	\begin{aligned}
		\ket{d \psi } &= \frac{1}{\hbar}(-{i}\hat{H} -\frac{1}{2 }\hat{L}^{\dagger}\hat{L}+\braket{\hat{L}^{\dagger}}\hat{L} - \frac{1}{2}\braket{\hat{L}^{\dagger}}\braket{\hat{L}})\ket {\psi} dt \\
		 &\phantom{=}+ \frac{1}{\sqrt{2 \hbar}}(\hat{L}-\braket{\hat{L}})\ket{\psi} (d\xi_R +i d\xi_I),
	\end{aligned}
\end{equation} 
where $d\xi_R$ and $d\xi_I$ are independent ($\mathbb{E}[d\xi_Rd\xi_I]=0$)  It\^{o} stochastic processes with mean zero ($\mathbb{E}[d\xi_R]=\mathbb{E}[d\xi_I]=0$) and normalisation $d\xi_R^2=d\xi_I^2=dt$. The SSE trajectories are driven with a continuous stochastic process and are often used to model systems undergoing weak continuous measurement such as heterodyne detection in quantum optics \cite{wiseman1996quantum} or quantum Brownian motion \cite{petruccione}. 
 
In the quantum-jump description, on the other hand, the pure-state trajectories deterministically evolve under an effective non-Hermitian Hamiltonian $\hat H-i\hat \Gamma$ with $\hat\Gamma=\frac{1}{2}\hat L^{\dagger} \hat L$, periodically interrupted by stochastic quantum jumps. These jumps may be used to represent random discrete measurements of quantum systems such as photodetection from a microwave cavity \cite{wiseman1996quantum}. The cumulative effect of these jumps when averaged over many trajectories induces the ``jump term" contribution $ \hat{L} \hat{\rho}(t) \hat{L}^{\dagger}$ in the density operator dynamics. Concretely, quantum-jump pure-state dynamics can be described by the dynamical equation
\begin{equation} \label{eq:QuantumJump}
	\begin{aligned}
		\ket{d \psi } &= \frac{1}{\hbar}\left(-{i}\hat{H} -\frac{1}{2 }\hat{L}^{\dagger}\hat{L} - \frac{1}{2}\braket{\hat{L}^{\dagger}\hat{L}}\right)\ket {\psi}(1-dN) dt \\
		 &\phantom{=}+ \left(\frac{\hat{L}}{\sqrt{\braket{\hat{L}^{\dagger}\hat{L}}}}-1\right)\ket{\psi} dN.
	\end{aligned}
\end{equation} 
 Here $dN$ is a Poisson process, taking the values 0 (no jump) or 1 (jump) with expectation value $\mathbb{E}[dN]=\braket{\hat{L}^{\dagger}\hat{L}} dt$. The following algorithm  \cite{wisemanBook} is equivalent to \cref{eq:QuantumJump}:
 \begin{enumerate} 
	\item Discretise the time interval $(t_0,t_N) \rightarrow \{t_k \} $ with $\Delta t=t_{k+1}-t_k$;
	\item Pick a random number $R$ from the uniform distribution on the interval $[0,1]$;
	\item For all $t_k$ with $t_0< t_k\leq t_N$ evaluate the following loop:
    \begin{lstlisting}
    for k=1:N-1
        if $R\geq ||\psi_k||^2$ %Jump
            $\ket{\psi_{k+1}}=\hat L \ket{\psi_{k}}/\braket{\hat L^{\dagger}\hat L}$ 
            $R$=rand$\sim U(0,1)$
        else %No jump
            $\ket{\psi_{k+1}}=\ket{\psi_{k}}- \frac{i}{\hbar}(\hat H-\frac{i}{2}\hat L^{\dagger}\hat L)\ket{\psi_{k}}\Delta t$ 
        end
    end
    \end{lstlisting}
	\item Normalise the entire trajectory; $\ket{\psi_{k}}\rightarrow \ket{\psi_{k}}/||\psi_{k}||$ for all $k$.
\end{enumerate}

Considering quantum-jump trajectories of a Markovian open system described by the Lindblad master equation and postselecting only trajectories in which no jumps have occurred, thus leads to effective non-Hermitian Hamiltonian dynamics \cite{daley2014quantum,klauck2019observation,graefenatphoton}. Quantum dynamics generated by non-Hermitian Hamiltonians is an active area of research on its own, and we will make use of some techniques developed in this context \cite{graefe2012complexified,lasser2018non}. It should be noted that $\hat L^{\dagger}\hat L$ is a positive operator and thus non-Hermitian Hamiltonians, $\hat H-\frac{i}{2}\hat L^{\dagger}\hat L$, arising in the context of postselection of Lindblad/quantum-jump dynamics may only describe loss (but not gain). 

Despite converging to the same ensemble dynamics, the SSE and quantum-jump trajectories can differ not just quantitatively but qualitatively. We will analyse these differences in detail for quadratic Hamiltonians and linear Lindbladians with initially Gaussian states. While for Lindblad and SSE dynamics the state remains Gaussian for all times for quadratic Hamiltonians and linear Lindbladians, this is in general not the case for quantum-jump dynamics, even though averaging over quantum-jump trajectories recovers the Gaussian Lindblad results. We will return to this issue after considering the Gaussian Lindblad and SSE dynamics. 
 
\section{Gaussian dynamics} \label{sec:GaussDyn}

Gaussian states are well suited for the analysis of quantum dynamics, since they are localised in phase space on order $\hbar$ (in appropriate coordinates they are minimum uncertainty states), and are the only states with a completely positive Wigner function \cite{hudson1974wigner}. In this regard, they may be thought of as the most ``classical" of quantum states \cite{Wignegnegativity}, and Gaussian approximations of the full quantum dynamics lead to simple phase-space dynamics. As has been observed by Schr\"odinger already in the early days of quantum mechanics, Gaussian wavepackets remain Gaussian in the dynamics of quantum harmonic oscillators and follow classical trajectories. For open systems described by Lindblad equations or Schr\"odinger dynamics generated by non-Hermitian Hamiltonians, a similar statement holds, which allows one to reduce the full quantum Hilbert space dynamics to a simple phase-space dynamics described by a handful of parameters for Gaussian states for quadratic Hamiltonians and linear Lindbladians \cite{graefe2012complexified,de2002propagation,graefebradley}. Gaussian evolution for stochastic Schr\"odinger equations with quadratic Hamiltonian and linear Lindbladians in position representation has been considered in \cite{halliwell1995quantum,strunz1997path,strunz1999quantum}, and here too, an initial Gaussian state remains Gaussian. 

In what follows we shall provide a brief review of the derivation and result for Lindblad dynamics, and then extend the idea to stochastic Schr\"odinger dynamics in quantum phase space, where we use the Wigner-Weyl formalism that illuminates the underlying phase-space geometry and allows for a better direct comparison with the Lindblad dynamics. At the heart of the Wigner-Weyl scheme is the Weyl transformation, a bijective map that maps observables on Hilbert spaces to their corresponding Weyl symbols. The Weyl symbol corresponding to an observable $\hat{O}(\hat{X},\hat{P})$ is a distribution on classical phase space and given by
\begin{equation}
	{O}(x,p)=\int_{-\infty}^{\infty}ds \bra{x-\frac{s}{2}}\hat{O}(\hat{X},\hat{P})\ket{x+\frac{s}{2}}e^{\sfrac{i p s}{\hbar}}.
\end{equation}
The Weyl symbol corresponding to the density operator is known as the Wigner function. We will consider Gaussian states with Wigner functions of the form 
\begin{equation} \label{eq:ansatz}
	\begin{aligned}
	W(z)=\frac{\sqrt{\det G}}{\pi \hbar}e^{-\frac{1}{\hbar} \delta z \cdot G \delta z}& &\text{with}& &\delta z =\begin{pmatrix}x-\braket{x} \\ p-\braket{p}\end{pmatrix} =z-\tilde{z},
	\end{aligned}
\end{equation}
with a real symmetric matrix $G$, that encodes the phase-space covariance matrix of the system as
\begin{equation} \label{eq:Covar}
    \Sigma_{ij}=\Delta(z_i z_j)^2=\frac{\hbar}{2}G^{-1}_{ij}.
\end{equation}
This describes a pure state if and only if $\det G=1$ \cite{adesso2014continuous}. Inserting an ansatz of the form (\ref{eq:ansatz}) with time-dependent (and in the SSE case stochastic) parameters $G$ and $\tilde z$ into the evolution equation for the Wigner function yields dynamical equations for the parameters. 

\subsection{Gaussian dynamics of the Lindblad equation} \label{sec:GaussLind}
The Lindblad equation in Weyl representation takes the form 
\begin{equation} \label{eq:LindWeyl}
	\begin{aligned}
	 \frac{dW}{dt}&=\frac{1}{\hbar}\left(-i(H \star W-W\star H)+   L\star W \star \bar{L}-\frac{1}{2}\bar{L}\star L\star W -\frac12W \star \bar{L}\star L  \right),
	\end{aligned}
\end{equation}
with the Moyal (star) product of Weyl symbols given by
\begin{equation}
	(A\star B)(q,p) = A(q,p)e^{\frac{i\hbar}{2}\overleftarrow{\nabla}\cdot \Omega \cdot \overrightarrow{\nabla}} B(q,p).
\end{equation}
Since we are considering only quadratic Hamiltonians and linear Lindbladians, the Moyal products in  \cref{eq:LindWeyl} can be fully expanded to yield \cite{graefebradley}
\begin{equation}
\label{eq:Wignerdyn}
 	\frac{dW}{dt}   =-i \nabla\bar{L}\cdot \Omega \nabla L W+ \nabla H\cdot\Omega\nabla W + \Im(L \nabla \bar{L})\cdot \Omega \nabla W-\frac{\hbar}{2}\Re(\nabla L\cdot \Omega W'' \Omega \nabla\bar{L}),
\end{equation}
where 
\begin{equation}
\Omega=\begin{pmatrix} 0& 1\\-1&0\end{pmatrix}
\end{equation}
is the symplectic matrix. 
An initial Gaussian state remains Gaussian for all times \cite{graefebradley}. 
It is useful to rewrite $H$ and $L$ as polynomials in $\delta z$
\begin{equation}
    \begin{aligned}
    H(z)&=H(\tilde{z})+\nabla H|_{z=\tilde{z}}\cdot \delta z+\frac12 \delta z\cdot H''|_{z=\tilde{z}}  \delta z \label{eq:Taylor}\\
    L(z)&=L(\tilde{z})+\nabla L|_{z=\tilde{z}}\cdot \delta z , 
    \end{aligned}
\end{equation}
where we have used a Taylor series (which is exact here) to identify the linear and quadratic coefficients. We substitute these expansions into \cref{eq:Wignerdyn} to obtain
\begin{multline}
 \label{eq:CenterWignerDyn}
     \frac{dW}{dt}=[-\frac{2}{\hbar}  (\nabla H+H''\delta z)\cdot\Omega G \delta z-\frac{2}{\hbar} \Im(L\nabla \bar{L})\cdot \Omega G \delta z-\frac{2}{\hbar} \Im(\nabla L\nabla \bar{L})\cdot \Omega G \delta z \\
      -i \nabla\bar{L}\cdot \Omega \nabla L+ \Re(\nabla L \cdot \Omega G \Omega \nabla \bar L) -\frac{2}{\hbar}\delta z\cdot G\Omega\Re(\nabla L \nabla \bar L^{T})\Omega G \delta z]W.
\end{multline}

On the other hand, taking the time derivative of \cref{eq:ansatz} we find
\begin{equation} \label{eq:LindLHS}
	\begin{aligned}
	 \frac{dW}{dt}&= \left(\frac12\Tr(G^{-1}\dot G)+\frac{2}{\hbar}\dot{\tilde{z}}\cdot G \delta z-\frac{1}{\hbar}\delta z \cdot \dot{G}\delta z \right)W.
	\end{aligned}
\end{equation}
Comparing coefficients of  $\delta z$ in (\ref{eq:CenterWignerDyn}) and (\ref{eq:LindLHS}) we obtain the dynamical equations 
\begin{equation} 
    \begin{aligned} \label{eq:LindSemi}
    \frac{d \tilde z}{d t} &= \Omega \nabla H+\Omega \text{Im} (L \nabla \bar{L})\\
  	\frac{dG}{dt} &=\left(H''+\text{Im}(\nabla L \nabla \bar{L}^{T})\right)\Omega G-G\Omega\left(H''-\text{Im}(\nabla L \nabla \bar{L}^{T})\right)\\
	&\quad +2 G \Omega \text{Re}(\nabla L \nabla \bar{L}^T) \Omega G
    \end{aligned}
\end{equation}
for the dynamics of the Gaussian parameters. The first order differential equation describing the central motion is linear and can be trivially integrated. The dynamical equation for the covariance matrix G decouples from the central motion as the Hamiltonian and Lindbladian dependent terms become constant. The general solution can be obtained as (see \cite{plastow2020semiclassical} and references therein)
\begin{equation}
G(t)=\left[2D(t)+(R^T(-t)G^{-1}(0)R(-t))^{-1}\right]^{-1},
\end{equation}
where 
\begin{equation}
    R(t)={\rm e}^{\left(\Omega H''+\text{Im}(\nabla L \nabla \bar{L}^{T})\Omega\right)t}
\end{equation} 
and
\begin{equation}
    D(t)=\int_0^t R(s) \text{Re}(\nabla L \nabla \bar{L}^T) R^T(s){\rm d}s.
\end{equation}

For Hermitian Lindbladians, the Lindblad term in the central dynamics vanishes leaving only Hamiltonian dynamics, as expected for these \textit{purely decohering} systems. In general non-Hermitian Lindbladians lead to both decoherence (that may be characterised by the evolution of $G$ in our case) and dissipation that leads to non-Hamiltonian dynamics of the centre of the Gaussian. 

\subsection{Gaussian Stochastic Schr\"odinger dynamics} \label{sec:GaussSSE}
Let us now use the same approach to derive parameter dynamics for the SSE. We begin by writing the SSE in projector form as
\begin{align} \label{eq:SSEProj}
	d(\ket{\psi}\bra{\psi})&=\ket{d \psi}\bra{\psi}+\ket{\psi}\bra{d\psi}+\ket{d \psi}\bra{d \psi} \nonumber \\ 
	&=\frac{1}{\hbar}(-{i}\hat{H} -\frac{1}{2 }\hat{L}^{\dagger}\hat{L}+\braket{\hat{L}^{\dagger}}\hat{L} - \frac{1}{2}\braket{\hat{L}^{\dagger}}\braket{\hat{L}})\ket {\psi} \bra{\psi}dt\\
		&\quad+\frac{1}{\sqrt{2 \hbar}}(\hat{L} \hat{\rho} +\hat{\rho} \hat{L}^{\dagger}-\braket{\hat{L}+\hat{L}^{\dagger}}\rho)d\xi_R+\frac{i}{\sqrt{2 \hbar}}(\hat{L} \hat{\rho} -\hat{\rho} \hat{L}^{\dagger}-\braket{\hat{L}-\hat{L}^{\dagger}}\rho)d\xi_I. \nonumber
\end{align}
The deterministic ($dt$) part of equation (\ref{eq:SSEProj}) is the same as that of the Lindblad \cref{eq:Wignerdyn} which can evolve pure states into mixed states. In the SSE case, however, the state remains pure for all times, since the stochastic terms conspire to conserve the purity of the state.
 
We translate \cref{eq:SSEProj} into the Wigner-Weyl representation to obtain
\begin{multline} \label{eq:MoyalSSE}
	dW=-\frac{i}{\hbar}[H \star W-W\star H+ i ( L\star W \star \bar{L}-\tfrac{1}{2}\bar{L}\star L\star W -\tfrac12W \star \bar{L}\star L ) ]dt \\
		+ \tfrac{1}{\sqrt{2\hbar}}[L\!\star\! W\!+W\!\star\! \bar{L}-\!\braket{\hat{L}+\hat{L}^{\dagger}}W]d\xi_R+\tfrac{ i}{\sqrt{2\hbar}}[L \star W\!-W \star\bar{L}-\!\braket{\hat{L}-\hat{L}^{\dagger}}W]d\xi_I. 
\end{multline}
Since the deterministic part is the same as that of the Lindblad system \cref{eq:Wignerdyn}, we need only calculate the stochastic terms. We have
\begin{equation} \label{eq:dxiR}
	 \frac{1}{\sqrt{2\hbar }}(L\star W+ W\star \bar{L}-\braket{\hat{L}+\hat{L}^{\dagger}}W)= \sqrt{\frac{2}{\hbar}}(L^{R}-\braket{L^R}) W+\sqrt{\frac{\hbar}{2}}\{W,L^I\},
\end{equation}
and
\begin{equation} \label{eq:dxiI}
	\frac{i}{\sqrt{2\hbar }}(L\star W-W\star\bar{L} -\braket{\hat{L}-\hat{L}^{\dagger}}W)=-\sqrt{\frac{2}{ \hbar}}(L^{I}-\braket{L^{I}})+\sqrt{\frac{\hbar}{2}}\{W,L^R\},
\end{equation}
where $\{,\}$ denotes the Poisson bracket and $L^R/L^I$ denote the real/imaginary parts of $L$. 
Using the expressions (\ref{eq:dxiR}) and (\ref{eq:dxiI}) equation (\ref{eq:MoyalSSE}) becomes 
\begin{multline}
		 dW = [ \{H,W\}  + \text{Im}(L\{\bar{L},W\}) -i  \{\bar{L},L \}W+\tfrac{\hbar}{2} \text{Re}\{L,\{\bar{L},W\}\} ]dt \\
		 \quad+( \sqrt{\tfrac{2}{\hbar}}(L^{R}-\!\braket{L^R}) W\!+\!\sqrt{\tfrac{\hbar}{2}}\{W,L^I \} )d\xi_R\! -\!(\sqrt{\tfrac{2}{ \hbar}}(L^{I}-\!\braket{L^{I}})-\sqrt{\tfrac{\hbar}{2}}\{W,L^R\})d\xi_I.
\end{multline}
As in the Lindblad case in \cref{sec:GaussLind} we replace the Weyl symbols with their finite Taylor series \cref{eq:Taylor}. 
Using
\begin{align}
	\begin{split}
	\sqrt{\frac{2}{\hbar}}(L^{R}-\braket{L^R}) W+\sqrt{\frac{\hbar}{2}}\{W,L^I \}&= \sqrt{\frac{2}{ \hbar}}(\nabla L^R\cdot \delta z+ \nabla L^I \cdot \Omega G \delta z)W,  \\
	 -\sqrt{\frac{2}{ \hbar}}(L^{I}-\braket{L^{I}})+\sqrt{\frac{\hbar}{2}}\{W,L^R\}&= -\sqrt{\frac{2}{ \hbar}}(\nabla L^I\cdot \delta z- \nabla L^R \cdot \Omega G \delta z)W,
	 \end{split}
\end{align}
we find
\begin{equation}
\begin{aligned} \label{eq:RHS}
		 dW &= \big[\tfrac{2}{\hbar} \left(\delta z \cdot G \Omega(H''+\text{Im}( \nabla \bar{L}\nabla{L}^{T})) \delta z  -\delta z \cdot G \Omega \text{Re}(\nabla \bar{L}  \nabla L^{T}) \Omega G \delta z	 \right.\\
		&\quad\left. -(\nabla H +\text{Im}(L\nabla \bar{L})) \cdot \Omega G \delta z -i \tfrac{\hbar}{2} \nabla \bar{L}\cdot \Omega \nabla L +\tfrac{\hbar}{2} \nabla L \cdot \Omega G\Omega \nabla \bar{L} \right)dt \\
		 &\quad +\sqrt{\tfrac{2}{ \hbar} }( \nabla L^R\cdot \delta z+\nabla L^I \cdot \Omega G \delta z) d\xi_R - \sqrt{\tfrac{2}{ \hbar} } ( \nabla L^I\cdot \delta z-\nabla L^R \cdot \Omega G \delta z) d\xi_I  \big]W. 
\end{aligned}
\end{equation}

On the other hand, using the It\^{o} chain rule \cite{bernt}
\begin{equation} \label{eq:itoChain}
	dW=\frac{\partial W}{\partial y_i}dy_i+\frac12\frac{\partial^2 W}{\partial y_i \partial y_j}dy_idy_j,
\end{equation}
where 
\begin{equation}
dy_j=\mu_j dt +\sigma_j^R d\xi_R +\sigma_j^I d\xi_I,
\end{equation}
we have
\begin{equation} 
\label{eqn:dWSSEIto}
	\begin{aligned}
	dW&=\frac{\partial W}{\partial \tilde{z}_k} d\tilde{z}_k+\frac{\partial W}{\partial A_{kl}} dA_{kl} \\
	&\quad+\frac{\partial^2 W}{\partial \tilde{z}_k\partial A_{mn}} d\tilde{z}_k dA_{mn}+\frac12\frac{\partial^2 W}{\partial \tilde{z}_k \partial \tilde{z}_l} d\tilde{z}_kd\tilde{z}_l+\frac12\frac{\partial^2 W}{\partial A_{kl}\partial A_{mn}} dA_{kl}dA_{mn}.
	\end{aligned}
\end{equation}
Here we have introduced the matrix $A$ with $G=\frac12(A+A^T)$, to circumvent complications arising from the symmetry condition on $G$. Here and for the rest of this section, we implicitly sum over repeated indices (Einstein summation convention). 

As we shall see, equation (\ref{eqn:dWSSEIto}) simplifies drastically as many of the terms vanish. In particular, we find that 
\begin{equation}
\label{eq:sigmaA}
	\begin{aligned}
	\sigma_A^R=0=\sigma_A^I,
	\end{aligned}
\end{equation}
and thus the second order derivatives evolving $A_{jk}$ in equation (\ref{eqn:dWSSEIto}) vanish. This can be seen as follows. Let us focus our attention on the term $\frac12\frac{\partial^2 W}{\partial A_{kl}\partial A_{mn}} dA_{kl}dA_{mn}$. This term would generate a term $O(\delta z^4)$
\begin{multline}
	\frac12\frac{\partial^2 W}{\partial A_{kl}\partial A_{mn}} dA_{kl}dA_{mn} =\frac{1}{2\hbar^2}((\delta z \cdot \sigma_A^R \delta z)^2+(\delta z \cdot \sigma_A^I \delta z)^2)dt \\+\text{terms of lower order in $\delta z$},
\end{multline}	
which would not be matched by any term on the right hand side of equation (\ref{eq:RHS}), from which we infer (\ref{eq:sigmaA}). Hence equation (\ref{eqn:dWSSEIto}) simplifies to 
\begin{equation} 
	\begin{aligned}
	dW=\frac{\partial W}{\partial \tilde{z}_k} d\tilde{z}_k+\frac{\partial W}{\partial A_{kl}} dA_{kl} +\frac12\frac{\partial^2 W}{\partial \tilde{z}_k \partial \tilde{z}_l} d\tilde{z}_kd\tilde{z}_k,
	\end{aligned}
\end{equation}
and we have explicitly
\begin{multline} \label{eq:LHS}
	dW = [\frac{2}{\hbar}(\mu_z dt +\sigma_z^R d\xi_R+\sigma_z^I d\xi_I) \cdot G \delta z +\frac{1}{2 } \text{Tr}( G^{-1} \mu_A)dt  -\frac{1}{\hbar} \delta z \cdot \mu_A \delta z dt \\
	-\frac{1}{\hbar}(\sigma_z^R \cdot G \sigma_z^R+\sigma_z^I \cdot G \sigma_z^I) dt+\frac{2}{\hbar^2}(\delta z \cdot G \sigma_z^R\sigma_z^{R^T}G \delta z+\delta z \cdot G \sigma_z^I\sigma_z^{I^T}G \delta z)dt ]W	.
\end{multline}
Using $dG=\frac12(dA+dA^T)$ and equating eqs (\ref{eq:LHS}) and (\ref{eq:RHS}) we obtain the (stochastic) dynamical equations
\begin{align} 
	\nonumber d\tilde z &= \left(\Omega \nabla H+\Omega \text{Im}(L \nabla \bar{L})\right)dt\\ 
	&\quad +\sqrt{\frac{\hbar}{2}}( G^{-1} \nabla L^R- \Omega \nabla L^I) d\xi_R-\sqrt{\frac{\hbar}{2}}( G^{-1}\nabla L^I+ \Omega \nabla L^R)d\xi_I\label{eq:SSEParamz}\\
	\frac{d G}{dt}&=-G\Omega H''+H''\Omega G+\text{Re}(\nabla \bar L \nabla L^T)+G \Omega \text{Re}(\nabla \bar{L} \nabla L^T) \Omega G \label{eq:SSEParamG}
\end{align}
for the Gaussian parameters.
 
We notice that the deterministic part of the dynamics of the centre $\tilde z$ is the same as that of the Lindblad equation, however, the SSE dynamics have an additional stochastic component as expected. This stochastic component contains covariance dependent terms and unlike in the Lindblad case, we can no longer simulate the centre trajectories without calculating the covariance dynamics. For quadratic systems the evolution of the covariance matrix $G$ is deterministic and independent of the motion of the centre but different from that of the Lindblad evolution. This difference is not surprising taking into account that the $G$ matrix in the SSE describes the covariances of the individual pure-state trajectories, while the $G$ matrix of the Lindblad evolution describes that of the total density matrix arising from the ensemble average. In fact, the dynamics of $G$ for the SSE are the same as those arising from deterministic non-Hermitian Hamiltonian dynamics, which we shall briefly review in the next section as the first phase of quantum-jump dynamics. 
The dynamical equation (\ref{eq:SSEParamG}) for $G$ can be explicitly solved by \cite{graefe2012complexified}
\begin{equation} \label{eq:GtS}
        G(t)=\left(-\Omega \Re(S(t)\Omega G(0) +\Omega \Im(S(t))\right)\left(\Im(S(t))\Omega G(0)+\Re (S(t)) \right)^{-1},
\end{equation}
where $S(t)$ is given by 
\begin{equation}
  S(t)=e^{\Omega K''t},
\end{equation}
With $K''=H''- i \Re(\nabla \bar{L} \nabla L^T)$. A short calculation also confirms that $\frac{d}{dt}\det G=0$, implying that the Gaussian state remains pure as expected. Having obtained the evolution equations for Gaussian dynamics according to the full Lindblad and SSE dynamics, we shall now turn towards the corresponding quantum-jump dynamics.

\section{Quantum-Jump dynamics} \label{sec:HagJumps}

 Quantum-jump dynamics do not preserve Gaussian states for arbitrary linear Lindbladians. As an example, consider the Lindblad operator $\hat L=\hat a^{\dagger}$. The first jump maps a state $\ket{\psi_g}$ to $ \hat a^{\dagger}\ket{\psi_g}$ and thus transforms a Gaussian state into a non-Gaussian one. We will show in what follows, that it is still possible to calculate the quantum-jump dynamics building on the propagation of Gaussian states, leading again to just a handful of time-dependent parameters. For this purpose, we adapt a method introduced for non-Hermitian dynamics in \cite{lasser2018non}. 
 
 In the quantum jump unravelling of the Lindblad dynamics, we propagate the initial state $\ket{\psi_0}$ with the time evolution generated by the non-Hermitian Hamiltonian 
$\hat H-\frac{i}{2}\hat L^{\dagger}\hat L$, and  intersperse it  with jumps at discrete times $t_j$, 
\begin{equation}
\ket{\psi}\mapsto\frac{\hat L\ket{\psi}}{\sqrt{\braket{\psi|\hat{L}^{\dagger}\hat{L}|\psi}}}\,\, . 
\end{equation}
The non-Hermitian nature of time evolution $U(t)=e^{-\frac{i}{\hbar}(\hat H-\frac{i}{2}\hat L^{\dagger}\hat L)}$ will lead to a decreasing norm of the propagated state, and the quantum jumps reset the norm to $1$ due to the inclusion of the denominator. 
After $k$ quantum jumps at times $t_1,t_2, \cdots, t_k$, i.e, for $t\in [t_k, t_{k+1})$,  the state will be of the form 
\begin{equation}\label{eq:jump-state-k}
\ket{\psi(t)}=\frac{1}{\lVert \psi(t_k)\lVert}U(t-t_k)LU(t_k-t_{k-1})LU(t_{k-1}-t_{k-2})L\cdots U(t_2-t_1)LU(t_1)\ket{\psi_0}
\end{equation}
To compute this expression we will introduce a basis which is moving with the state $\hat U(t)\psi_0$, the so called Hagedorn basis, which we will recall in the next subsection. We then show how to compute $\hat U(t)$ in the moving bases for the case that the non-Hermitian Hamiltonian $\hat H-\frac{i}{2}\hat L^{\dagger}\hat L$ is of no higher than quadratic order in $\hat p$ and $\hat q$. Combining these results will allow us to evaluate \eqref{eq:jump-state-k} explicitly by an iterative algorithm.  

\subsection{Hagedorn Basis}

It is well known that a coherent state creates an 
orthonormal basis by applying powers of creation operators to it. In \cite{hagedorn1998raising} Hagedorn introduced a parametrization of coherent states and their associated raising and lowering operators which is particularly well adapted to the study of the time evolution of wavepackets. This has been used in numerical analysis of the time dependent Schr{\"o}dinger equation in \cite{lubich08,lasser20} and it has been adapted to non-Hermitian evolution in \cite{lasser2018non}. We will now recall some of the notions we use in the following, for simplicity we will restrict ourselves to one degree of freedom systems. 

A coherent state centred at the origin is characterised by its annihilation operator which is defined in terms of a complex vector $a\in \mathbb{C}^2$ as
\begin{equation}
\hat A(a):=\frac{i}{\sqrt{2\hbar}} a\cdot \Omega \hat z\,\, .
\end{equation}
The corresponding coherent state $\ket{0,a}$ is defined up to normalisation by $\hat A(a)\ket{0,a}=0$, and one can show that in position representation it is given by 
\begin{equation}\label{eq:ground-state-0}
\psi(a,x)=(\pi\hbar)^{-1/4}(a_q)^{-1/2}e^{\frac{i}{2\hbar} \frac{a_p}{a_q} x^2}\,\, , \quad \text{where}\quad a=(a_q,a_p)\,\, .
\end{equation}
An important role is played by the Hermitian form defined as 
\begin{equation}\label{eq:def-h}
h_{\Omega}(a,b):=\frac{1}{2i} a^{\dagger}\Omega b\,\, .
\end{equation}
The state \eqref{eq:ground-state-0} is normalizable  if $h_{\Omega}(a,a)>0$ and has norm $1$  if $h_{\Omega}(a,a)=1$. Notice as well that $h_{\Omega}( \bar{a},a)=0$ and $h_{\Omega}( \bar{a},\bar{a})=-h_{\Omega}(a,a)$, and with these relations it follows that if $h_{\Omega}(a,a)=1$ then we have for any complex vector $b$ the expansion
\begin{equation}\label{eq:expansion-h}
b=h_{\Omega}(a,b)a-h_{\Omega}( \bar{a}, b) \bar{a}\,\, .
\end{equation}
That is, $a, \bar{a}$ form a basis of $\mathbb{C}^2$, similar to an orthonormal basis.  

The creation operator is the adjoint of $\hat A(a)$ and we have 
\begin{equation} \label{eq:ADagHag}
\hat A^{\dagger}(a)=-\hat A(\bar{a})\,\, ,
\end{equation}
and the corresponding orthonormal basis is defined  as 
\begin{equation}
\ket{n,a}:=\frac{1}{\sqrt{n!}}\big[\hat A^{\dagger}(a)\big]^n\ket{ 0,a}\,\, , \quad n=0,1,2,3,\cdots \,\, .
\end{equation}
We can move this basis by applying  the phase space translation operators $\hat T(z):=e^{-\frac{i}{\hbar} z\cdot \Omega \hat z}$, $z\in \mathbb{R}^2$, to the  creation and annihilation operators, $\hat A(a,z):=\hat T(z)\hat A(a)\hat T^{\dagger}(z)$, $\hat A^{\dagger}(a,z):=\hat T(z)\hat A^{\dagger}(a)\hat T^{\dagger}(z)$ and to the basis states
\begin{equation}\label{def:basis-shifted}
\ket{n,a,z}:=\hat T(z)\ket{n,a}=\frac{1}{\sqrt{n!}}\big[\hat A^{\dagger}(a,z)\big]^n\ket{ 0,a,z}\,\, .
\end{equation}
We have as well the explicit representation 
\begin{equation}\label{eq:shifted-A}
\hat A(a,z)=\frac{i}{\sqrt{2\hbar}} a\cdot \Omega (\hat z-z) \,\, , \quad \hat A^{\dagger}(a,z)=-\hat A(\bar{a},z),
,\end{equation}
and we note that 
\begin{equation}\label{eq:actA}
\hat A(a,z)\ket{n,a,z}=\sqrt{n}\ket{n-1,a,z}\,\, \quad \hat A^{\dagger}(a,z)\ket{n,a,z}=\sqrt{n+1}\ket{n+1,a,z}\,\, .
\end{equation}
In the following we will encounter operators of the form $\hat A(b,\chi)=\frac{i}{\sqrt{2\hbar}} b\cdot \Omega (\hat z-\chi)$ for some $b, \chi\in \mathbb{C}^2$. It is convenient to write them as linear combinations of $\hat A(a,z)$ and $\hat A^{\dagger}(a,z)$ to determine their action on the basis vectors  $\ket{n,a,z}$. To that end, we note that
\begin{equation}
\hat A(b,\chi)-\hat A(b,z)=\frac{i}{\sqrt{2\hbar}} b\cdot \Omega (z-\chi)=\sqrt{2/\hbar} \, h_{\Omega}( \bar z-\bar \chi, b)\,\, , 
\end{equation}
and with the expansion \eqref{eq:expansion-h} and \eqref{eq:shifted-A} we then find
\begin{equation}\label{eq:expansion-A}
\hat A(b,\chi)=h_{\Omega}(a,b) \hat A(a,z)+h_{\Omega}( \bar{a},b)\hat A^{\dagger}(a,z)+\sqrt{2/\hbar} \, h_{\Omega}( \bar z- \bar \chi, b)\,\, .
\end{equation}

\subsection{Non-Hermitian evolution}
 
 The propagation of a coherent state under a time evolution generated by a non-Hermitian operator has been studied in \cite{graefe2011wave} and the special case of quadratic Hamiltonians was analysed in \cite{graefe2012complexified} and in the context of Hagedorn wave-packets in \cite{lasser2018non}. In order to apply these results we need first to rewrite the Hamiltonian $\hat H -\frac{i}{2} \hat L^{\dagger}\hat L$ slightly; since we assume that $L$ is linear, we have $(\bar{L}\star L)(z)=\bar{L}(z)L(z)+\frac{i\hbar}{2} \{\bar{L},L\}$ and the term $\{\bar{L},L\}=\nabla \bar{L}\cdot \Omega\nabla L$ is constant. This gives us $\hat L^{\dagger}\hat L=\widehat {\bar{L}L}+\frac{i\hbar }{2} \{\bar{L},L\}$ and so we obtain
\begin{equation}\label{eq:split_sub-part}
\hat U(t)=e^{-\frac{i}{\hbar} ( \hat H -\frac{i}{2} \hat L^{\dagger}\hat L)t}=e^{-\frac{i}{4 }\nabla \bar{L}\cdot \Omega\nabla Lt}e^{-\frac{i}{\hbar} \hat K t} \quad \text{where}\quad K(z)=H(z)-\frac{i}{2}|L(z)|^2\,\, .
\end{equation}

In \cite{graefe2011wave}, the dynamics of a Gaussian wavepacket under a non-Hermitian Hamiltonian were derived following a similar procedure to the one we have outlined for the Lindblad and SSE cases. Substituting the effective non-Hermitian Hamiltonian $K(z)$ \eqref{eq:split_sub-part} into the results from \cite{graefe2011wave} yields the parameter dynamics 
\begin{eqnarray} 
\label{eq:NHVNsemiz}
    \frac{d \tilde z}{d t} &= \Omega \nabla H-G^{-1} \text{Re}(\bar L \nabla L) \\
\label{eq:NHVNsemiG}	\frac{d G}{dt} &=-G\Omega H''+ H'' \Omega G+\text{Re}(\nabla \bar L  \nabla L^T ) +G \Omega \text{Re}(\nabla \bar L  \nabla L^T ) \Omega G.
\end{eqnarray}
Where the evolution equation for $G$ is the same as the one for the SSE case. 
As expected this fulfils $\frac{\rmd}{\rmd t}\det G=0$ for $\det G=1$, and an initially pure state remains pure.

As has been discussed in \cite{graefebradley} the dissipative part of the central motion of the non-Hermitian dynamics can appear either quite different or very similar to that of the Lindblad case, depending on the structure of the Lindblad operator. For a Lindblad operator that is an analytic function of $\hat a$ or $\hat a^\dagger$, for example, the dissipative term in the central dynamics in equation (\ref{eq:LindSemi}), given by $\Omega \text{Im} (L \nabla \bar{L})$ can be rewritten as $-\text{Re}(\bar L\nabla L)$, which is very similar to the non-Hermitian dissipation, with the difference that the latter is modulated by the changing covariance metric $G$. An example for which Lindblad and non-Hermitian central dynamics are very different, are Hermitian Lindbladians, for which the dissipative term in the Lindblad dynamics vanishes entirely. The quantum-jump evolution turns the non-Hermitian behaviour into the Lindblad one, by averaging over different quantum jumps, that in general do not leave an initially Gaussian state Gaussian.

In practice it is often useful to use instead of \eqref{eq:NHVNsemiz} and \eqref{eq:NHVNsemiG} the complex classical dynamics created by $K(z)$. If we write  $K(z)=\frac{1}{2}z\cdot K_2z+k_1\cdot \Omega z+k_0$, where $K_2$ is a symmetric complex matrix, $k_1$ is a complex vector, and $k_0$ is a constant, then the corresponding solution to Hamilton's equations in phase space is given by 
\begin{equation}
    \Phi(t,z)=S(t)z+v(t)\,\, ,
\end{equation}
 with 
 \begin{equation}
 S(t)=e^{t\Omega K_2}\,\, \quad\text{and}\quad v(t)=\int_0^tS(t-s) k_1\, d s\,\, ,    
\end{equation}
where  $S(t)$ is complex and symplectic, i.e., $S^T\Omega S=\Omega$. 
It turns out that $\hat U(t)\ket{0,a_0,z_0}$ can be described entirely in term of $\Phi(t), S(t)$ and its action on $a_0$ and $z_0$, \cite{graefe2012complexified,lasser2018non}, to that end  let us first define 
\begin{equation} \label{eq:NDef}
N(t):=\frac{1}{\sqrt{h_{\Omega}(S(t)a_0,S(t)a_0)}} \quad \text{and}\quad a_t:=N(t) S(t) a_0\,\, , 
\end{equation}
so that $a_t$ is normalised again. Now we introduce $J_t:=-\Re(a_ta_t^{\dagger})\Omega$, where $a_t^{\dagger}$ denotes the transposed and the complex conjugate so that $a_t a_t^{\dagger}$ is a $2\times 2$ matrix, then the solution to \eqref{eq:NHVNsemiz} is given by
\begin{equation}
z_t:=\Re \Phi(t,z)+J\Im \Phi(t,z)\in \mathbb{R}^2\,\, ,
\end{equation}
and this vector is real-valued, in contrast to the complex centre $z_t^{(\mathbb{C})}=\Phi(t,z)$. 
There is as well a corresponding expression for $G$ in \eqref{eq:NHVNsemiG} 
\begin{equation} \label{eq:GSol}
    G(t)=\Omega J_t=\Omega^T\Re(a_ta_t^{\dagger})\Omega\,\, .
\end{equation}

 Using these definitions we can write the propagated coherent state as
\begin{equation}
\hat U(t)\ket{0,a_0,z_0}=e^{\frac{i}{\hbar}\alpha(t)}\sqrt{N(t)} \ket{0,a_t,z_t}
\end{equation}
where
\begin{equation}
\alpha(t)=\int_0^t \dot q_sp_s-\frac{1}{2}z_sKz_s\,{\rm d} s-\frac{\hbar}{4 }\nabla \bar{L}\cdot \Omega\nabla Lt
\end{equation}
with $z_s=(q_s,p_s)$ and we have incorporated the factor containing $\nabla \bar{L}\cdot \Omega\nabla L$ from \eqref{eq:split_sub-part} into $\alpha$. The state $\ket{0,a_t,z_t}$ is now normalised, so all the information about the decay of the norm of $\hat U(t)\ket{0,a,z}$ is contained in 
\begin{equation}
\sqrt{N(t)}|e^{\frac{i}{\hbar}\alpha(t)}|=\sqrt{N(t)}e^{-\frac{1}{2\hbar}\int_0^t|L(z_s)|^2{\rm d} s+\frac14\nabla \bar{L}\cdot \Omega\nabla Lt}\,\, .
\end{equation}
In the following we will use the moving basis $\{\ket{n,a_t,z_t}\, :\, n\in \mathbb{N}_0\}$ associated with $a_t$ and $z_t$ by \eqref{def:basis-shifted}, which is an orthonormal basis centred around the moving state $U(t)\ket{0,a_0,z_0}$. 

We want to emphasise that if $\hat U(t)$ is non-unitary then $\hat U(t)\ket{n,a_0,z_0}$ is in general not proportional to $\ket{n,a_t,z_t}$ but will acquire contributions from lower order excited states. In order to compute the expansion of $\hat U(t)\ket{n,a_0,z_0}$ into the basis $\ket{n,a_t,z_t}$ at $t$ we use \eqref{eq:transported-annihilation} from \ref{app:A}, $\hat U(t)\hat A^{\dagger}(a_0,z_0)\hat U(-t)=-\hat A(S(t)\bar a_0,S(t)z_0)$
\begin{equation} \label{eq:HagProp}
\begin{split}
\hat U(t)\ket{n,a_0,z_0)}&=\frac{1}{\sqrt{n!}} \hat U(t)\big[\hat A^{\dagger}(a_0,z_0)\big]^n\ket{0,a_0,z_0}\\
&=\frac{1}{\sqrt{n!}} \big[-\hat A(S(t)\bar a_0,S(t)z_0)\big]^n \hat U(t)\ket{0,a_0,z_0}\,\, .
\end{split}
\end{equation}
Now we expand $\hat A(S(t)\bar a_0,S(t)z_0)$ in terms of the annihilation and creation operators of our moving basis at time $t$ using \eqref{eq:expansion-A} which gives
\begin{equation}\label{eq:expansion-A-t}
-\hat A(S(t)\bar a_0,S(t)z_0)=h_+ \hat A^{\dagger}(a_t,z_t)+h_-\hat A(a_t, z_t)+h_0
\end{equation}
where $h_+=-h_{\Omega}( \bar{a}_t,S(t)\bar{a}_0)$, $h_-=-h_{\Omega}(a_t,S(t) \bar{a}_0)$ and $h_0=-\sqrt{2/\hbar} h_{\Omega}(z_t- \bar{S}(t)z_0,S(t)\bar{a}_0)$. 
It will be useful to introduce 
\begin{equation}\label{eq:HagM}
    M(t):=\frac{h_{\Omega}(S(t)a_0,S(t)\bar{a}_0)}{h_{\Omega}(S(t)a_0,S(t)a_0)}
\end{equation}
and then we have $h_+=N(t)$,  $h_-=-M(t)/N(t)$ and  
\begin{equation}\label{eq:h0-m-N} 
    h_0(t)=\sqrt{\frac{2}{\hbar}}\ 
\big[h_{\Omega}(a_0,z_0)(N(t)^2-1)+h_{\Omega}(\bar a_0,z_0)M(t)\big]\,\, ,
\end{equation}
valid if $h_{\Omega}(a_t,a_t)=1$.

In \eqref{eq:transported-annihilation} in  \ref{app:B} we show that 
\begin{equation}\label{eq:binom-exp}
\frac{1}{\sqrt{n!}} \big[h_+ \hat A^{\dagger}(a_t,z_t)+h_-\hat A(a_t, z_t)+h_0\big]^n\ket{0,a_t,z_t}=
\sum_{m=0}^n  B_{nm}(t)\ket{m,a_t,z_t}
\end{equation}
with 
\begin{equation}\label{eq:matrix-elements-A} 
 B_{nm}(t)=N(t)^m\sum_{k=0}^{[\frac{n-m}{2}]}\sqrt{\frac{n!}{m!}} \frac{(-M(t))^{k} h_0^{n-m-2k}}{2^k(n-m-2k)!k!} \,\, .  
\end{equation}
where $[\frac{n-m}{2}]$ denotes the floor of $\frac{n-m}{2}$, and hence we get an explicit expression for the matrix elements
\begin{equation}\label{eq:matrix-elements-U}
\bra{m,a_t,z_t} \hat U(t) \ket{n,a_0,z_0}=\begin{cases} 0 & m>n\\
e^{\frac{i}{\hbar}\alpha(t)}\sqrt{N(t)}\, B_{mn}(t) & m\leq n \end{cases}\,\, .
\end{equation}
For the first few states, we find explicitly 
\begin{equation}
\hat U(t)\ket{1,a_0,z_0}=e^{\frac{i}{\hbar}\alpha(t)}\sqrt{N(t)}\big[N(t)\ket{1,a_t,z_t}+h_0\ket{0,a_t,z_t}\big]\,\, ,
\end{equation}
\begin{equation}
\begin{split}
\hat U(t)\ket{2,a_0,z_0}&=e^{\frac{i}{\hbar}\alpha(t)}\sqrt{N(t)}\bigg[N(t)^2\ket{2,a_t,z_t}+\sqrt{2}\, h_0N(t) \ket{1,a_t,z_t}\\
&\hspace*{5cm}+\frac{1}{\sqrt{2}}\big(h_0^2-2M(t)\big)\ket{0,a_t,z_t}\bigg]\,\, .
\end{split}
\end{equation}

We can generalise \eqref{eq:matrix-elements-U} to the situation that the initial state is at a time $t_0\neq 0$. To that end, it is useful to introduce a notation for $N(t)$ and $M(t)$ which makes the dependence on the vector $a$ explicit, 
\begin{equation}
    N(t,a_0):=\big[h_{\Omega}(S(t)a_0,S(t)a_0)\big]^{-1/2}\,\, ,\quad M(t,a_0):=N^2(t,a)h_{\Omega}(S(t)a_0,S(t)\bar a_0)
\end{equation}
and we assume that $a$ satisfies $h_{\Omega}(a,a)=1$. It is not hard to show that $N(t,a)$ satisfies 
\begin{equation}
    N(t_2-t_0,a_{t_0})=N(t_2-t_1,a_{t_1})N(t_1-t_0,a_{t_0})
\end{equation}
for $t_2\geq t_1\geq t_0\geq 0$ and where $a_t:=N(t,a_0)S(t)a_0$. We then use \eqref{eq:h0-m-N} to extend this notation to $h_0$, 
\begin{equation}
    h_0(t,a_0,z_0):=\sqrt{\frac{2}{\hbar}}\ 
\big[h_{\Omega}(a_0,z_0))(N(t,a_0)^2-1)+h_{\Omega}(\bar a_0,z_0)M(t,a_0)\big]\,\, .
\end{equation}
With these notations we have for  $\mathbf{U}_{mn}(t_2,t_1):=\bra{m,a_{t_2},z_{t_2}}\hat U(t)\ket{n,a_{t_1},z_{t_1}}$
\begin{equation}\label{eq:matrix-elements-U-t}
    \mathbf{U}_{mn}(t_2,t_1)=\begin{cases} 0 & m>n\\
e^{\frac{i}{\hbar}[\alpha(t_2)-\alpha(t_1)]}\sqrt{N(t_2-t_1,a_{t_1})}\, B_{mn}(t_2-t_1,a_{t_1},z_{t_1}) & m\leq n \end{cases}\,\, ,
\end{equation}
where 
\begin{equation}\label{eq:matrix-elements-A-t} 
 B_{mn}(t,a_0,z_0):=N(t,a_0)^m\sum_{k=0}^{[\frac{n-m}{2}]}\sqrt{\frac{n!}{m!}} \frac{(-M(t,a_0))^{k} h_0(t,a_0,z_0)^{n-m-2k}}{2^k(n-m-2k)!k!} \,\, .  
\end{equation}
\subsection{Quantum jumps in a Hagedorn basis}

To implement the quantum jumps generated by the Lindblad operator $\hat L$ at time $t$ we need to represent  $\hat L$ in the basis at time $t$. If we parametrize $\hat L$ as
 $\hat L=\frac{i}{\sqrt{2 \hbar}}l\cdot\Omega(\hat z-\chi)=\hat A(l,\chi)$ then \eqref{eq:expansion-A} gives immediately 
\begin{equation} \label{eq:Lhat-Decomp}
    \hat L=h_{\Omega}(a_t,l)\hat A(a_t,z_t)+h_{\Omega}(\bar a_t,l)\hat A^{\dagger}(a_t,z_t)
+\sqrt{2/\hbar}\, h_{\Omega}(z_t-\bar \chi, l)\, ,
\end{equation}
hence the corresponding matrix of $\hat L$ in the basis at time $t$ is given by
\begin{equation} \label{eq:LOrtho}
    \mathbf{L}_{nm}(t)=h_{\Omega}(a_t,l)\sqrt{n}\, \delta_{n-1,m}+h_{\Omega}(\bar a_t,l)\sqrt{n+1}\delta_{n+1,m}
+\sqrt{2/\hbar}\, h_{\Omega}(z_t-\bar \chi, l)\delta_{n,m}\,\, .
\end{equation}
With the explicit expression \eqref{eq:matrix-elements-U-t} and \eqref{eq:LOrtho} we can now use the following scheme to compute the quantum-jump dynamics. 
\begin{enumerate}
	\item Choose a time interval $[t_0,t_{end}]$ and initial parameters $z_{0}$, $a_{0}$ defining the Hagedorn basis.
	\item Expand the initial state $\ket{\psi_0}$ in the basis $\ket{n, a_0,z_0}$, this defines a normalised vector $\mathbf{c}(t_0)$.
    \item Pick a random number $R$ from the uniform distribution on the interval $[0,1]$
    \item Solve $$\mathbf{c}^{\dagger}(t_0)\mathbf{U}^{\dagger}(t_J,t_0)\mathbf{U}(t_J,t_0)\mathbf{c}(t_0)-R=0$$
    for $t_{J}$
    \item Calculate new coefficients $$\mathbf{c}'(t_J)\to\frac{\mathbf{L}(t_{J}) \mathbf{c}(t_J)}{\sqrt{\mathbf{c}(t_J)^{\dagger}\mathbf{L}^{\dagger}(t_{J}) \mathbf{L}(t_{J})\mathbf{c}(t_J)}}$$
    \item While the chosen end time $t_{end}\geq t_{J}$, set $t_0=t_J$ and repeat steps (iii)-(vi)
    \item Evaluate the end state $$\ket{\psi(t)}=\sum_n c_n\ket{n,a_{t},z_t}$$
    at $t=t_{end}$
\end{enumerate}
The above scheme is especially efficient if the initial state $\ket{\psi_0}$ is a Gaussian as our coefficient vector $\mathbf{c}(t_0)$ will only have one non-zero element. We also have that at each jump only increases the number of non-zero elements of $\mathbf{c}(t)$ by one in the Hagedorn basis, since any linear $\hat L$ may be written as a linear combination of a $\hat A(a_t,z_t)$, $\hat A^{\dagger}(a_t,z_t)$ and the identity by \cref{eq:Lhat-Decomp}. Further efficiency gains over standard methods are made by dynamically changing the basis size and using a root finding algorithm to solve step (v) since we do not need to calculate the propagator at many points. This scheme is most useful for studying the long-term behaviour in systems with long times in between jumps (weakly coupled Lindbladians).

We also will give an alternative scheme that calculates the time evolved states $\hat{U}(t)\ket{n,a_0,z_0}$ at all times and uses them as a non-orthogonal basis. This allows one to calculate all deterministic Hamiltonian evolution a priori greatly reducing the computational cost to calculate repeated trajectories. We start by calculating the norms and overlaps of the time evolved states using \cref{eq:matrix-elements-A-t}
\begin{equation}
\begin{aligned}
    O_{mn}(t)&=\bra{m,a_0,z_0}\hat{U}^{\dagger}(t)\hat{U}(t)\ket{n,a_0,z_0}\\
    &=e^{-\frac{2}{\hbar}\Im(\alpha(t))} N(t)\sum_{i,j=0}B^{\dagger}_{im}(t)\braket{i,a_t,z_t|j,a_t,z_t}B_{nj}(t)\\
    &=e^{-\frac{2}{\hbar}\Im(\alpha(t))} N(t)(\overline{\mathbf{B}}\mathbf{B}^{T})_{mn}(t).
\end{aligned}
\end{equation}
We also find a matrix representation of $\hat L$ in the non-orthogonal basis. We use a formalism found in \cite{brody2013biorthogonal} and find a set of states $\hat{\widetilde {U}}(t)\ket{m,a,z}$ dual to $\hat{U}(t)\ket{n,a_0,z_0}$ such that
\begin{equation} \label{eq:biorthog}
	\bra{m,a_0,z_0}\hat{\widetilde{U}}\,^{\dagger}(t)\hat{U}(t)\ket{n,a_0,z_0}=\delta_{m n}.
\end{equation}
We thus have $\hat{\widetilde{U}}(t)=(\hat{U}^{-1})^{\dagger}(t)$.
\begin{equation}\label{eq:DualBasis}
\hat{\widetilde{U}}(t)\ket{n,a_0,z_0}=
\sum_{m=0}\frac{e^{\frac{i}{\hbar}\bar{\alpha}(t)}\widetilde{B}_{nm}(t)}{\sqrt{N(t)}} \ket{m,a_t,z_t}
\end{equation}
with 
\begin{equation} 
    \widetilde{B}_{nm}(t)=N(t)^{-m}\sum_{k=0}^{[\frac{m-n}{2}]}\sqrt{\frac{m!}{n!}} \frac{(-1)^{m+n}\overline{M}(t)^{k} h_0^{m-n-2k}}{2^k(m-n-2k)!k!}.
\end{equation}
Using this basis the matrix elements of $\hat L$ are given simply by
\begin{equation} \label{eq:LMel}
\begin{aligned}
	\mathbb{L}_{mn}(t)&=\bra{m,a_0,z_0}\hat{\widetilde{U}}\,^{\dagger}(t)\hat L \hat{U}(t)\ket{n,a_0,z_0}\\
	&=\sum_{i,j=0}\widetilde{B}^{\dagger}_{i m}(t)\bra{i,a_t,z_t}\hat L \ket{j,a_t,z_t}B_{nj}(t)\\
	&=\sum_{i,j=0}\widetilde{B}^{\dagger}_{i m}(t)\mathbf{L}_{ij} B_{nj}(t)
\end{aligned}
\end{equation}
Substituting \cref{eq:LOrtho} into the above we obtain
\begin{multline}
	\mathbb{L}_{mn}(t)=\sum_{i,j=0}\overline{\widetilde{B}}_{mi}(t)\big(\sqrt{j+1} h_{\Omega}(\bar{a}_t,l) \delta_{i,j+1}\\+\sqrt{i}h_{\Omega} (a_t, l)\delta_{i,j-1}+\sqrt{2/\hbar}  h_{\Omega}(z_t-\bar{\chi},l)\delta_{i,j}\big)B^T_{j n}(t).
\end{multline}
Using this matrix representation of the Lindblad operator we may calculate the effect of a jump by acting directly on the coefficient vector $\mathbf{c}$. The jump maps 
\begin{equation} \label{eqn:Hagjumpcoeffmap}
\ket{\psi}=\sum_n c_n \hat{U}(t)\ket{n,a_0,z_0}\mapsto \ket{\psi'}=\sum_n  c'_n \hat{U}(t)\ket{n,a_0,z_0}, \quad\text{with}\quad \mathbf{c}'=\frac{\mathbb{L} \mathbf{c}}{\sqrt{\mathbf{c}^{\dagger}\mathbb{L}^{\dagger}\mathbf{O} \mathbb{L} \mathbf{c}}}.
\end{equation}
The coefficients after the jump are then used as the new input coefficients for the next stretch of non-Hermitian evolution until the next jump. This can be implemented using the following algorithm, although we may propagate any arbitrary state initial state some simplifications occur when using an initially Gaussian state. 
\begin{enumerate} 
	\item Discretise the time interval $(t_0,t_N) \rightarrow \{t_k \} $ with $\Delta t=t_{k+1}-t_k$;
	\item Choose $l(t_0)$ and $z_0$ such that the Gaussian we wish to propagate is given by \cref{eq:ground-state-0}. With this choice of parameters we have $c_0(0)=1$ and $c_{k\neq0}(0)=0$
	\item Calculate $B_{kl}(t_n),\widetilde{B}_{kl}(t_n),O_{kl}(t_n)$ and $L_{kl}(t_n)$ for all $t_n$.
	\item Pick a random number $R$ from the uniform distribution on the interval $[0,1]$;
	\item For all $t_k$ with $t_0< t_k\leq t_N$ evaluate the following loop:
    \begin{lstlisting}
    for k=1:N-1
        if $R\geq \textbf{c}(t_{n})^{\dagger}\textbf{O}(t_{n})\textbf{c}(t_{n})$ 
            $\textbf{c}(t_{n+1})=\mathbb{L}(t_{n})\textbf{c}(t_{n})/\sqrt{\textbf{c}^{\dagger}(t_{n})\mathbb{L}^{\dagger}(t_{n})\textbf{O}(t_{n})\mathbb{L}(t_{n})\textbf{c}(t_{n})}$
            $R$=rand 
        else 
            $\textbf{c}(t_{n+1})=\textbf{c}(t_{n})$ 
        end
    end
    \end{lstlisting}
    \item Normalise the trajectory; for all $t_n$ $$\mathbf{c}(t_n)\to \frac{\mathbf{c}(t_n)}{\sqrt{\mathbf{c}^{\dagger}(t_n)\mathbf{O}(t_n)\mathbf{c}(t_n)}}$$.
    \item Construct state trajectory $\ket{\psi(t_n)}=\sum_{k=0} c_k(t_n)\hat{U}(t_n)\ket{k,a_{t_0},z_{t_0}}$ for all $t_n$ 
\end{enumerate}
An advantage of using a Hagedorn basis to simulate quantum jumps over the standard method described in \cref{sec:unravell} is that a much smaller basis size may be used to accurately propagate an initially Gaussian state as each jump only increases the number of non-zero elements of the state vector $\mathbf{c}$ by one. In addition one can see from \cref{eq:matrix-elements-A} that the state $\hat{U}(t)\ket{n,a_0,z_0}$ is damped by a factor $N(t)^n$ and as $N(t)\leq1$ and strictly decreasing, the higher states have diminishing contribution to the overall dynamics. For a single trajectory, it may still be more efficient to avoid the extra effort of propagating the whole basis set in time and calculating the overlap matrices, however, this may be compensated for if the scheme is used to generate a large number of quantum-jump trajectories since the time-dependent basis is the same for every realisation. In the examples discussed, we will use the above algorithm to implement quantum-jump trajectories numerically as we wish to obtain observable expectation values over the full trajectory.

 In summary, the quantum-jump dynamics for a quadratic Hamiltonian and linear Lindbladian while in general non-Gaussian, can be understood and simulated almost entirely on the grounds of the single dynamical quantity $S(t)$ describing the linearised complexified flow intercepted with discrete quantum jumps. 
In what follows we shall explore the resulting dynamics for two instructive examples.

\section{Examples} \label{sec:examples}
To illustrate the results above, let us consider the Lindblad dynamics and the two unravellings for a harmonic oscillator Hamiltonian $\hat H = \frac{\omega}{2} \left(\hat p^2 +\hat x^2\right)$ with two different Lindblad operators, one Hermitian and one non-Hermitian. 
\FloatBarrier
\subsection{Example 1: Position measurement} 
Let us first consider a quantum harmonic oscillator with a Hermitian Lindblad operator
\begin{equation} \label{eq:PosModel}
   \hat L=\sqrt{\gamma} \hat x,
\end{equation}
which can be thought of as modelling a position measurement. Since the Lindbladian is Hermitian (purely decohering), in the Lindblad dynamics it yields no contribution to the dynamics of the expectation values $z_t$, which simply follow the familiar harmonic oscillator trajectories. This is different for individual SSE and quantum-jump trajectories.

\begin{figure}
  \centering
 	  \includegraphics[width=0.4\textwidth]{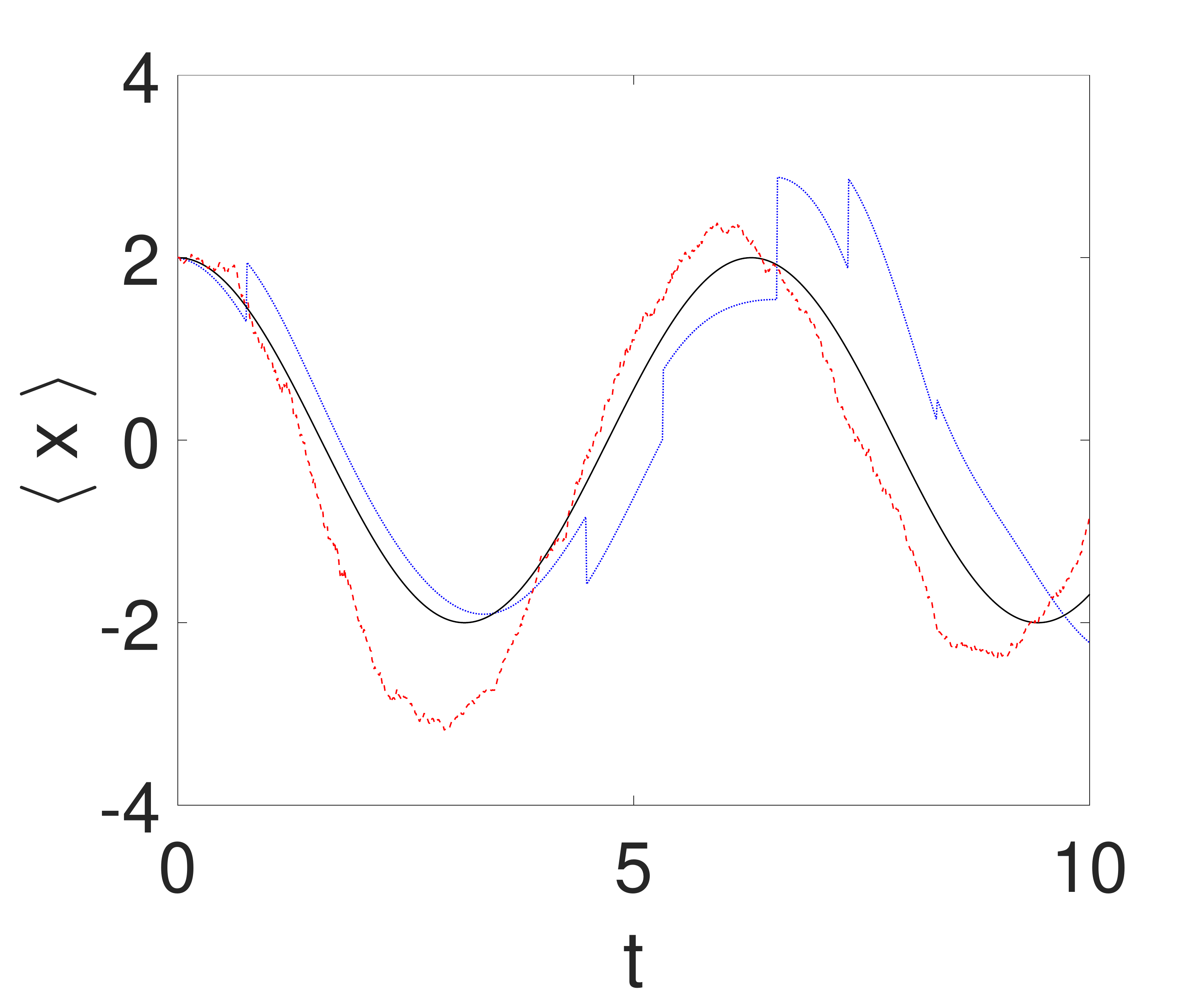}
	  \includegraphics[width=0.4\textwidth]{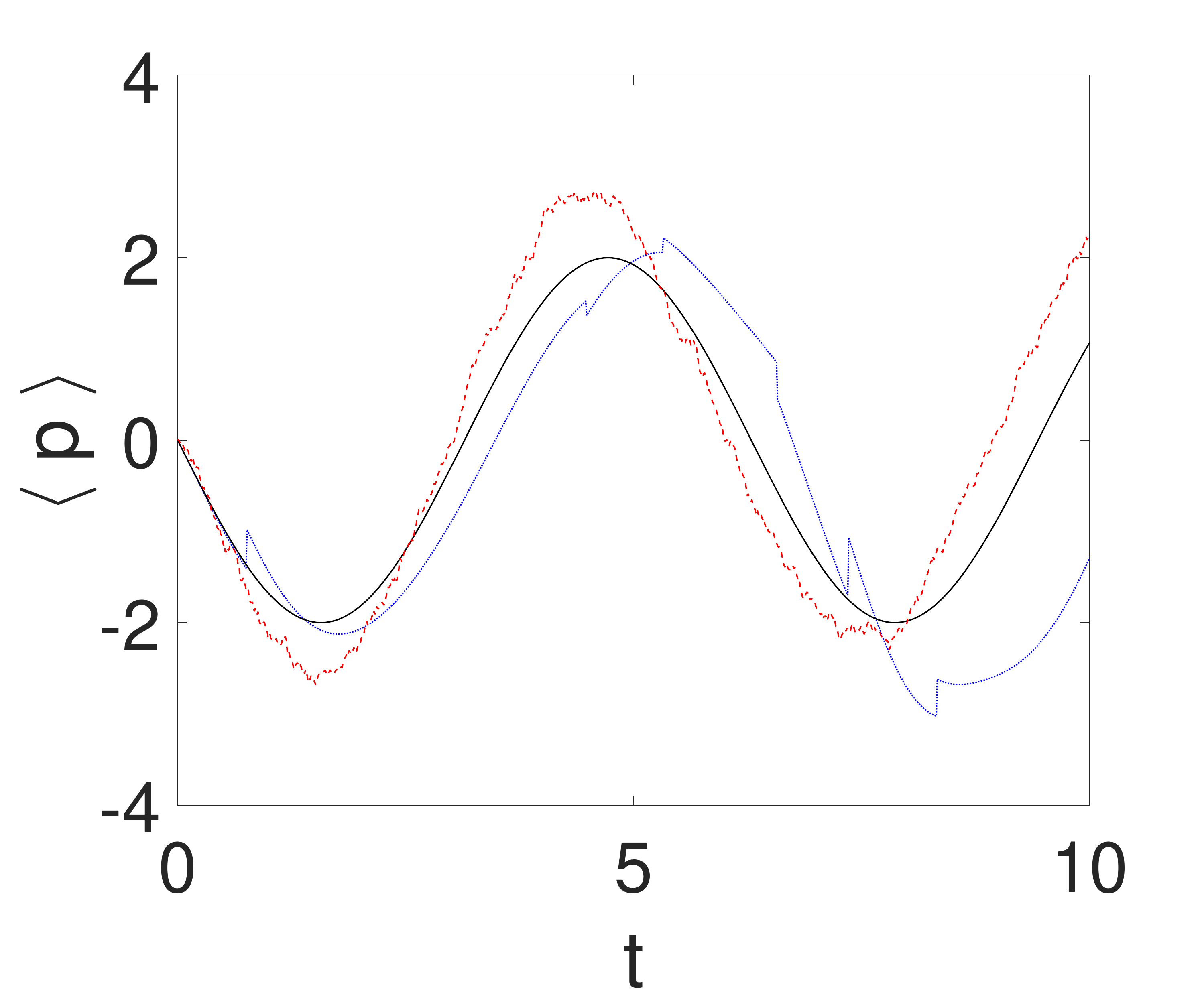}\\
	  \includegraphics[width=0.4\textwidth]{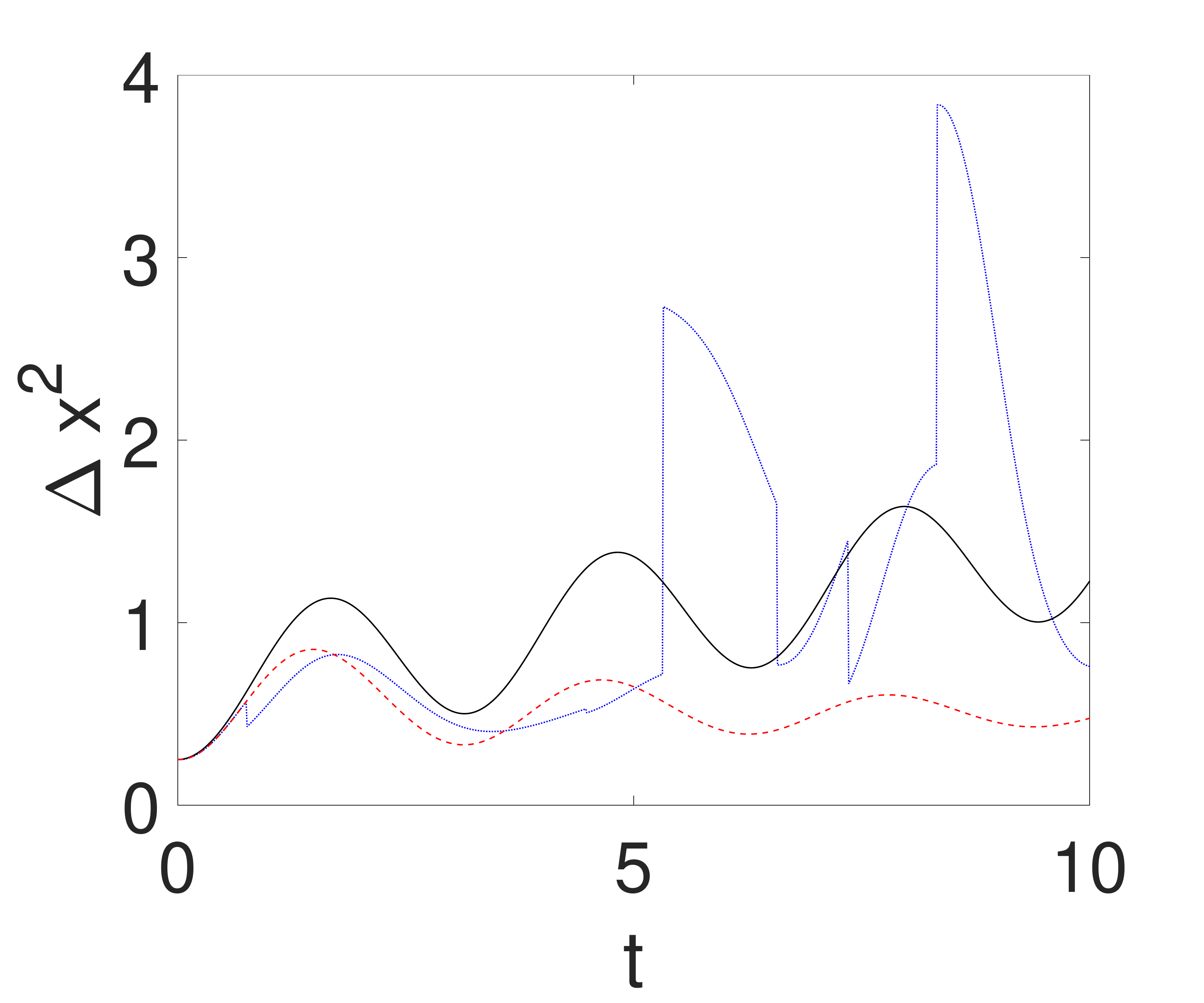}
	  \includegraphics[width=0.4\textwidth]{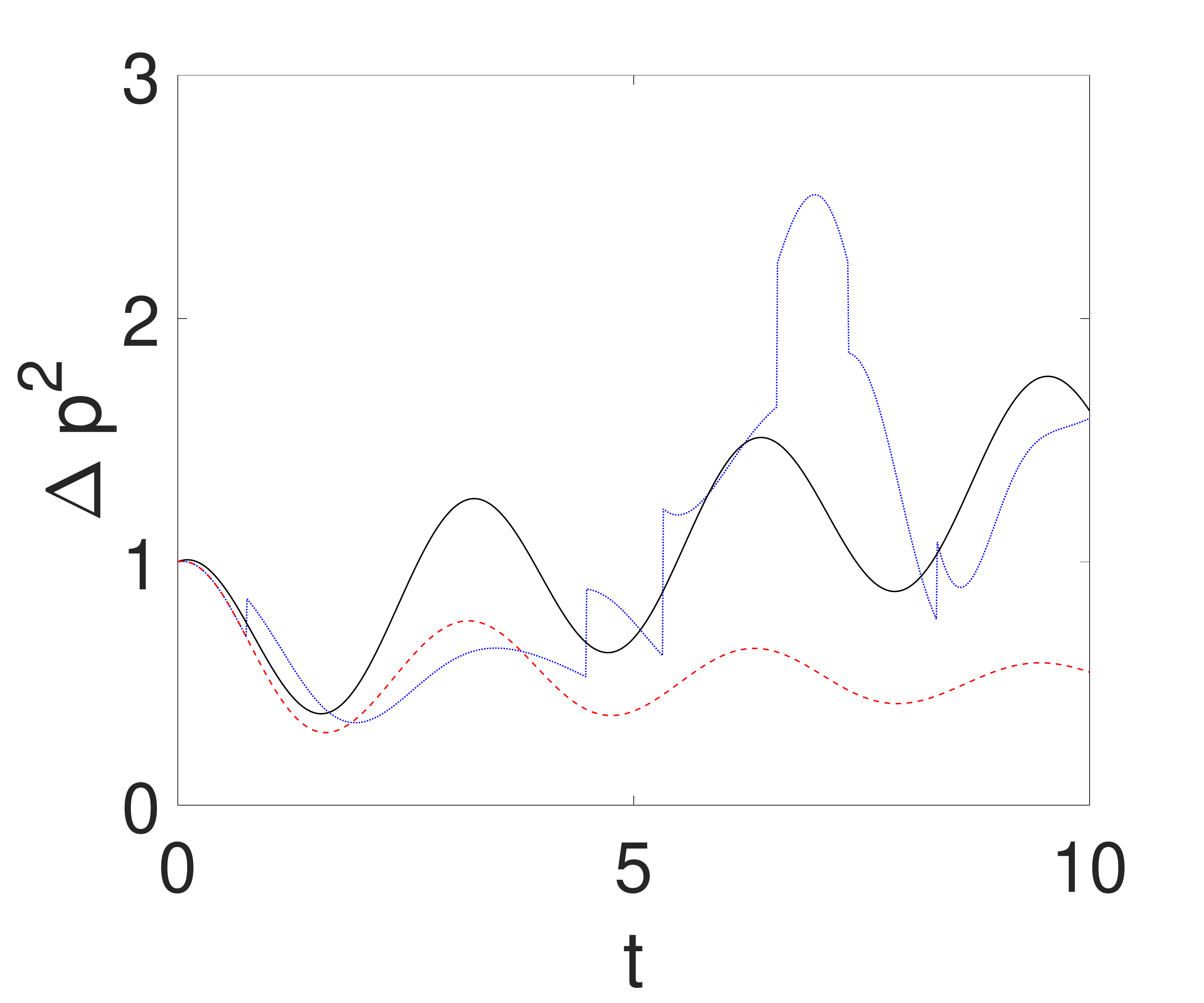}
\caption{ Lindblad dynamics (solid black line) compared with single trajectories of the SSE (dashed red line) and quantum-jump method (dotted blue line) for the position measurement model \cref{eq:PosModel}, with $\omega=1$ and $\gamma=0.2$. The initial Gaussian is a squeezed state (i.e. $a_0=(\tfrac{1}{\sqrt{2}}, i\sqrt{2})^T$ or $G=\left(\begin{smallmatrix}2&0\\0&\sfrac{1}{2}\end{smallmatrix}\right)$), centered at $\tilde z=\left(2,0\right)^T$. We show the time dependence of the position expectation $\braket{\hat x}$ (top left), the momentum expectation $\braket{\hat p}$ (top right), the positional variance $\Delta x^2$ (bottom left) and the momentum variance $\Delta p ^2$ (bottom right).}\label{fig:SingleBehaviorPos}
\end{figure}

The dynamical equations for the Gaussian parameters in the Lindblad dynamics \cref{eq:LindSemi} simplify to
\begin{align} 
    \frac{d \tilde z}{d t} &= \omega\Omega \tilde z,\\
  	\frac{dG}{dt} &=\omega(\Omega G-G\Omega)+2 G \Omega \Gamma\Omega G, \label{eq:CovarPos}
\end{align}
where we have defined 
\begin{equation}
   \Gamma=\Re(\nabla L\nabla\bar{L}^{T})=\begin{pmatrix}\gamma && 0 \\ 0 && 0 \end{pmatrix}.
\end{equation}
While the central dynamics is that of the unitary harmonic oscillator, the dynamics of the covariances encoded by $G(t)$, are influenced by the position measurement. Let us consider the simple example of an initially squeezed state with, $G(0)=\left(\begin{smallmatrix}\zeta&&0\\0&&\sfrac{1}{\zeta}\end{smallmatrix}\right)$. Solving \cref{eq:CovarPos} and substituting the result into \cref{eq:Covar} we find the physical variances as
\begin{equation}
\begin{aligned}
    \Delta x^2(t)&=
\frac{\hbar}{4 }\left(
\frac{\zeta ^2+2 \gamma  \zeta  t+1}{\zeta}-\frac{\gamma}{\omega} \sin (2 \omega t)-\frac{\zeta ^2-1}{\zeta}  \cos (2 \omega t )\right)\\
    \Delta p^2(t)&=\frac{\hbar}{4}\left(\frac{\zeta ^2+2
   \gamma  \zeta  t+1}{\zeta}+\frac{\gamma}{\omega} \sin (2 \omega t )+\frac{\zeta ^2-1}{\zeta}\cos (2 \omega t )\right)\\
   \Delta xp(t)&=\frac{\hbar}{4} \left(\frac{\gamma}{ \omega^2\zeta}+\frac{\zeta^2\omega-\omega- \gamma\zeta}{\omega^2\zeta^2}\cos(2 \omega t)\right) 
\end{aligned}
\end{equation}
That is, we observe the typical harmonic oscillations with frequency $2\omega$ in the covariances as they appear in the unitary harmonic oscillator, accompanied by a linear growth of the position and momentum uncertainties $\Delta x^2$ and $\Delta p^2$, associated with the effect of the position measurement. This behaviour is illustrated in figure \ref{fig:SingleBehaviorPos} which depicts the expectation values of position and momentum and their uncertainties as a function of time for an example with $\omega=1$ and $\gamma=0.2$ for an initially squeeze state centred at $\tilde z_t=(2, 0)^T$.
The Lindblad dynamics are depicted as solid black lines. 

The SSE dynamical equations (\ref{eq:SSEParamz}) and (\ref{eq:SSEParamG}) become
\begin{equation} \label{eq:SSEPos}
   \begin{aligned}
	d\tilde z &= \omega \Omega \tilde z dt+\sqrt{\frac{\hbar \gamma}{2}}\, G^{-1} \begin{pmatrix}1\\0\end{pmatrix}d\xi_R+\sqrt{\frac{\hbar \gamma}{2}} \begin{pmatrix}0\\ 1\end{pmatrix}d\xi_I,\\
	\frac{d G}{dt}&=\omega(\Omega G-G\Omega)+\Gamma+G \Omega \Gamma\Omega G.
\end{aligned} 
\end{equation} 
That is, for the central motion we again have the familiar unitary Hamiltonian flow term, and no damping term, but now there is an additional width-dependent stochastic noise. The equation for the covariances differs from that in the Lindbladian case, as expected. As discussed above $G(t)$ can be solved analytically by equation (\ref{eq:GtS}) or (\ref{eq:GSol}). The linearised flow $S(t)$ is given by 
\begin{equation}
\label{eqn:ex1S}
    S(t)=\begin{pmatrix}
    \cosh(\sqrt{\frac{\omega}{2}}\Lambda t) &
    \sqrt{\frac{\omega}{2 \lambda^2}}\Lambda^{*}\sinh(\sqrt{\frac{\omega}{2}}\Lambda t) \\
    \sqrt{\frac{1}{2 \omega}}\Lambda \sinh(\sqrt{\frac{\omega}{2}}\Lambda t)
    & \cosh(\sqrt{\frac{\omega}{2}}\Lambda t)
    \end{pmatrix},
\end{equation}
where we have defined 
\begin{equation} 
\lambda=\sqrt{\gamma^2+\omega^2}\quad\text{and}\quad\Lambda=\sqrt{\lambda-\omega}+i\sqrt{\lambda+\omega}.
\end{equation}
That is, we have an oscillatory contribution with frequency $\sqrt{\frac{\omega(\lambda+\omega)}{2}}$, which reduces to oscillations with frequency $\omega$ in the limit $\gamma=0$, and an additional exponential growth with rate $\sqrt{\lambda-\omega}$. The asymptotic behaviour for large times is given by 
\begin{equation}
    \lim_{t\to\infty}S(t)=e^{\sqrt{\frac{\omega}{2}}\Lambda t}\begin{pmatrix}
    1 &
    \sqrt{\frac{\omega}{2 \lambda^2}}\Lambda^{*} \\
    \sqrt{\frac{1}{2 \omega}}\Lambda 
    & 1
    \end{pmatrix}.
\end{equation}

As a result, independent of the initial value $G(0)$, $\Sigma(t)$ tends to a fixed point as $t\to\infty$ given by
\begin{equation} \label{eq:GFixedLind}
    \Sigma(t)\to\frac{\hbar}{2 \gamma}\begin{pmatrix}
    \sqrt{2 \omega  (\lambda-\omega )}
    &&\lambda-\omega \\
 \lambda-\omega&& \lambda\sqrt{\frac{2 (\lambda-\omega)}{\omega}}
    \end{pmatrix}.
\end{equation}
This is in stark contrast to the behaviour of the Lindblad covariances, with their linear growth in $\Delta x^2$ and $\Delta p^2$. We can see this in \cref{fig:SingleBehaviorPos} which shows the SSE dynamics as red dashed lines. For the parameter choices corresponding to the example in \cref{fig:SingleBehaviorPos} ($\omega=1,\gamma=0.2$) we have $\lambda=\frac{\sqrt{26}}{5}$, and the covariances approach
\begin{equation}
\begin{aligned}
    \Delta x^2(t)&\to \frac{5\sqrt{2(\lambda-1)}}{2}\approx 0.4975\\
  \Delta p^2(t)&\to\frac{5\lambda\sqrt{2(\lambda-1)}}{2}\approx0.5074\\
    \Delta xp(t)&\to\frac{5(\lambda-1)}{2}\approx 0.0495,
            \end{aligned}
\end{equation}
that is, the final state is very close to a coherent state, due to the relatively small value of $\gamma$.
For the central dynamics we observe stochastic fluctuations around the average Lindblad dynamics. 
\begin{figure}
\begin{centering}
	  \includegraphics[width=0.3\textwidth]{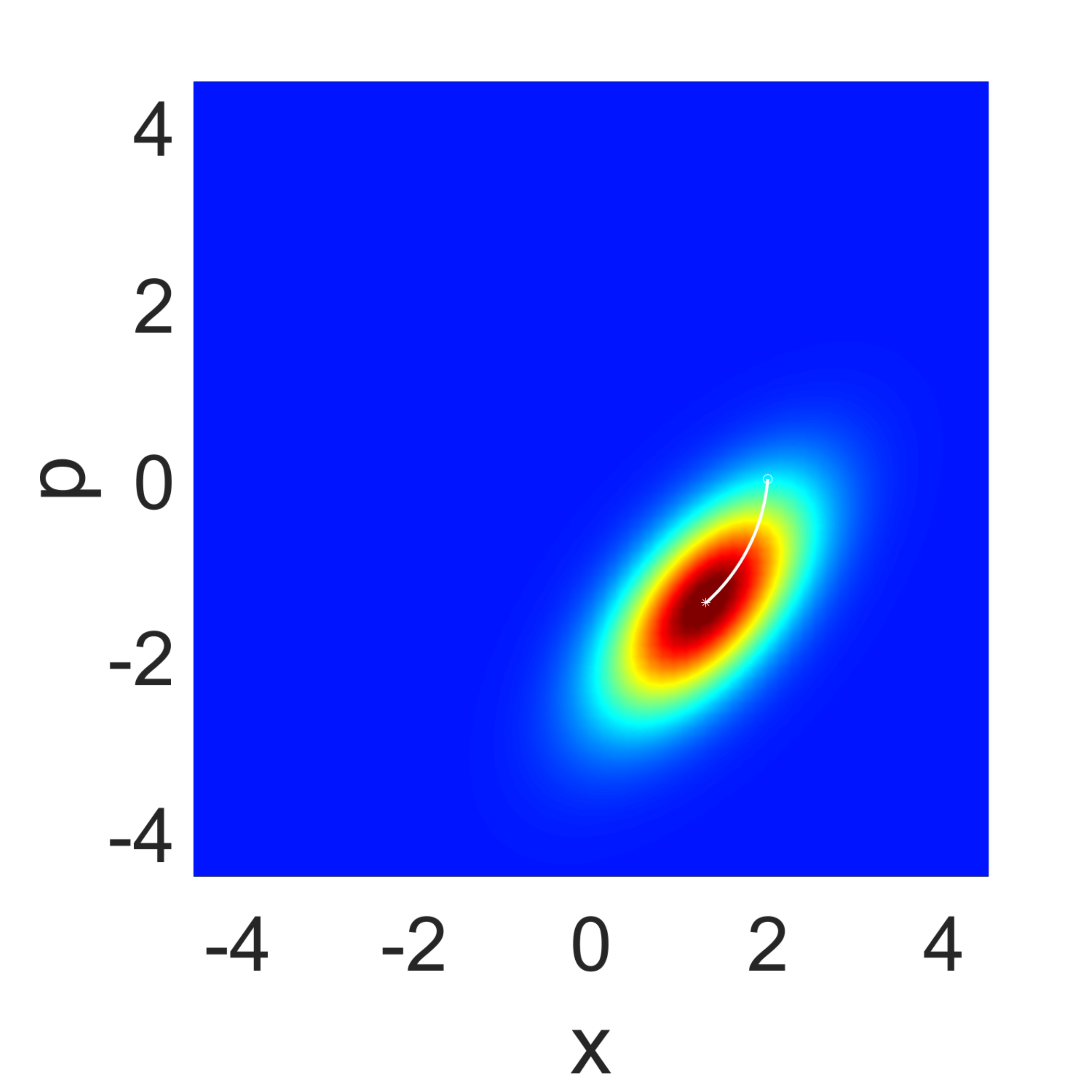}
	  \includegraphics[width=0.3\textwidth]{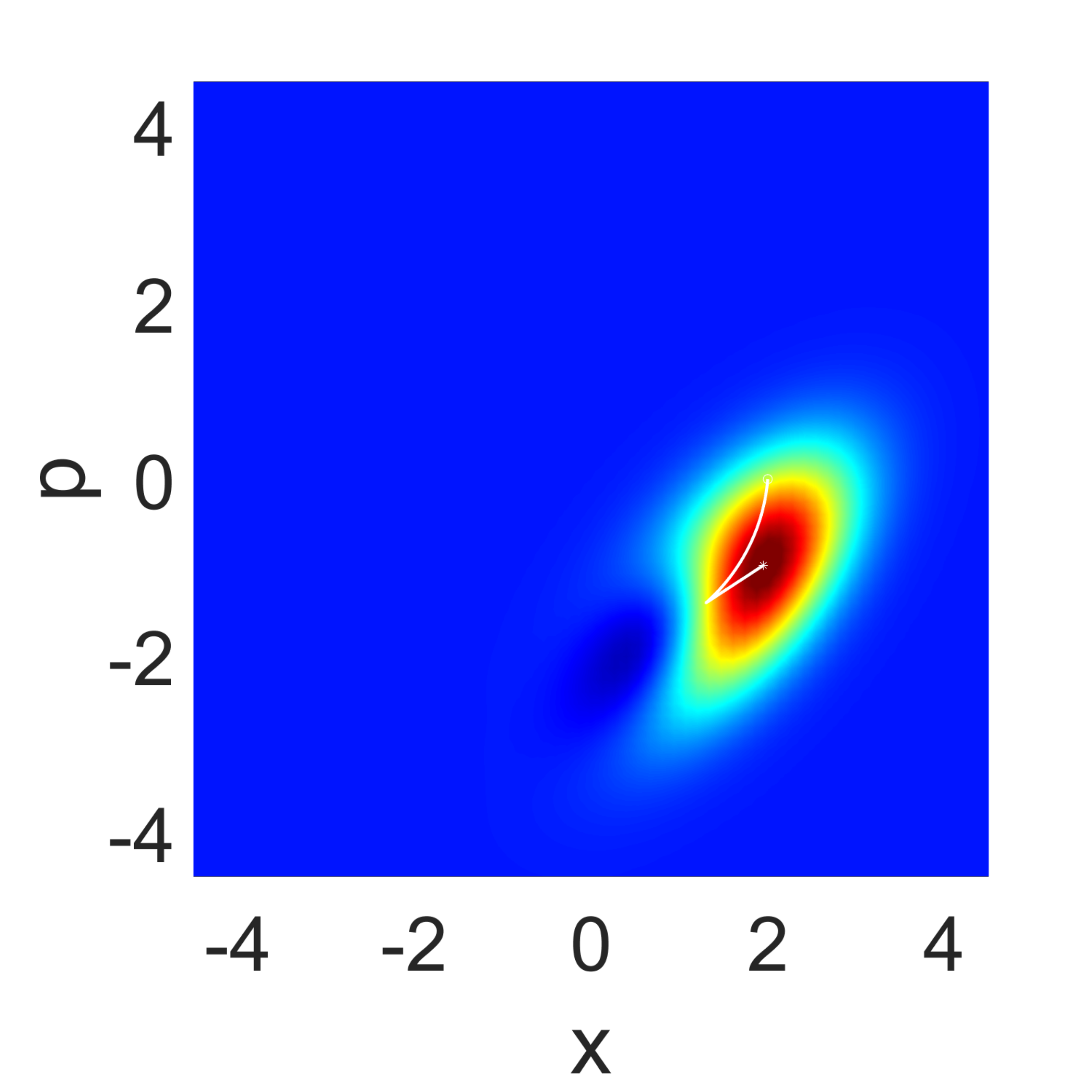}
	  \includegraphics[width=0.3\textwidth]{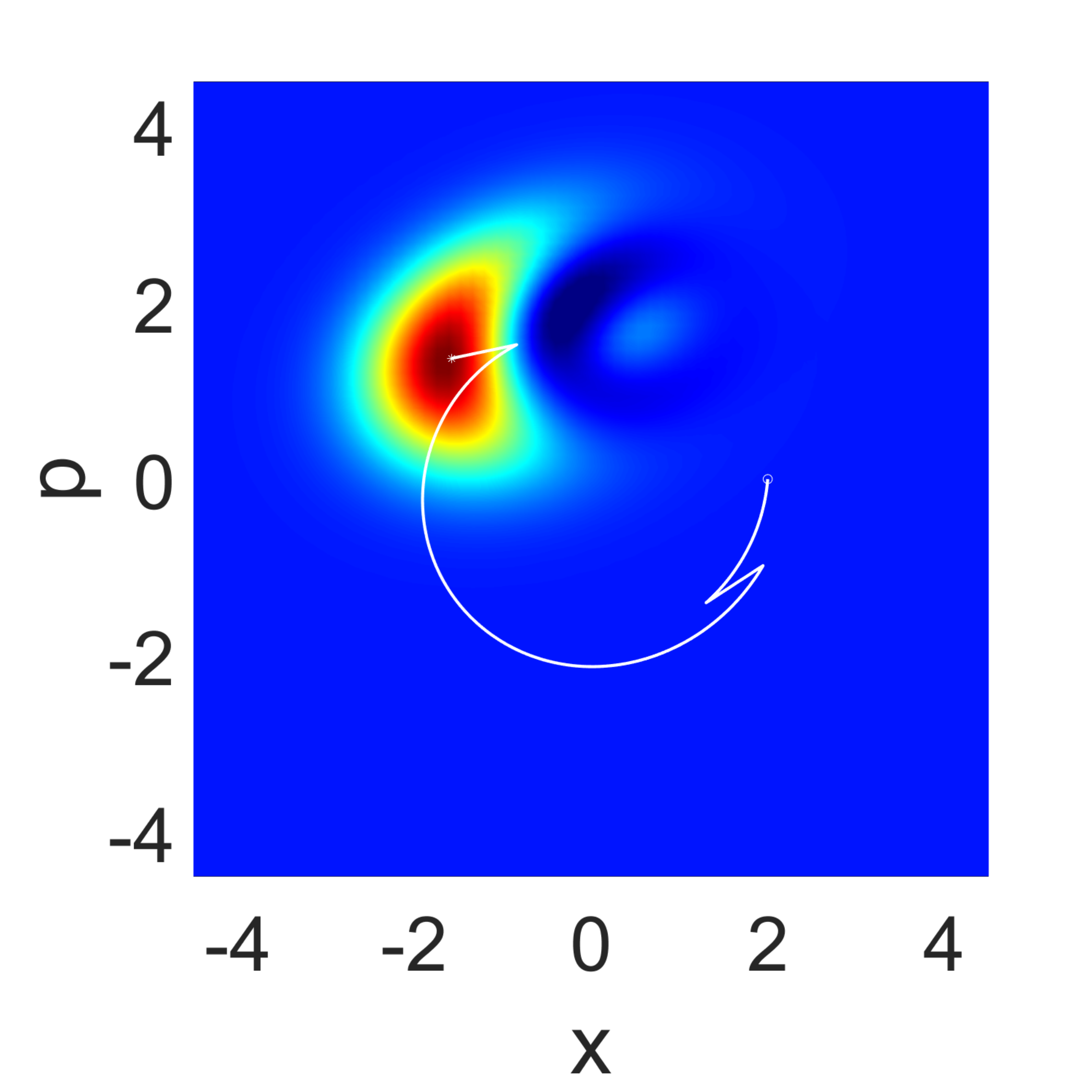} \\
	  \includegraphics[width=0.3\textwidth]{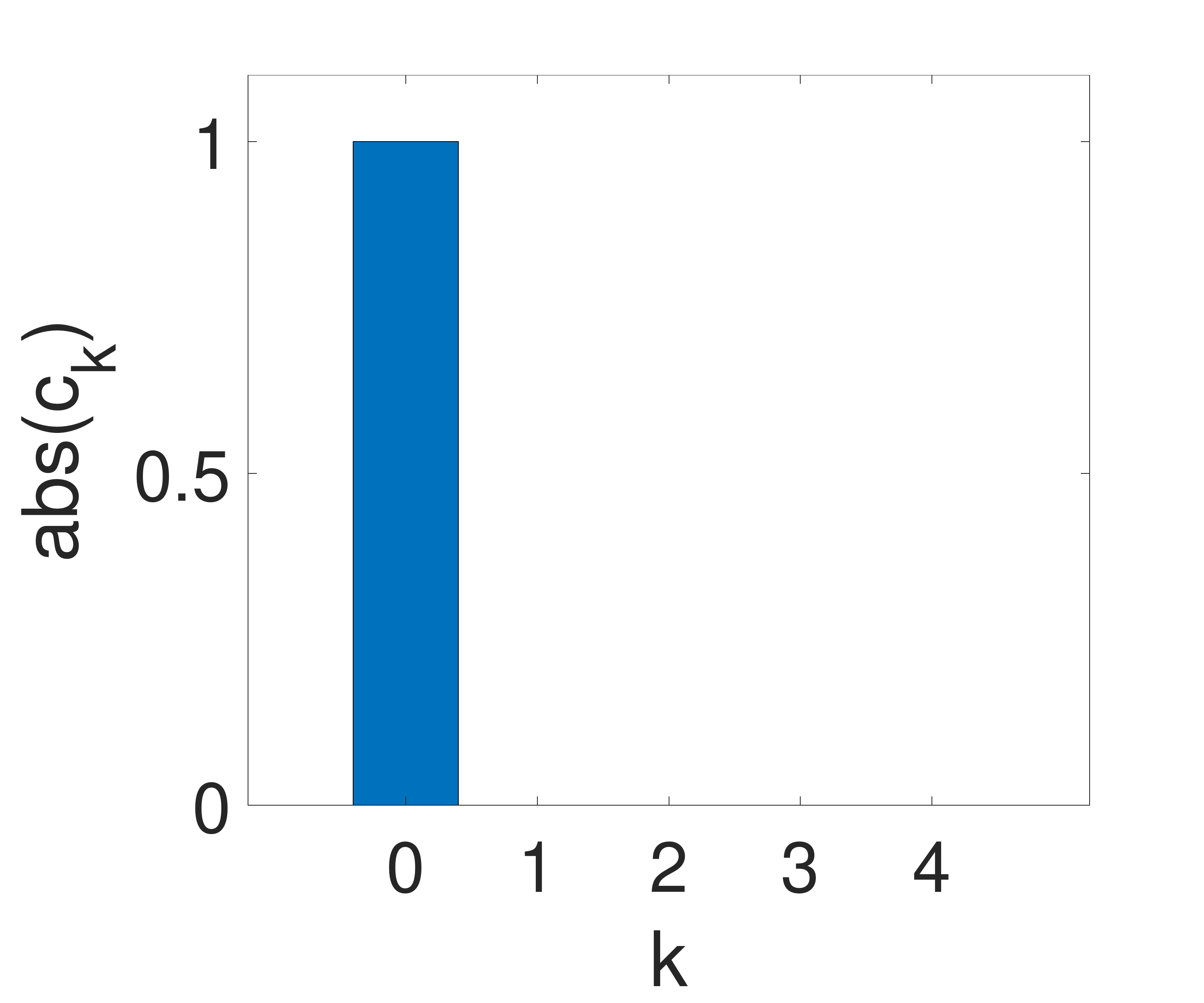}
	  \includegraphics[width=0.3\textwidth]{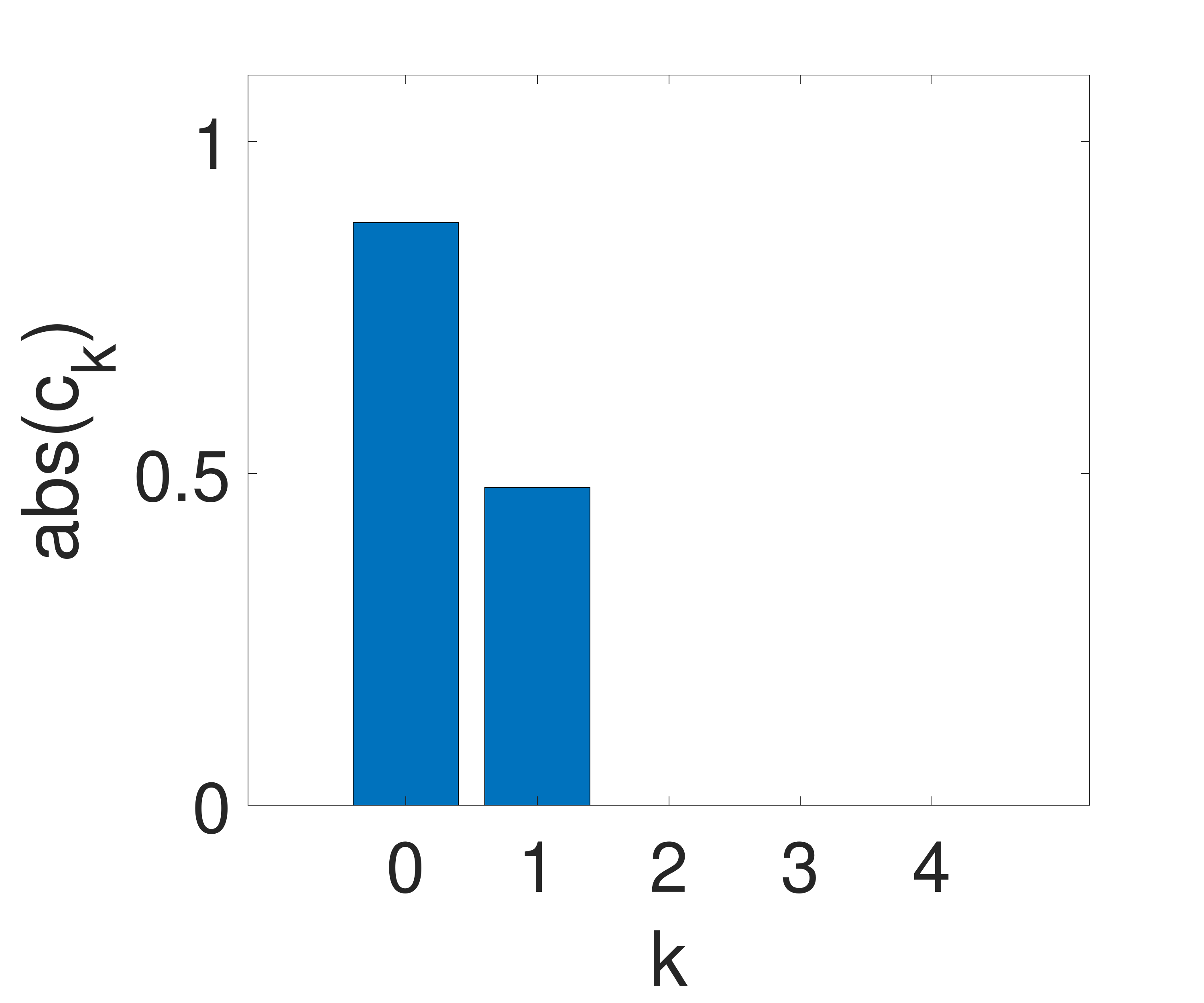}
	  \includegraphics[width=0.3\textwidth]{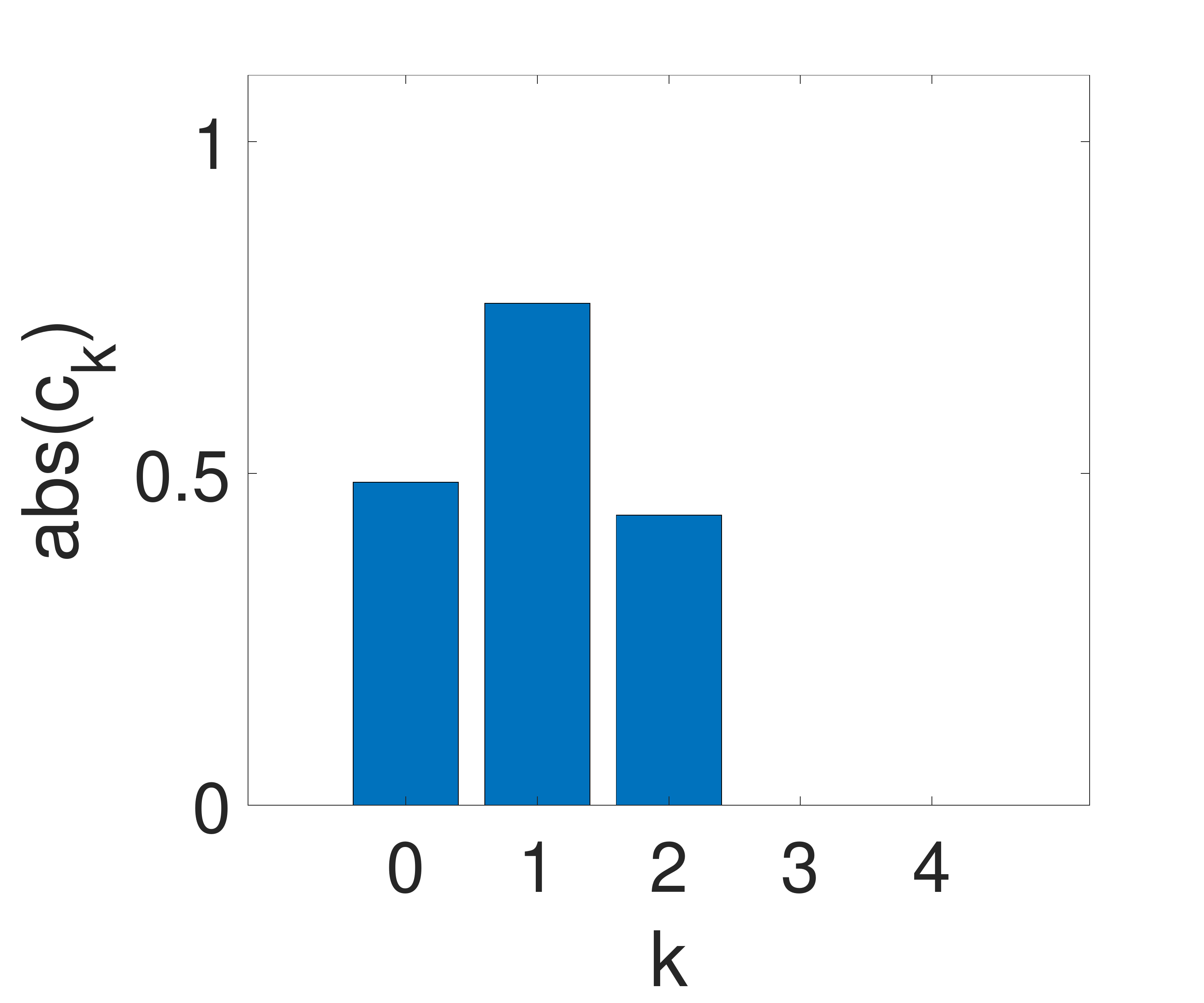}
\caption{Top row: Wigner functions of the single quantum jump trajectory in figure \ref{fig:SingleBehaviorPos} at selected times (from left to right: t=3.36 (shortly before the first jump) and t=3.37 (just after the first jump) and t=6.47 (just after the second jump)). The white line traces the preceding central motion. Bottom row: Relative magnitudes of the coefficients of the state in the time evolved basis ($\hat U(t)\ket{n,a_0,z_0}$) at the same times as in the top row.} \label{fig:WigPlotsPos_QJ}
\end{centering}
\end{figure}
The quantum-jump trajectories, on the other hand, depicted for an example run as blue dotted lines in figure \ref{fig:SingleBehaviorPos}, show very different behaviour. Here up to the first jump, the centre of the Gaussian state follows the non-Hermitian dynamics, which in the present case reduce to 
\begin{equation}
  \begin{pmatrix}\dot{ x_t} \\ \dot{p_t} \end{pmatrix}
  =\omega \begin{pmatrix}  p_t \\ -x_t \end{pmatrix} - \frac{2\gamma x_t }{\hbar} \begin{pmatrix}\Delta x^2(t) \\\Delta xp(t)\end{pmatrix},
\end{equation} 
where $G(t)$ evolves dynamically as in the SSE case. That is, there is an additional position dependent damping term in the evolution, modulated by the covariances of the state. This damping in comparison to the Lindblad evolution is visible in the example depicted in figure \ref{fig:SingleBehaviorPos}.
We also observe in figure \ref{fig:SingleBehaviorPos} that, as expected, the dynamics of position and momentum variances agree between the quantum jump and the SSE dynamics up to the first jump. 
What is not shown here, but has been numerically verified, is that averaging over many quantum jump trajectories simulated in the Hagedorn basis does indeed recover the Lindblad dynamics, the same is true of the SSE parameter dynamics as expected.   

In figure \ref{fig:WigPlotsPos_QJ} we show the Wigner functions of the quantum-jump trajectory for the same realisation as in figure \ref{fig:SingleBehaviorPos} at three selected times, where the central trajectory up to the respective time is depicted as a solid white line. The damped Gaussian motion is visible in the first figure, just before the first jump in this realisation. The remaining figures illustrate the effect of the quantum jumps, resulting in a sudden displacement of the centre as well as the expected deviations from a Gaussian state. 
At the jump, the state is acted on by the position operator and the resulting state is no longer Gaussian. We clearly observe interference patterns in the Wigner function corresponding to the excitation of higher harmonic oscillator states. It is remarkable that averaging over these non-classical excited states results in the same Gaussian state as the Lindblad equation.
\begin{figure} 
\begin{centering}
	  \includegraphics[width=0.4\textwidth]{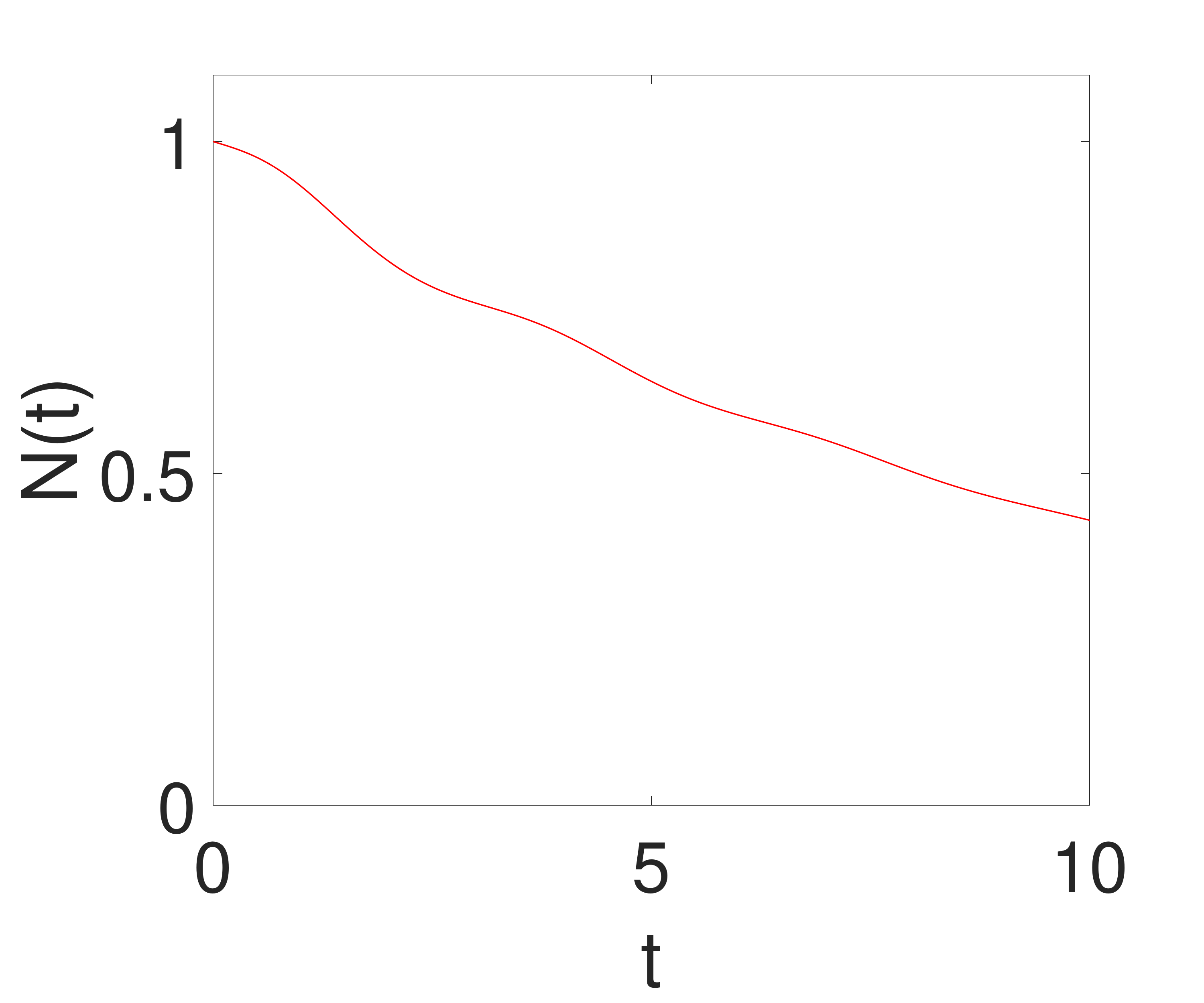}
	  \includegraphics[width=0.4\textwidth]{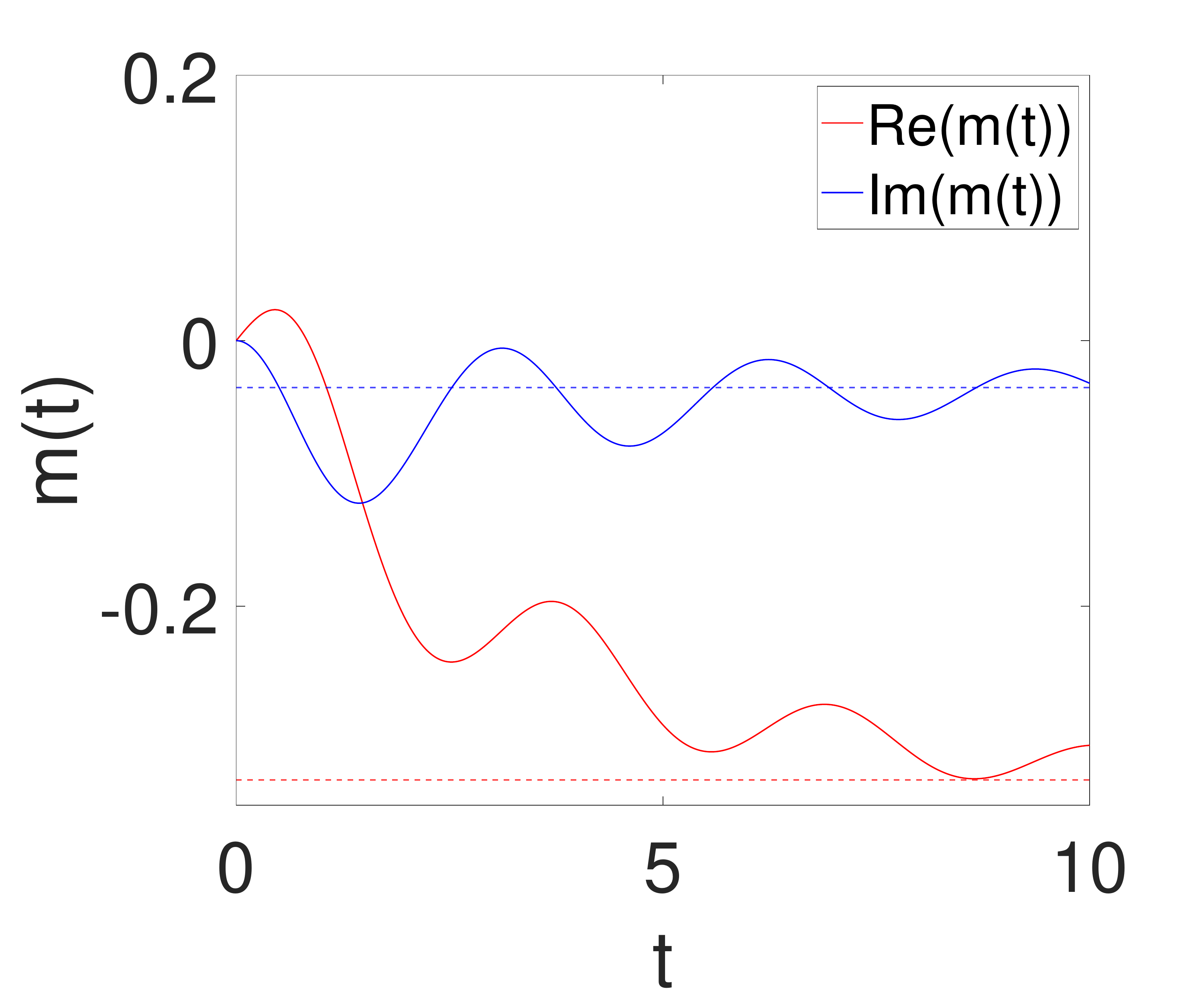}
\caption{Hagedorn basis parameter evolution for the position measurement model \cref{eq:PosModel}. The figure on the left shows the norm $N(t)$ of the ground state, and the right figure the evolution of the parameter $M(t)$ with dotted lines depicting the asymptotic fixed values of $M(t)$.} \label{fig:HagedornVarsPos}
\end{centering}
\end{figure}
The coefficients of the state in the moving Hagedorn basis are depicted in the histograms in the bottom panel of figure \ref{fig:WigPlotsPos_QJ}, for the same times as in the upper panel. We have chosen the initial ground state of the Hagedorn basis to coincide with the initial state, and thus there is no contribution from higher states before the first jump. We observe how each jump leads to a contribution from the next higher basis states, as expected.
\begin{figure} 
\begin{centering}
  	  \includegraphics[width=0.24\textwidth]{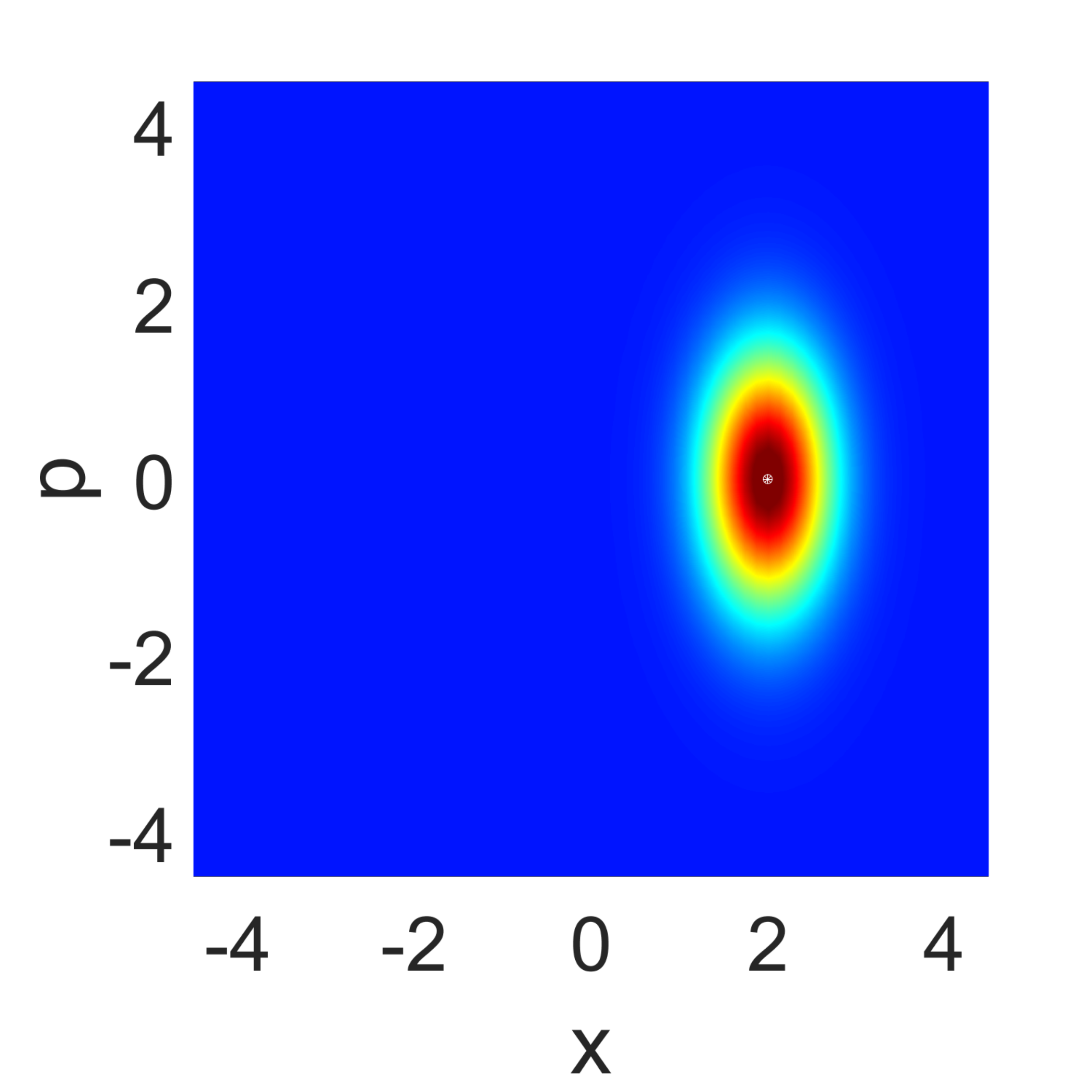}
	  \includegraphics[width=0.24\textwidth]{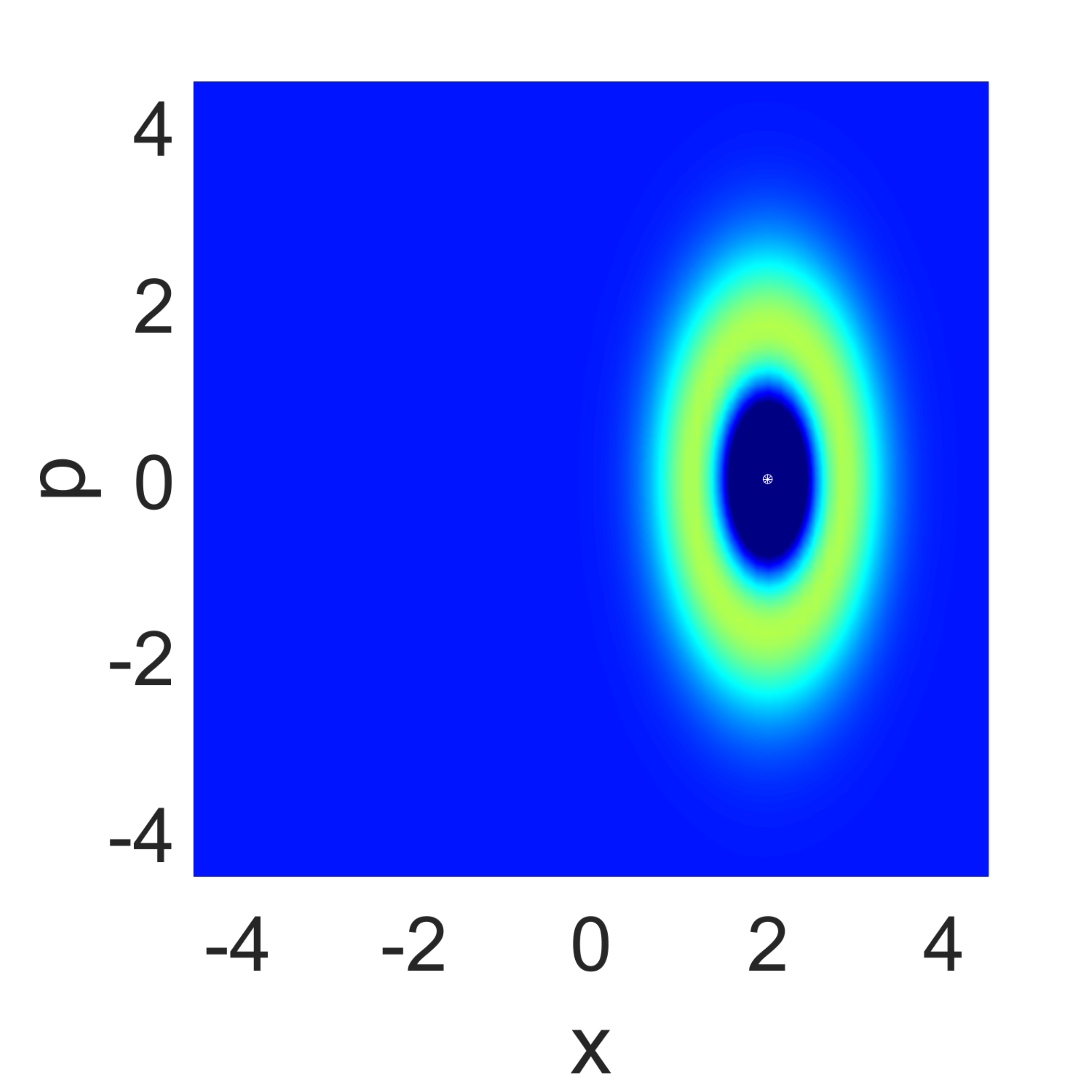}
	  \includegraphics[width=0.24\textwidth]{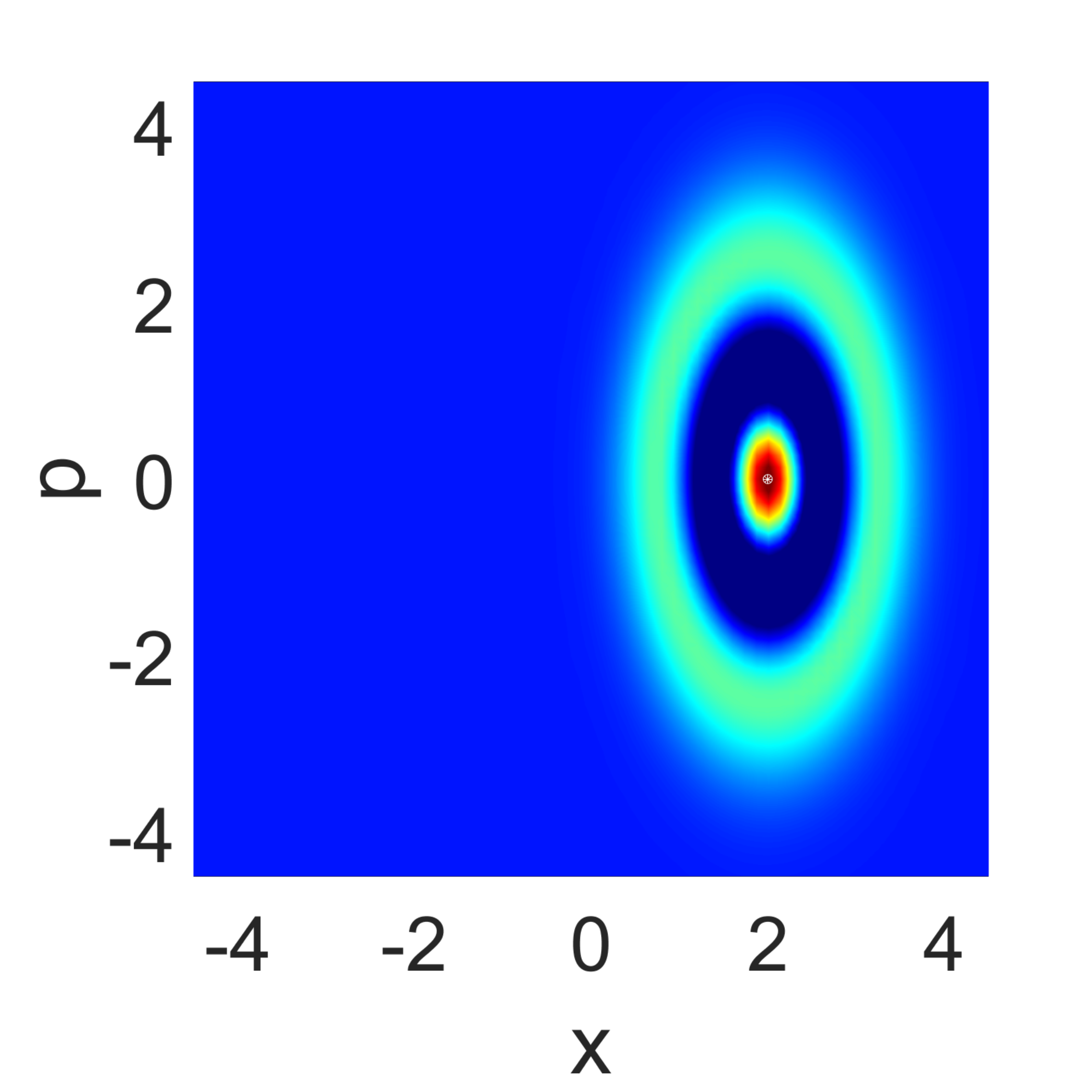}
	  \includegraphics[width=0.24\textwidth]{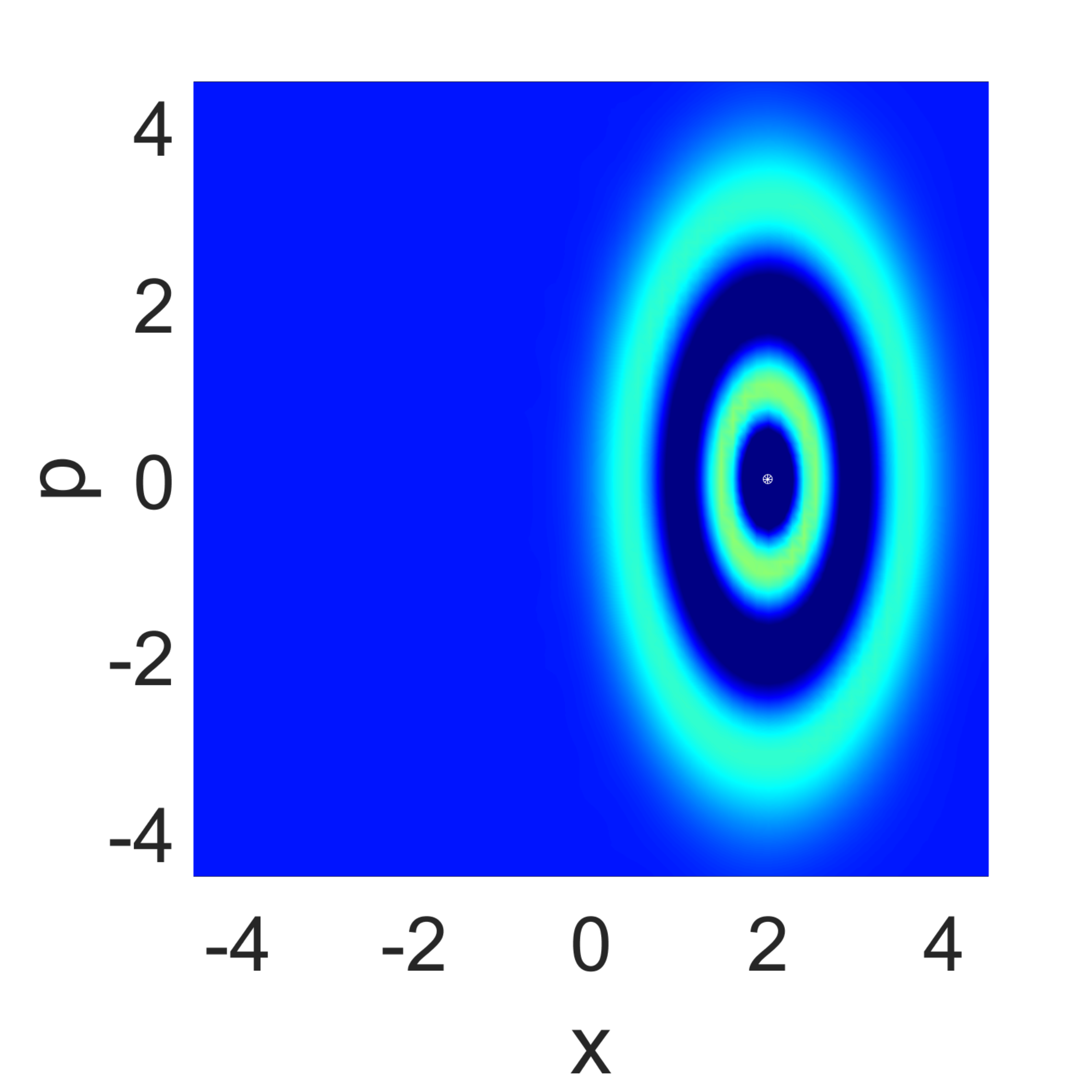}\\
	  \includegraphics[width=0.24\textwidth]{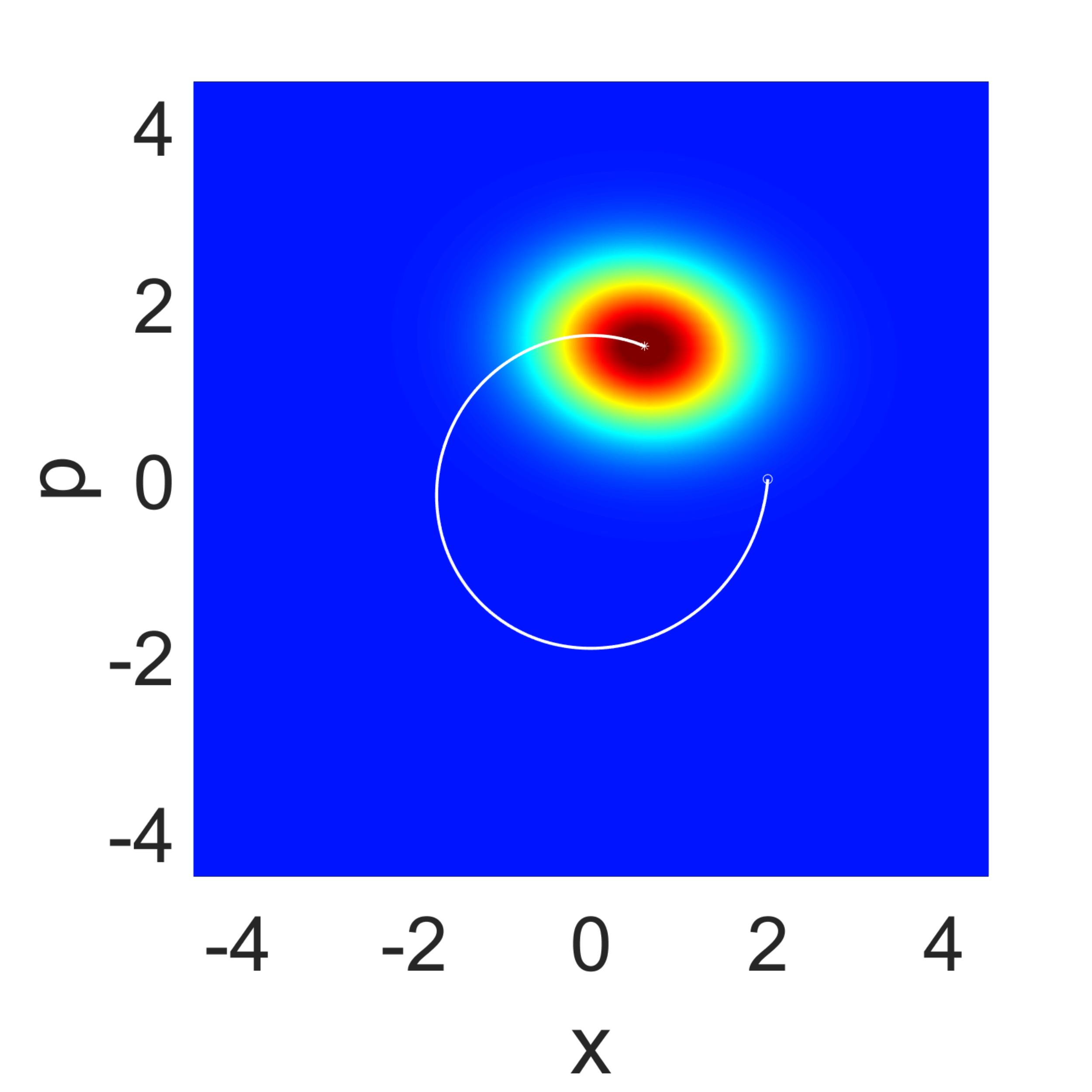}
	  \includegraphics[width=0.24\textwidth]{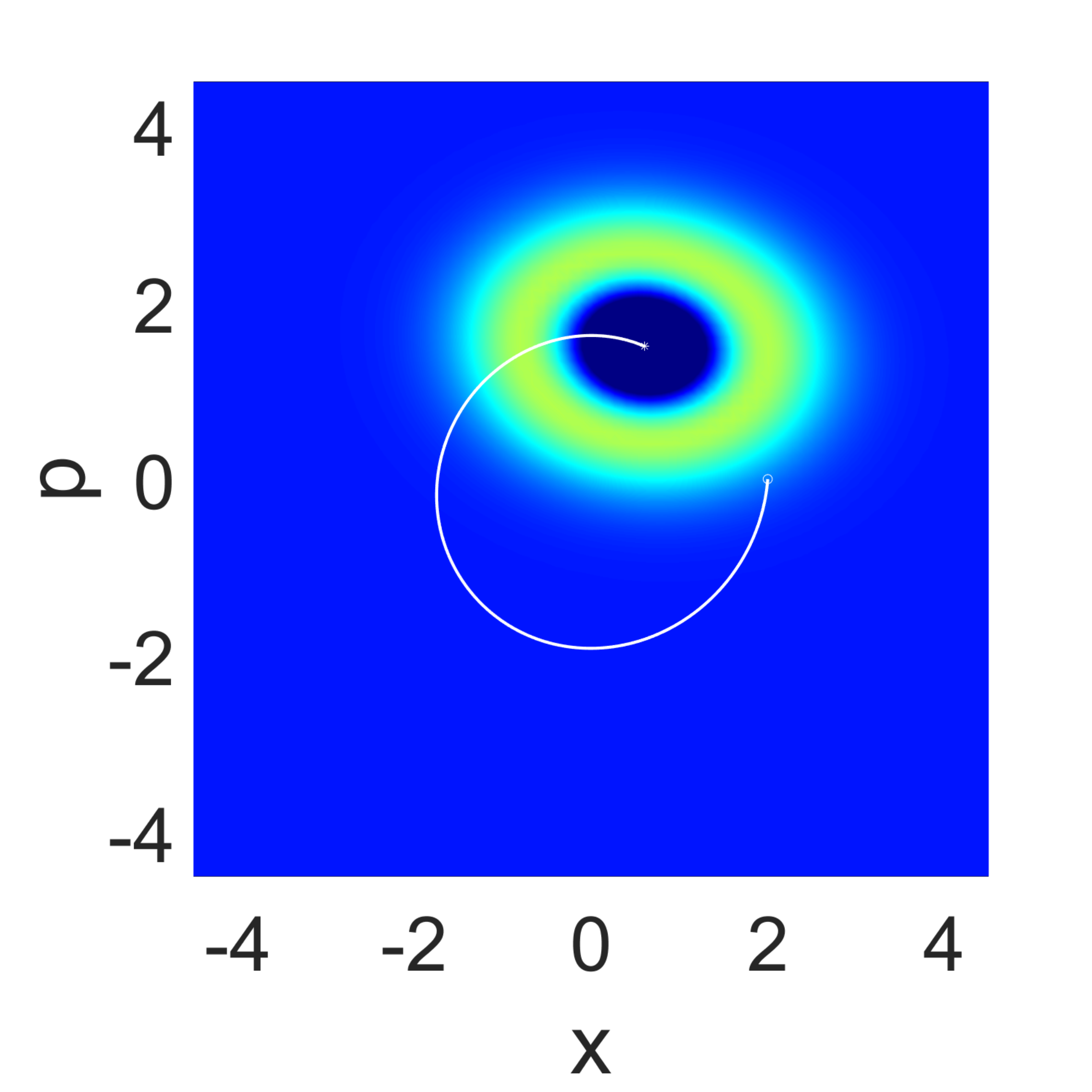}
	  \includegraphics[width=0.24\textwidth]{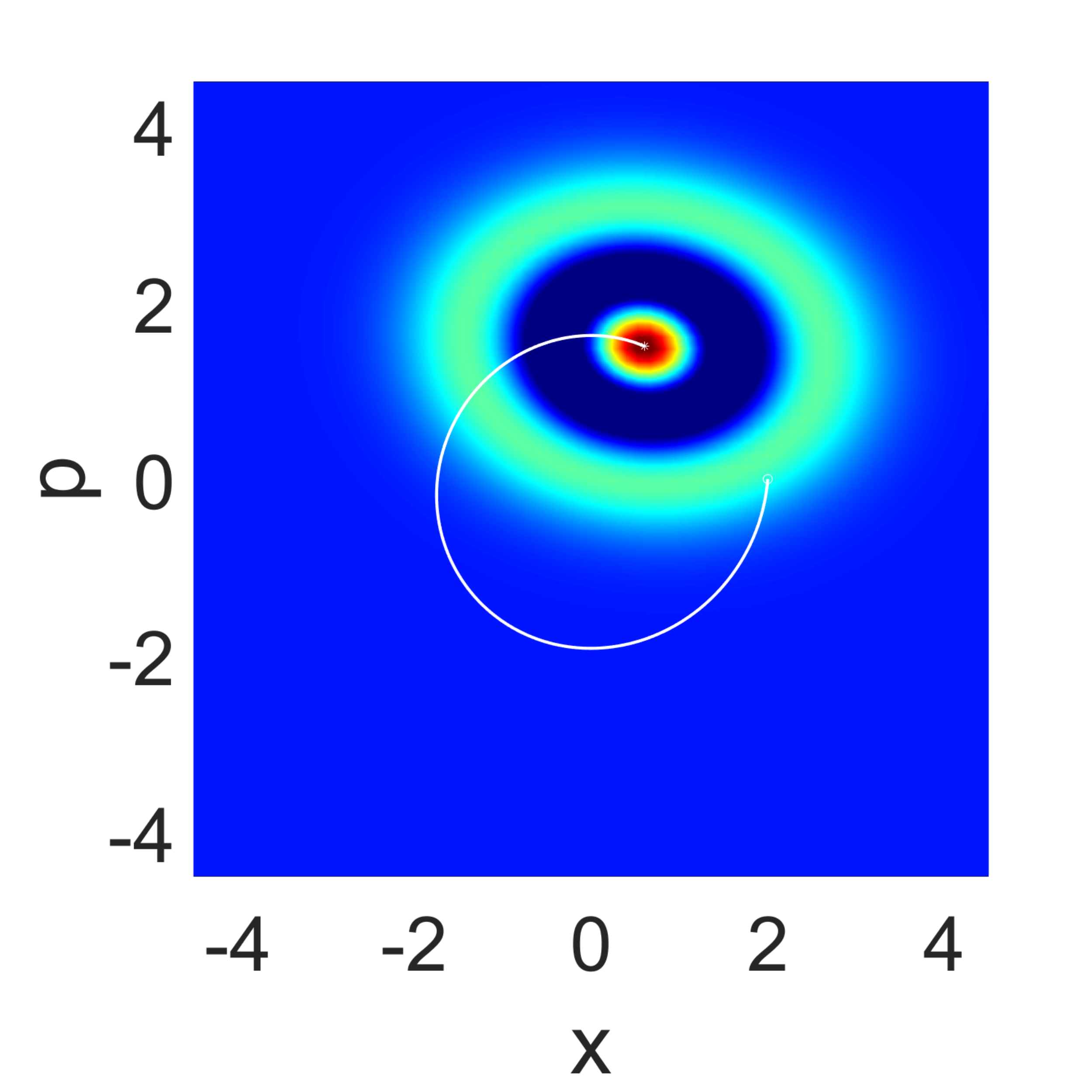} 
	  \includegraphics[width=0.24\textwidth]{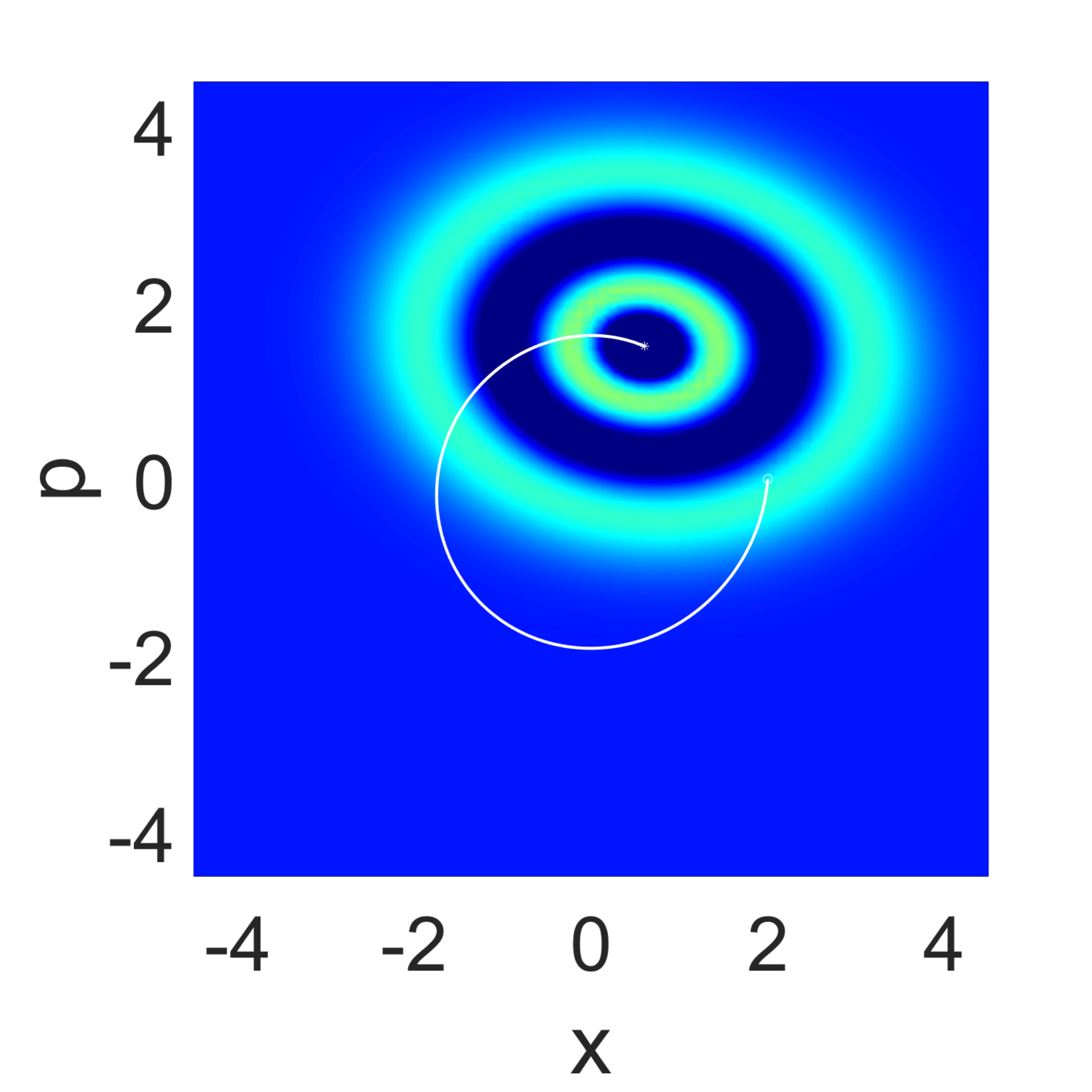}\\
	  \includegraphics[width=0.24\textwidth]{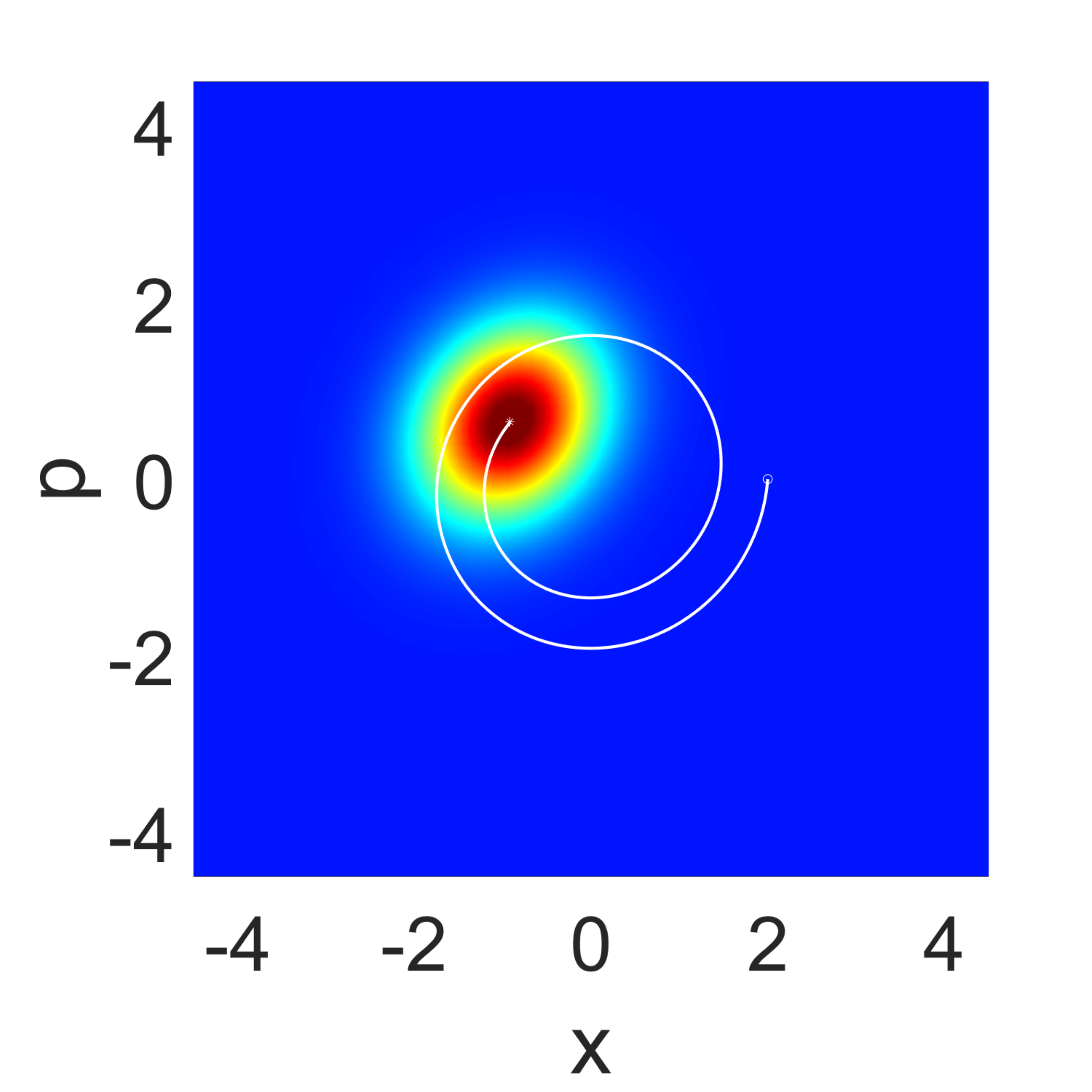}
	  \includegraphics[width=0.24\textwidth]{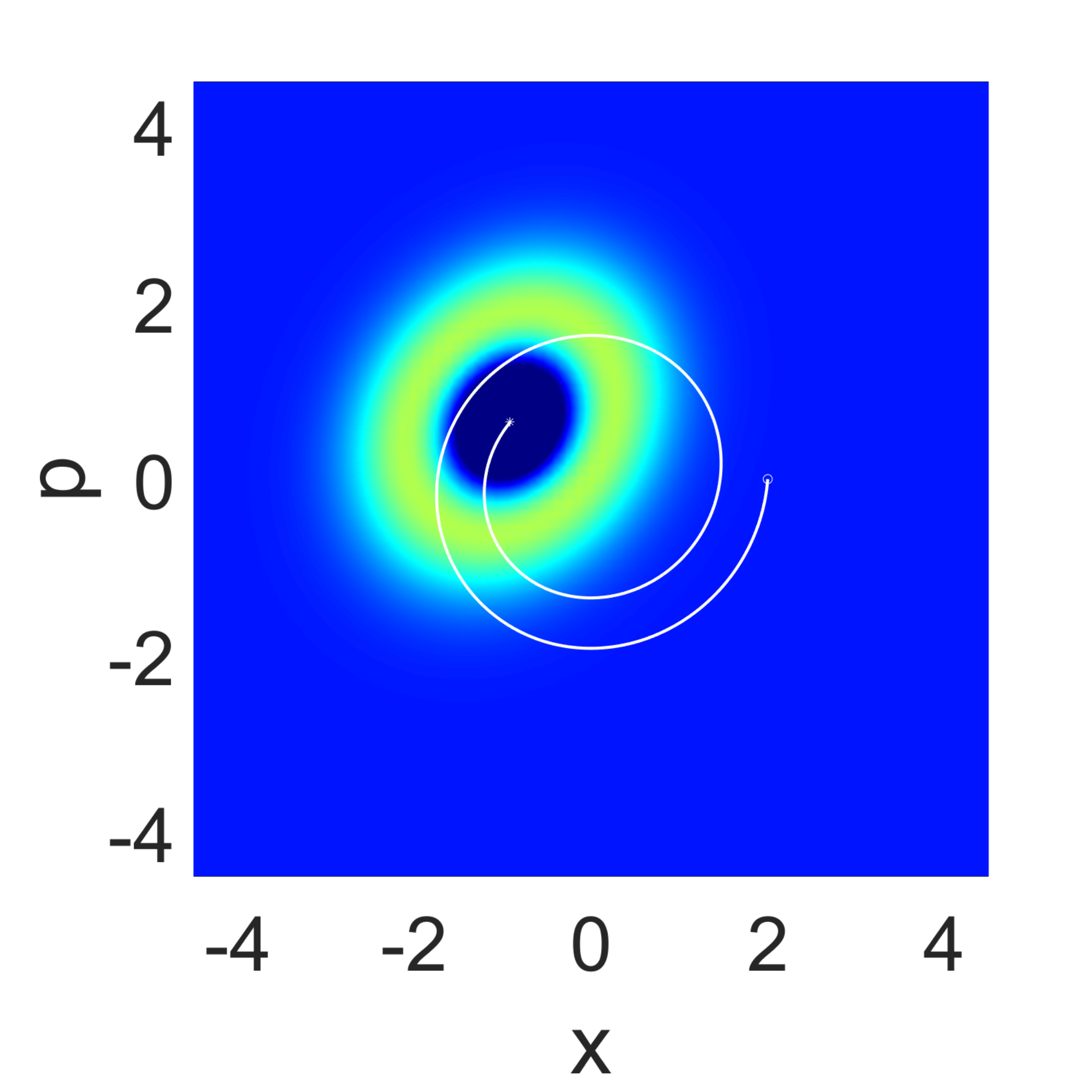}
	  \includegraphics[width=0.24\textwidth]{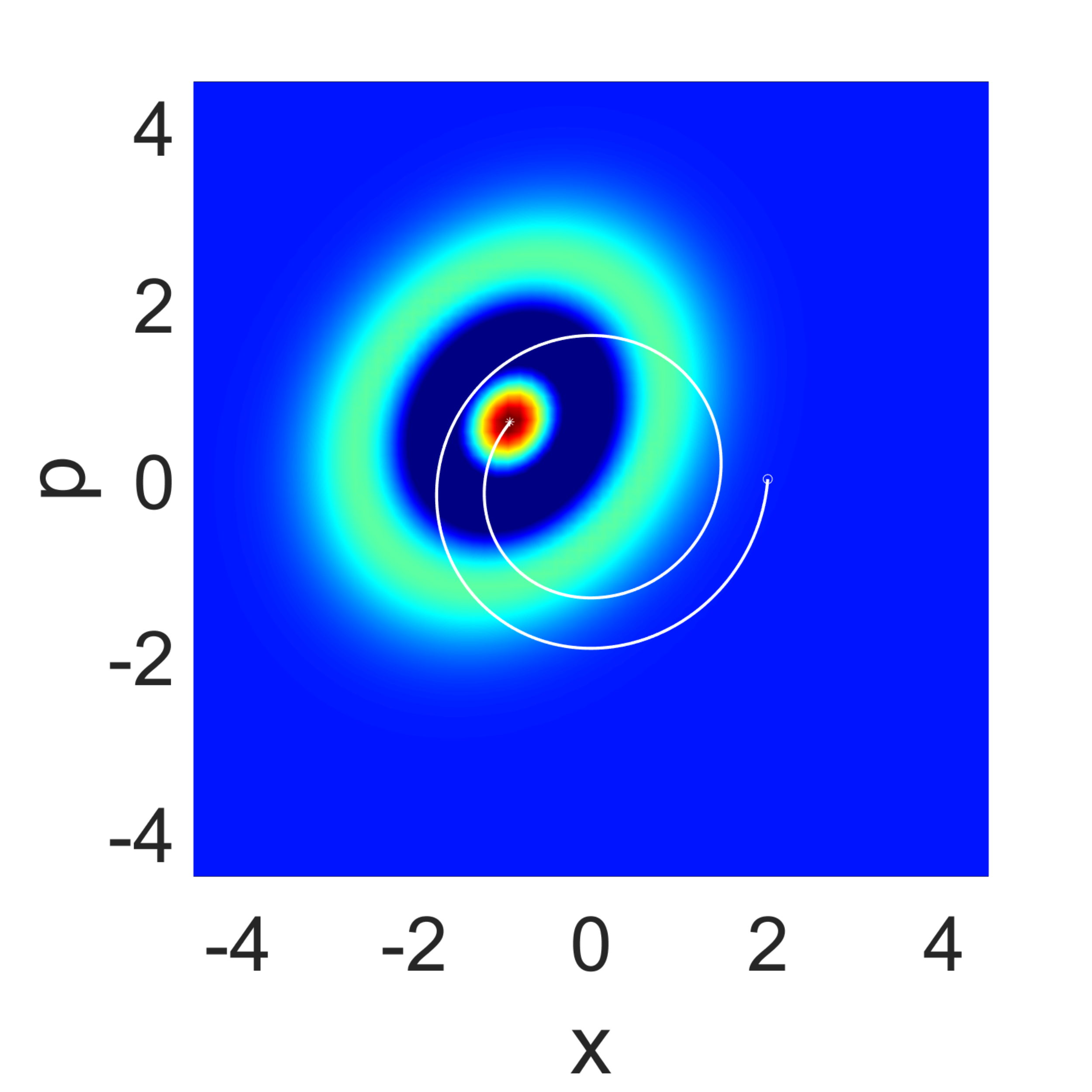}
	  \includegraphics[width=0.24\textwidth]{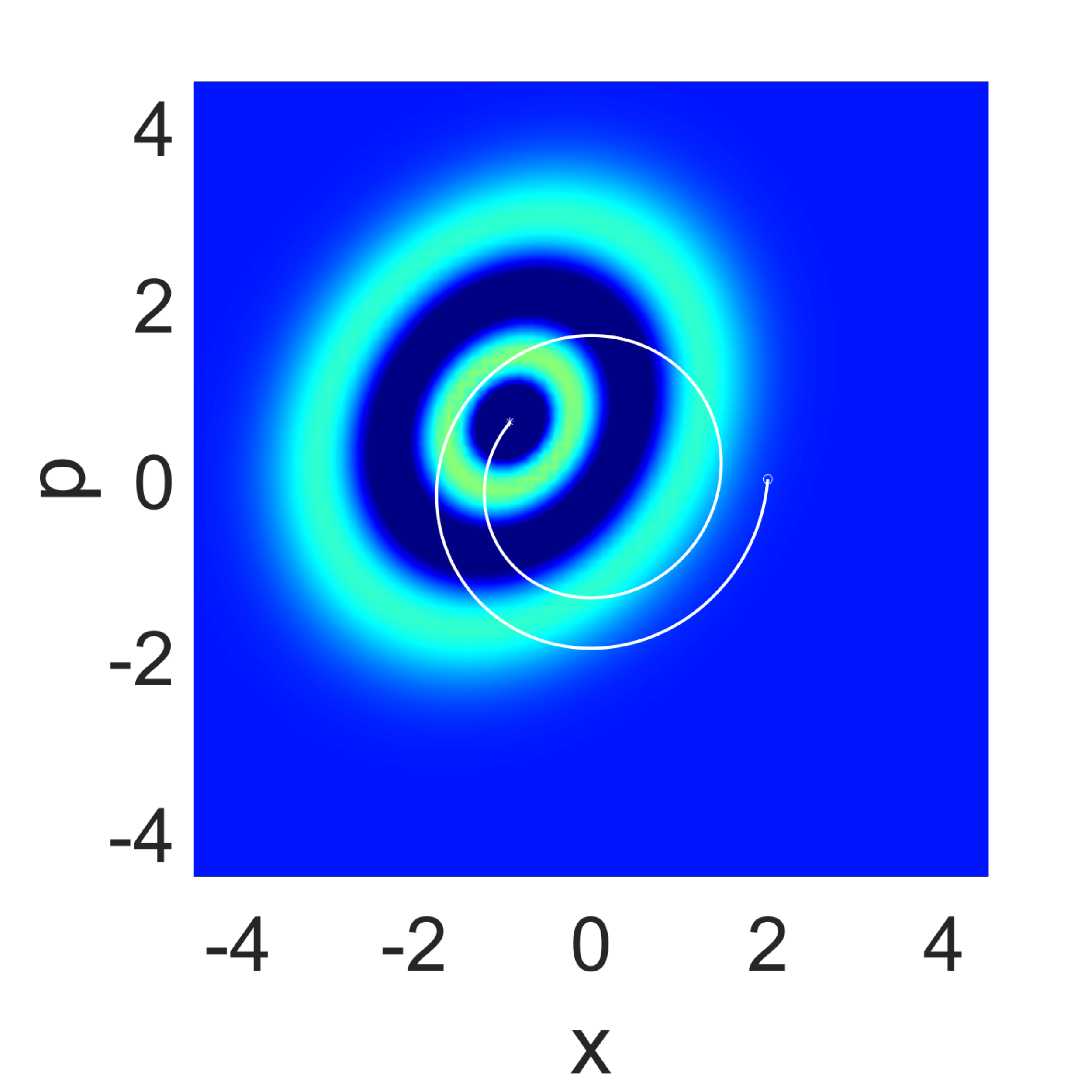} \\
\caption{Wigner function for the first four Hagedorn states $\ket{n,a_t,z_t}$ (n from 0-4 from left to right) for the position measurement model \cref{eq:PosModel}, at times $t=0,5,10$ (top to bottom)}\label{fig:HagBasis}
\end{centering}
\end{figure}
Analytically the evolution of the basis states is solely described by the linearized flow in equation (\ref{eqn:ex1S}). We may use this expression together with \cref{eq:NDef,eq:HagM} to calculate $N(t)$ and $M(t)$. Note that the dynamics of $N(t)$ and $M(t)$ do not depend on the initial position of the wavepacket, but solely on the values of $\omega$ and $\gamma$. 
Using the asymptotic behaviour of $S(t)$ we find that $N(t)$ for long times follows a simple exponential decay
\begin{equation}
    N(t)\to N_{\infty}e^{- \sqrt{\frac{\omega  (\lambda -\omega )}{2}}t},\quad\text{with}\quad N_{\infty}=\sqrt{\frac{8 \zeta  \lambda  \omega }{2 \zeta  \omega  (\lambda+\omega)+ \left(\zeta ^2 \omega +\lambda \right) \sqrt{2 \omega  (\lambda +\omega )} }}.
\end{equation}
$M(t)$ on the other hand tends to the fixed value
\begin{equation}
    M(t)\to \frac{\lambda-\zeta ^2 \omega -i  \zeta  \sqrt{2 \omega  (\lambda -\omega )} }{\lambda+\zeta ^2 \omega + \zeta  \sqrt{2\omega  (\lambda +\omega
   )} }.
\end{equation}
The dynamics of $N(t)$ and $M(t)$ for the example considered here are depicted in figure \ref{fig:HagedornVarsPos}. We observe the expected decay in the ground state normalisation and the approach to the fixed point value in $M(t)$. 

The Hagedorn raising operator $\hat{A}^{\dagger}(a_t,z_t)$ is also determined by $S(t)$ and in the long time limit it takes the form
\begin{equation}
    \hat{A}^{\dagger}\to \frac{N_{\infty}\e^{-i\sqrt{\frac{\omega(\lambda+\omega)}{2}}t}}{\sqrt{2 \hbar}}\left(\left(\sqrt{\zeta}+\frac{i}{\sqrt{\zeta \omega}}\Lambda^{*} \right)(\hat  x-x_t )-i\left(\frac{1}{\sqrt{\zeta}}-i\sqrt{\frac{\zeta \omega}{2 \lambda^2}}\Lambda \right) (\hat p-p_t) \right).
\end{equation}
Which we recognise as squeezed and shifted harmonic oscillator creation operator with a rotating phase of angular frequency $\sqrt{\tfrac{\omega(\lambda+\omega)}{2}}$.
\begin{figure}
\begin{centering}
  	  \includegraphics[width=0.24\textwidth]{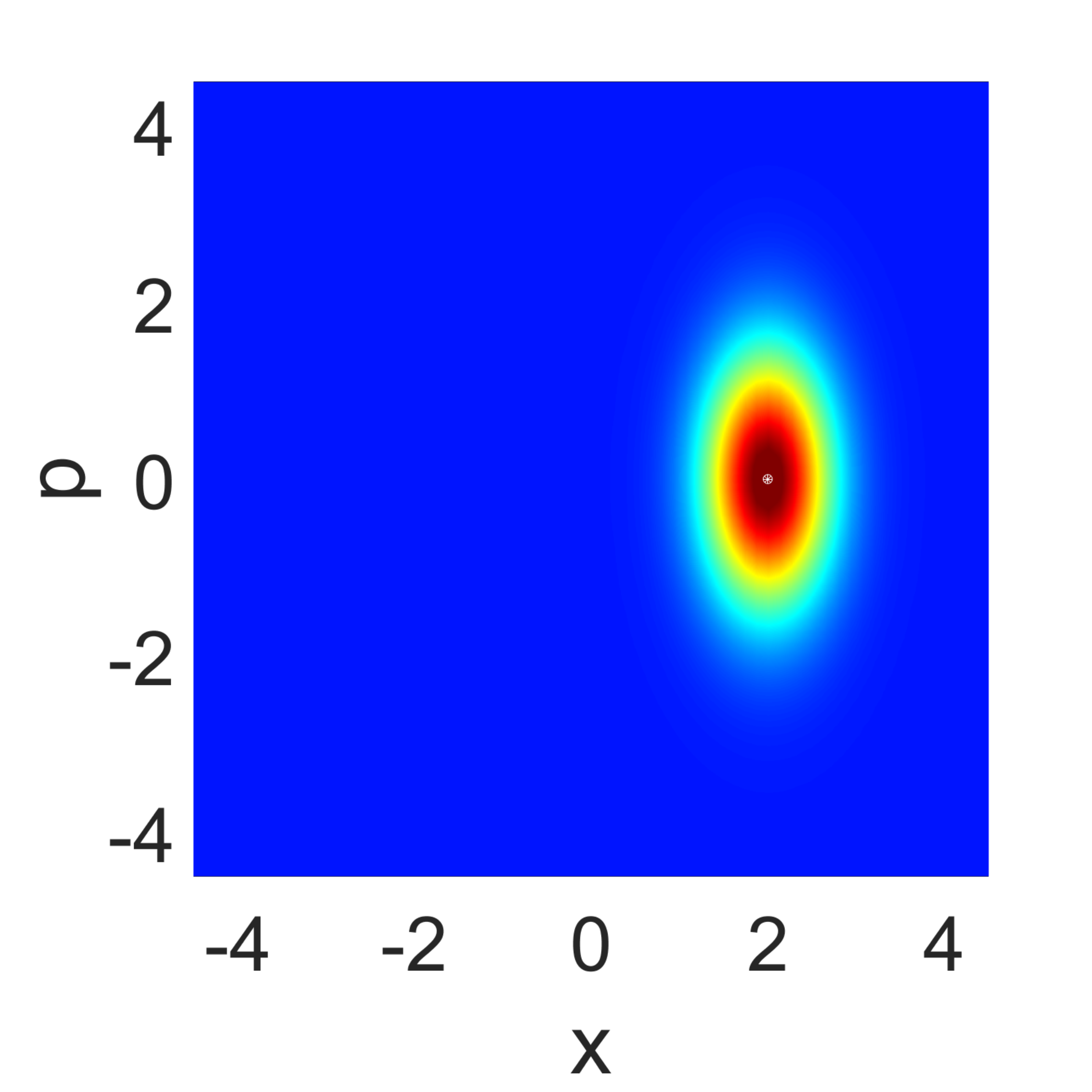}
	  \includegraphics[width=0.24\textwidth]{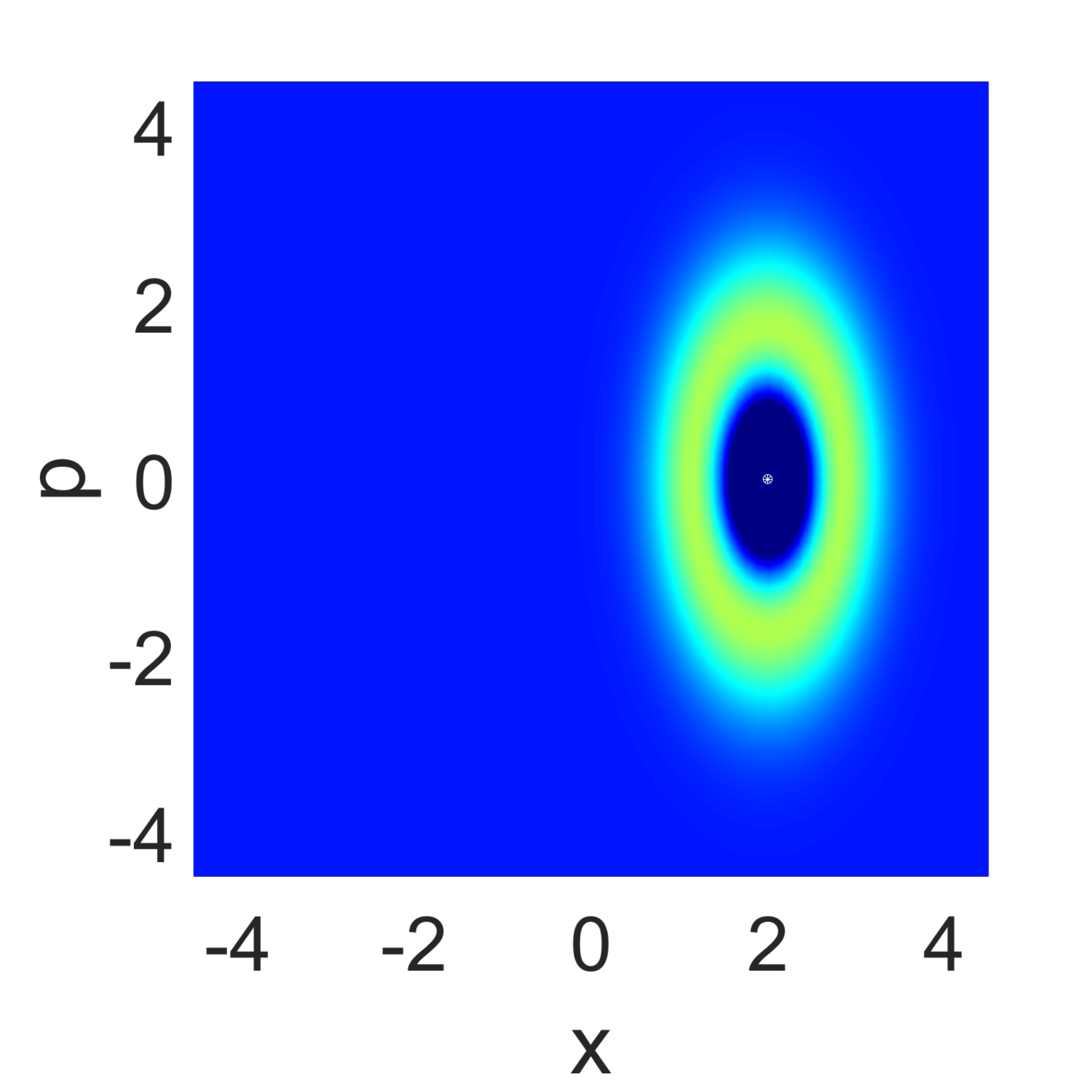}
	  \includegraphics[width=0.24\textwidth]{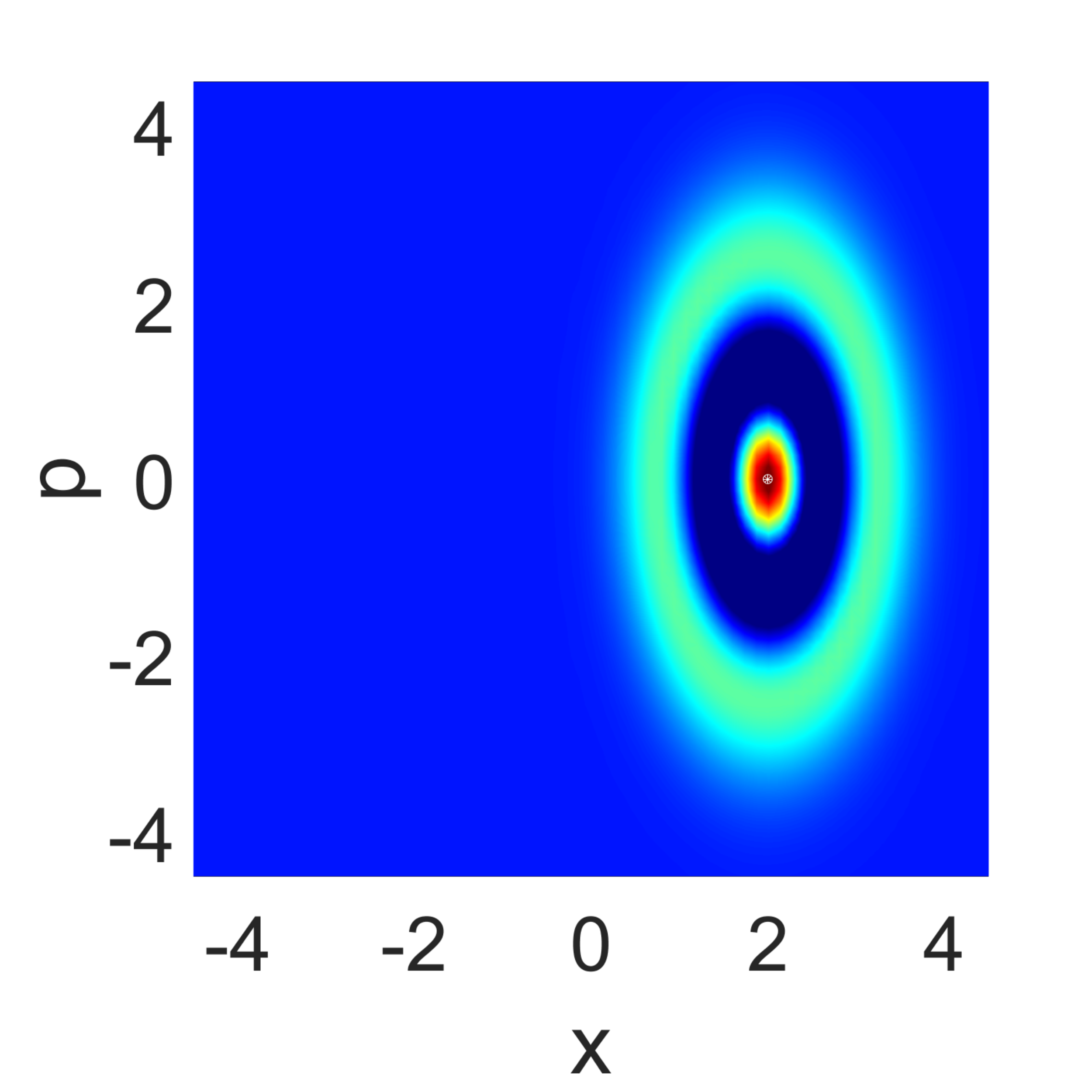}
	  \includegraphics[width=0.24\textwidth]{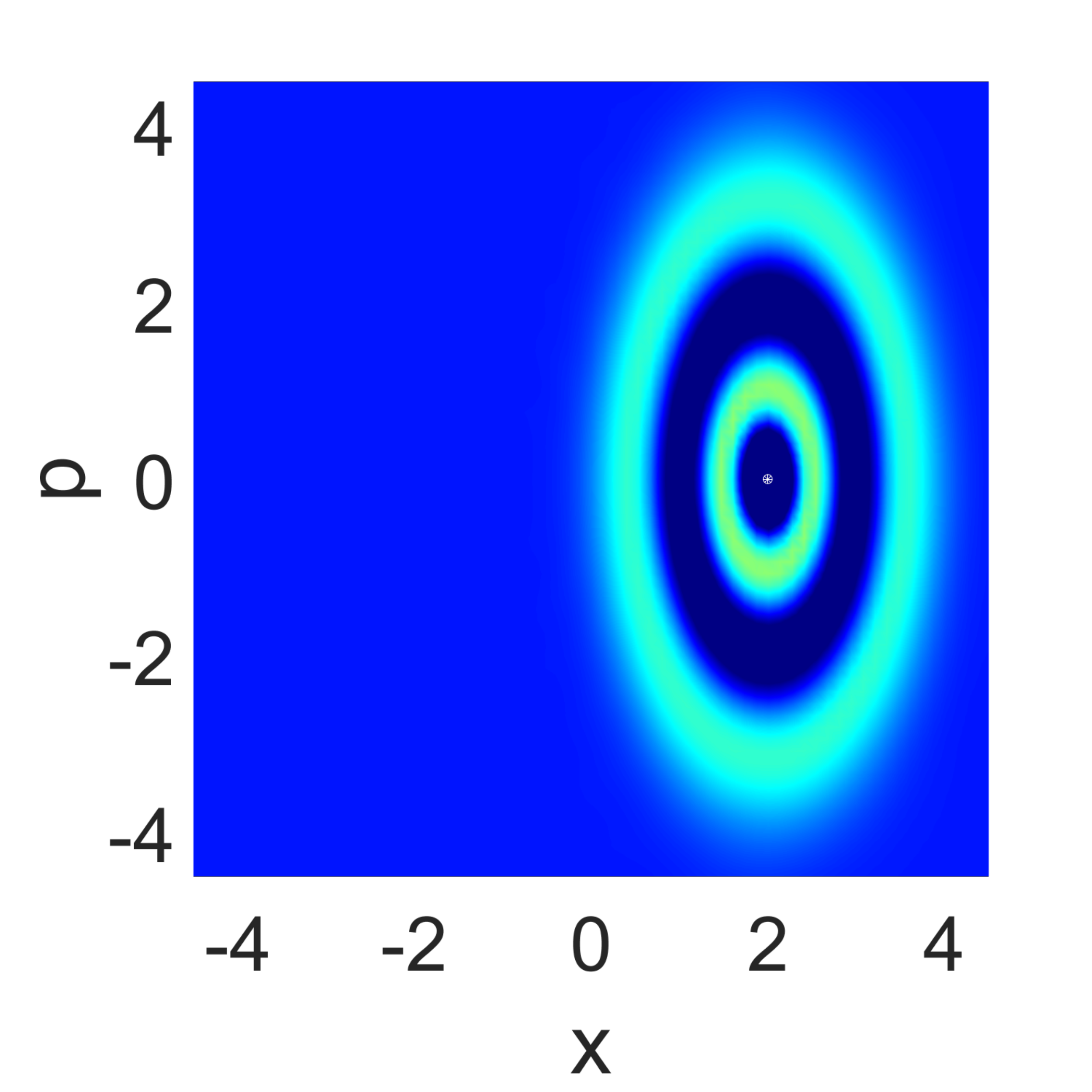}\\
	  \includegraphics[width=0.24\textwidth]{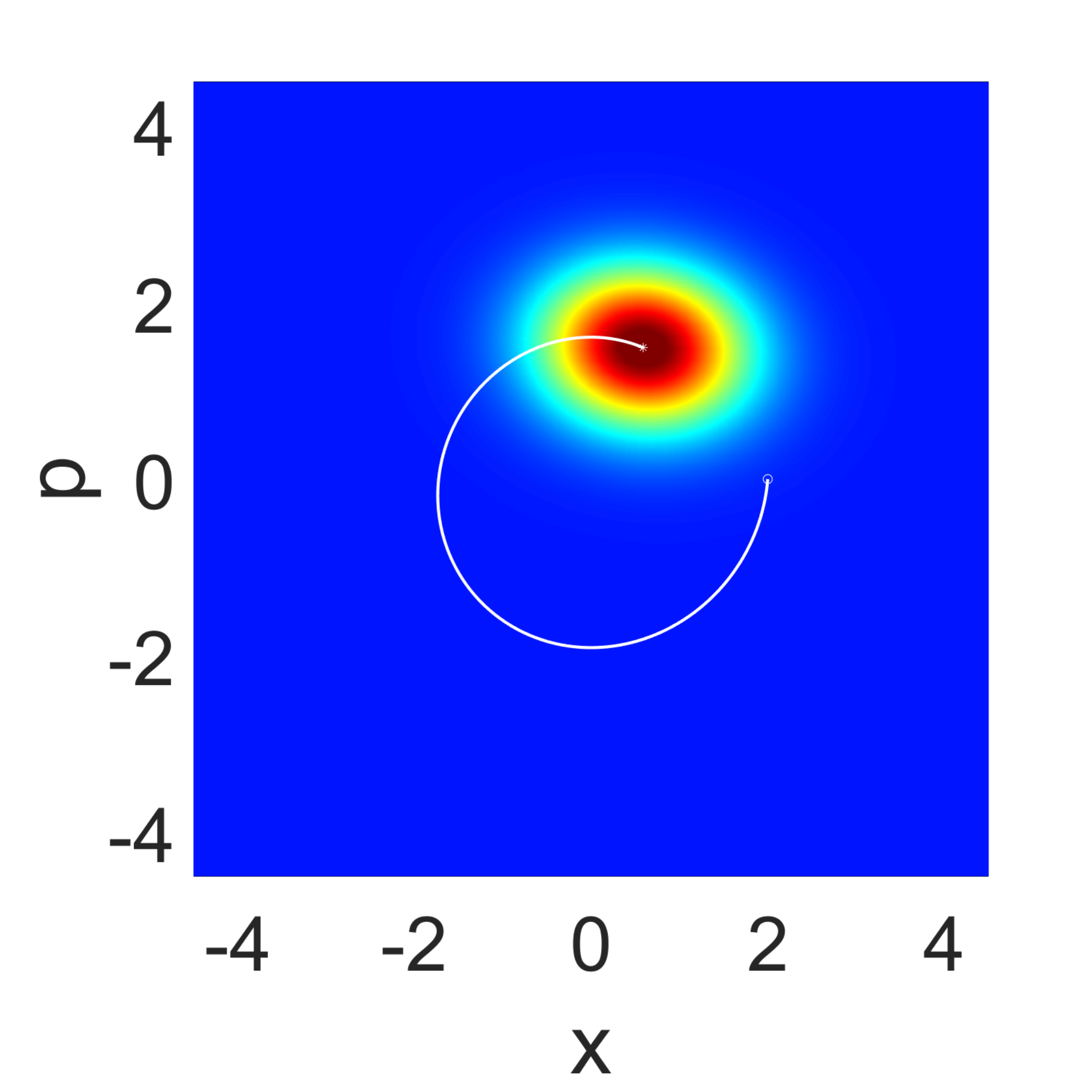}
	  \includegraphics[width=0.24\textwidth]{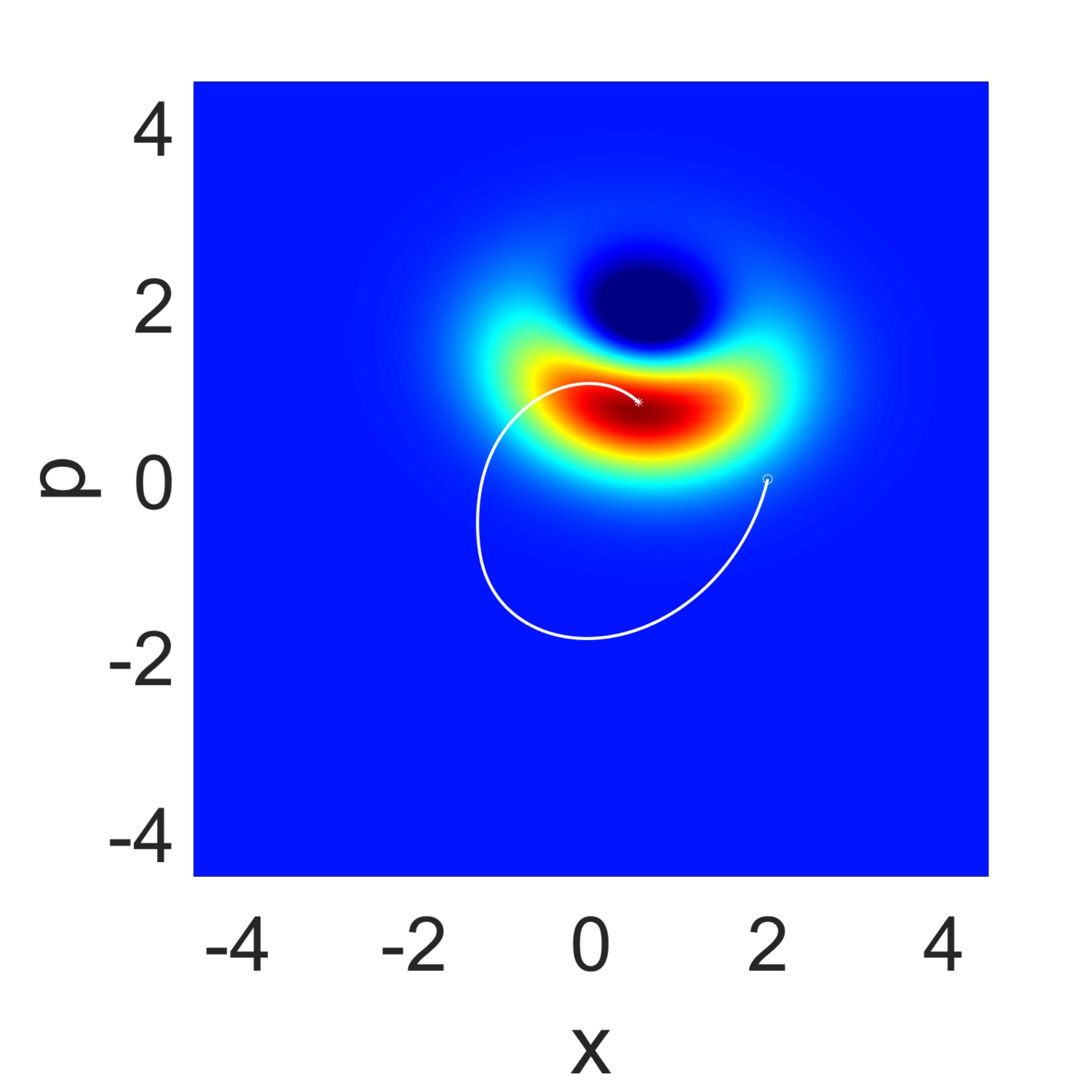}
	  \includegraphics[width=0.24\textwidth]{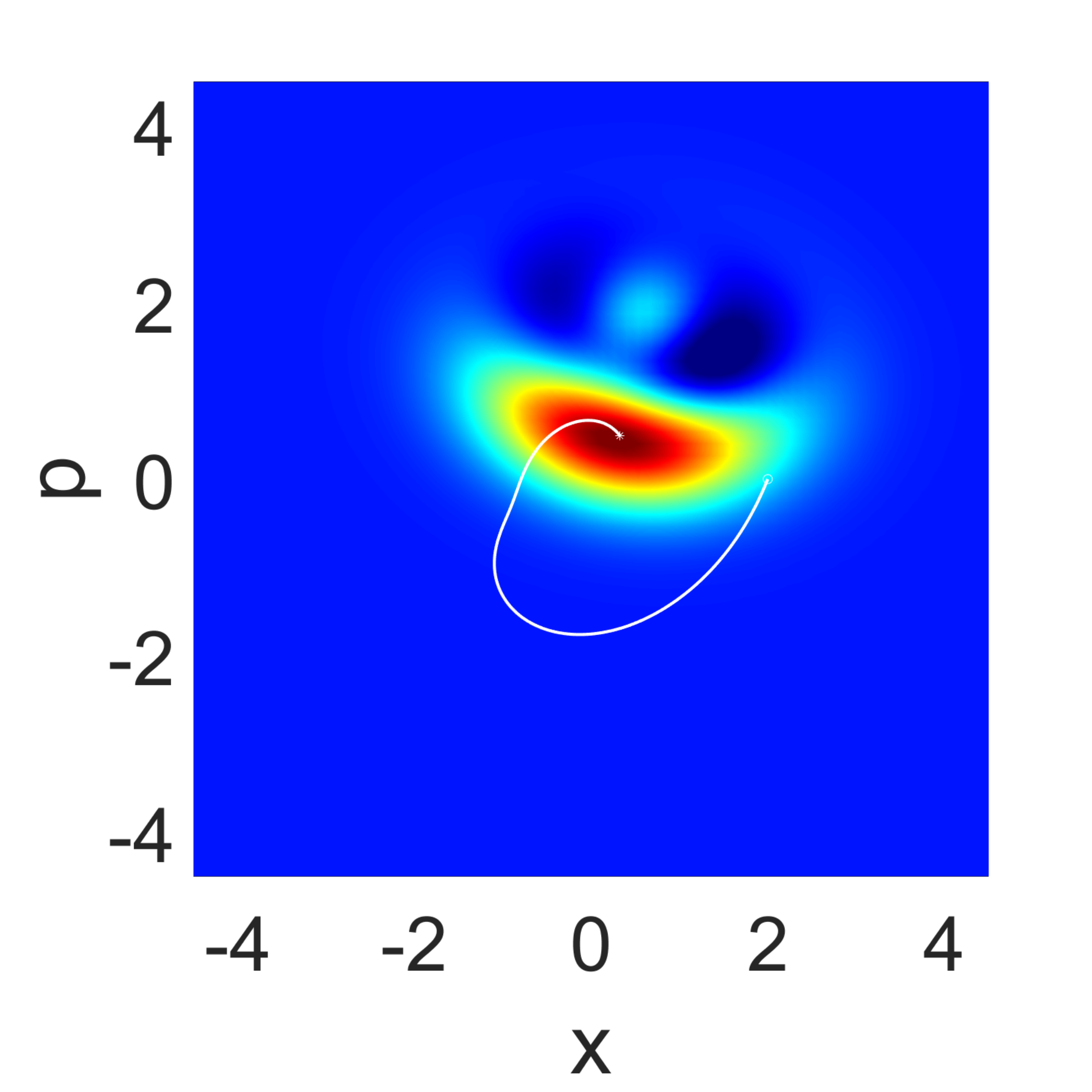} 
	  \includegraphics[width=0.24\textwidth]{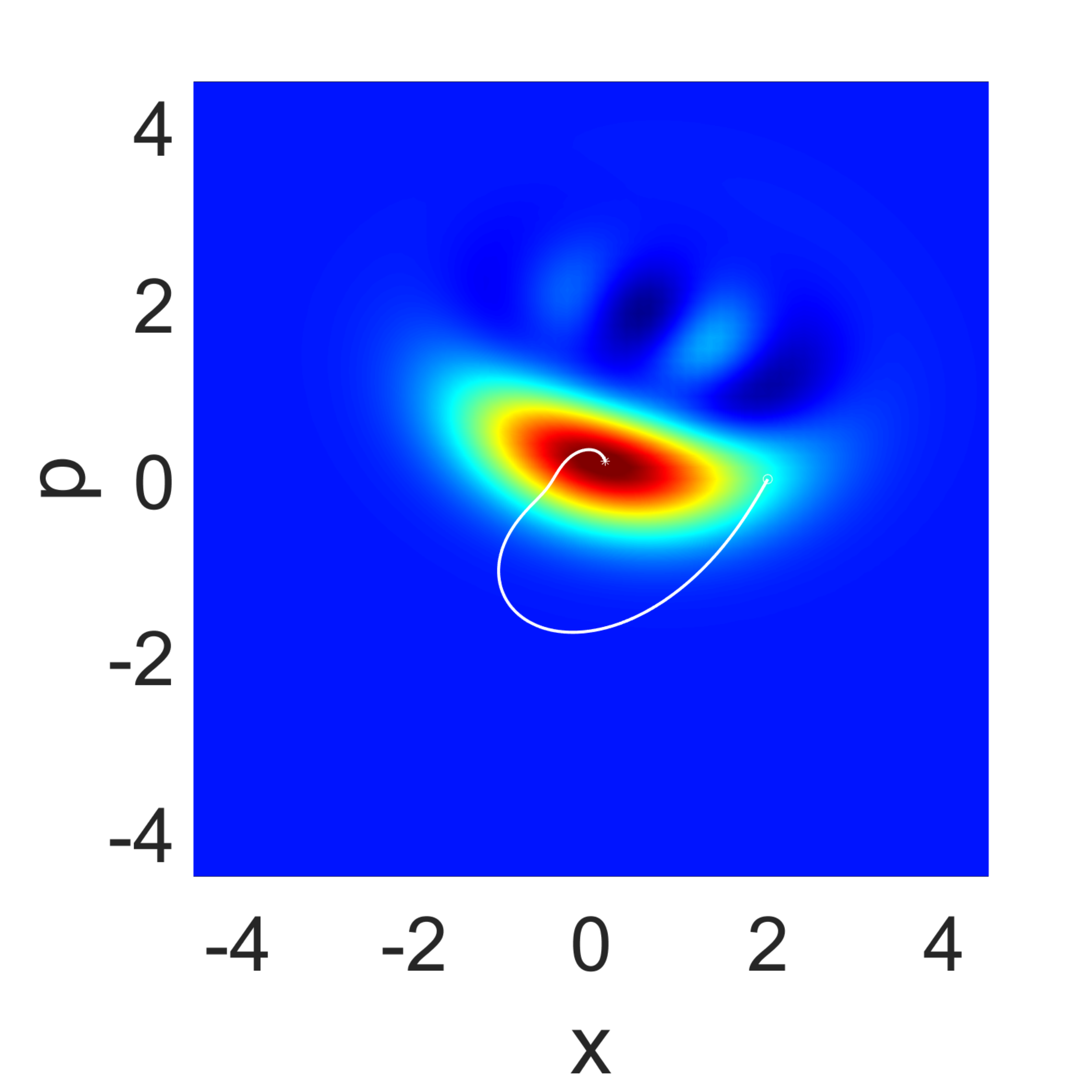}\\
	  \includegraphics[width=0.24\textwidth]{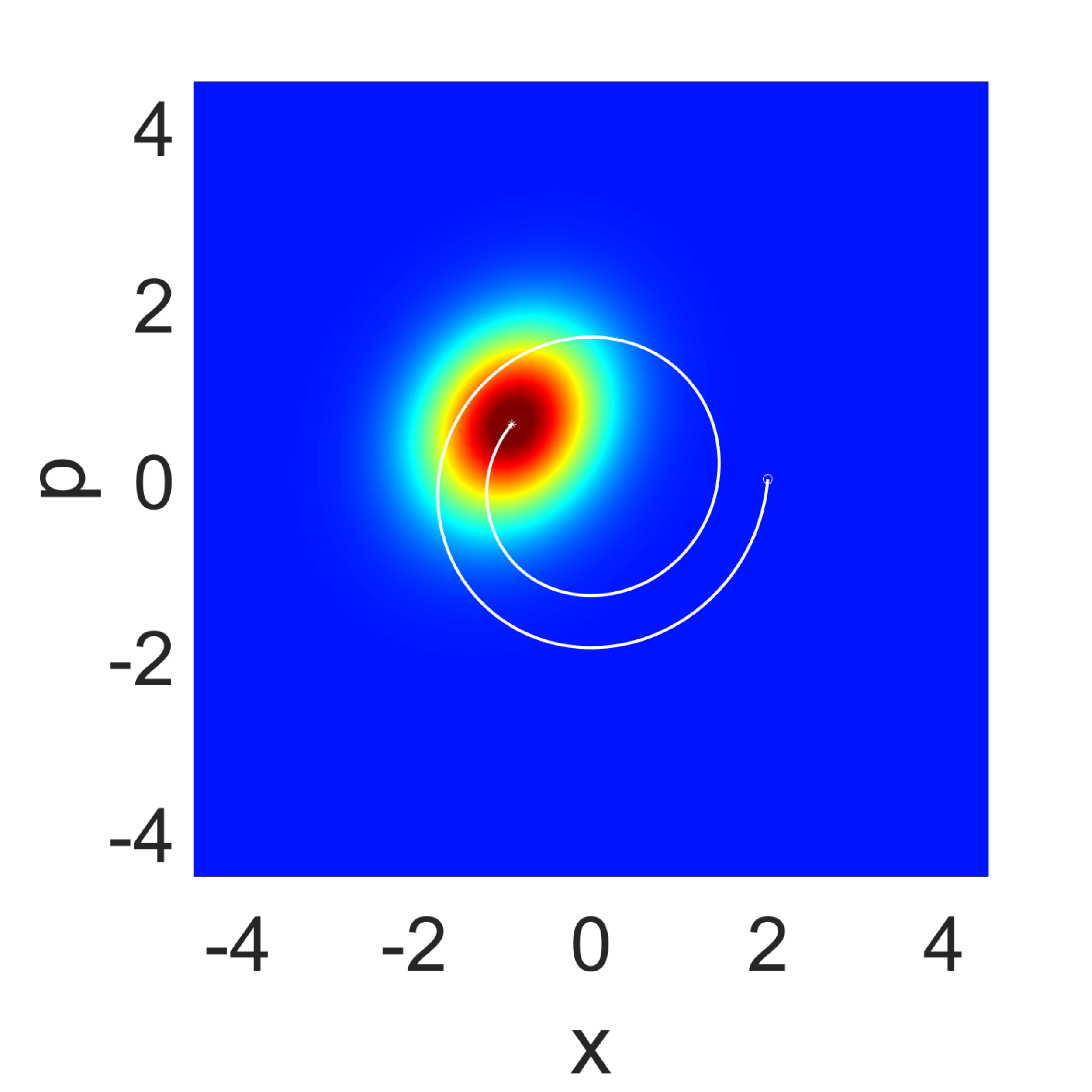}
	  \includegraphics[width=0.24\textwidth]{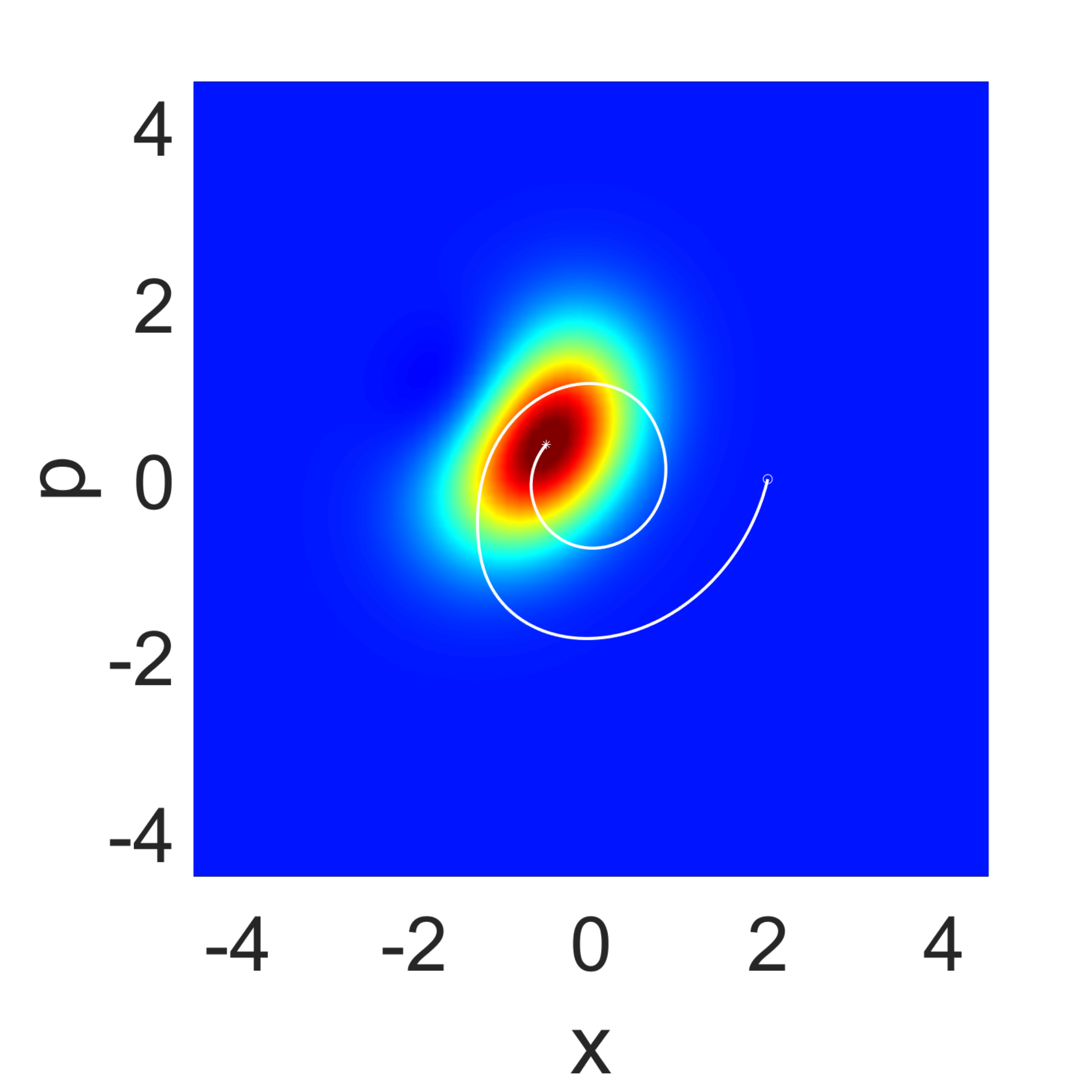}
	  \includegraphics[width=0.24\textwidth]{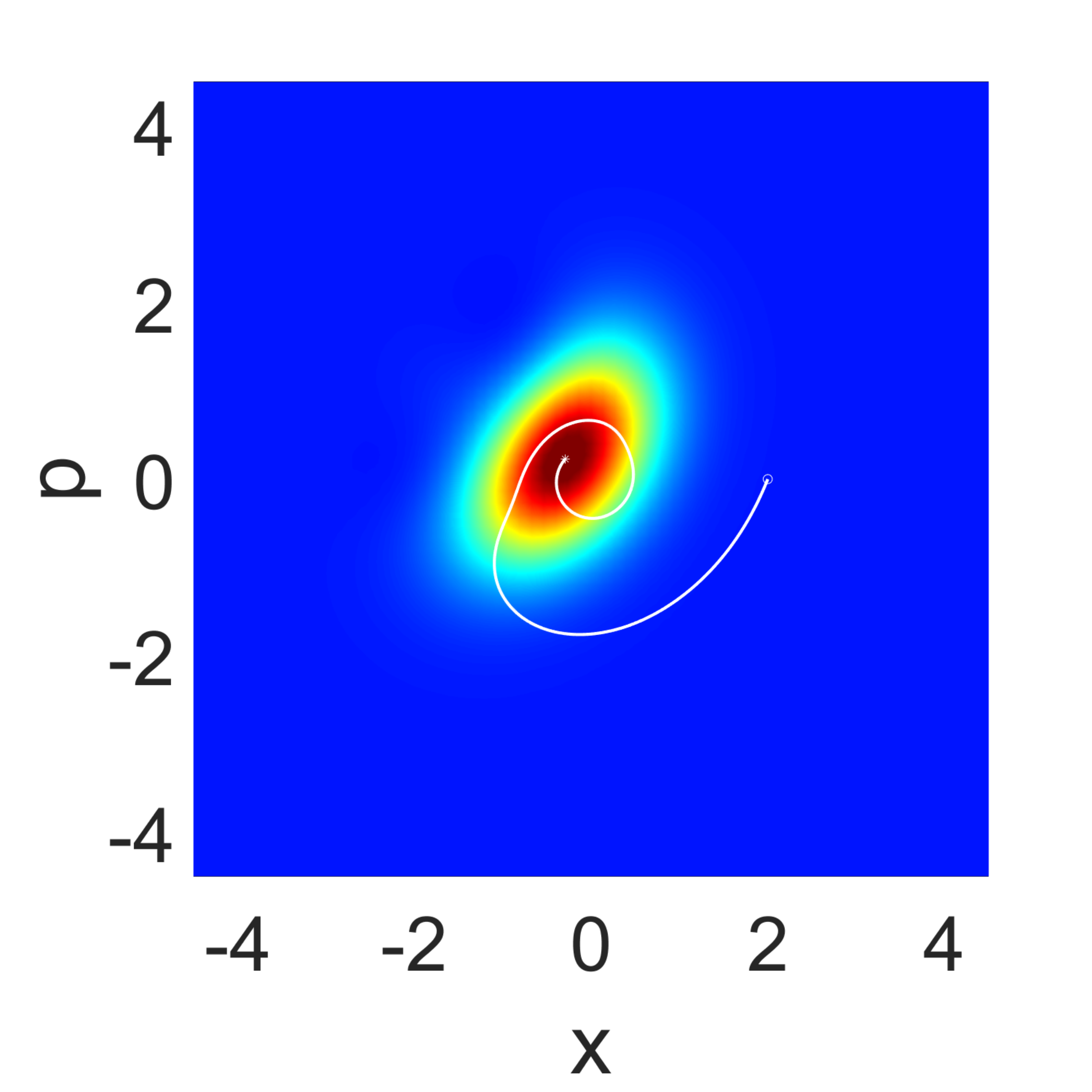}
	  \includegraphics[width=0.24\textwidth]{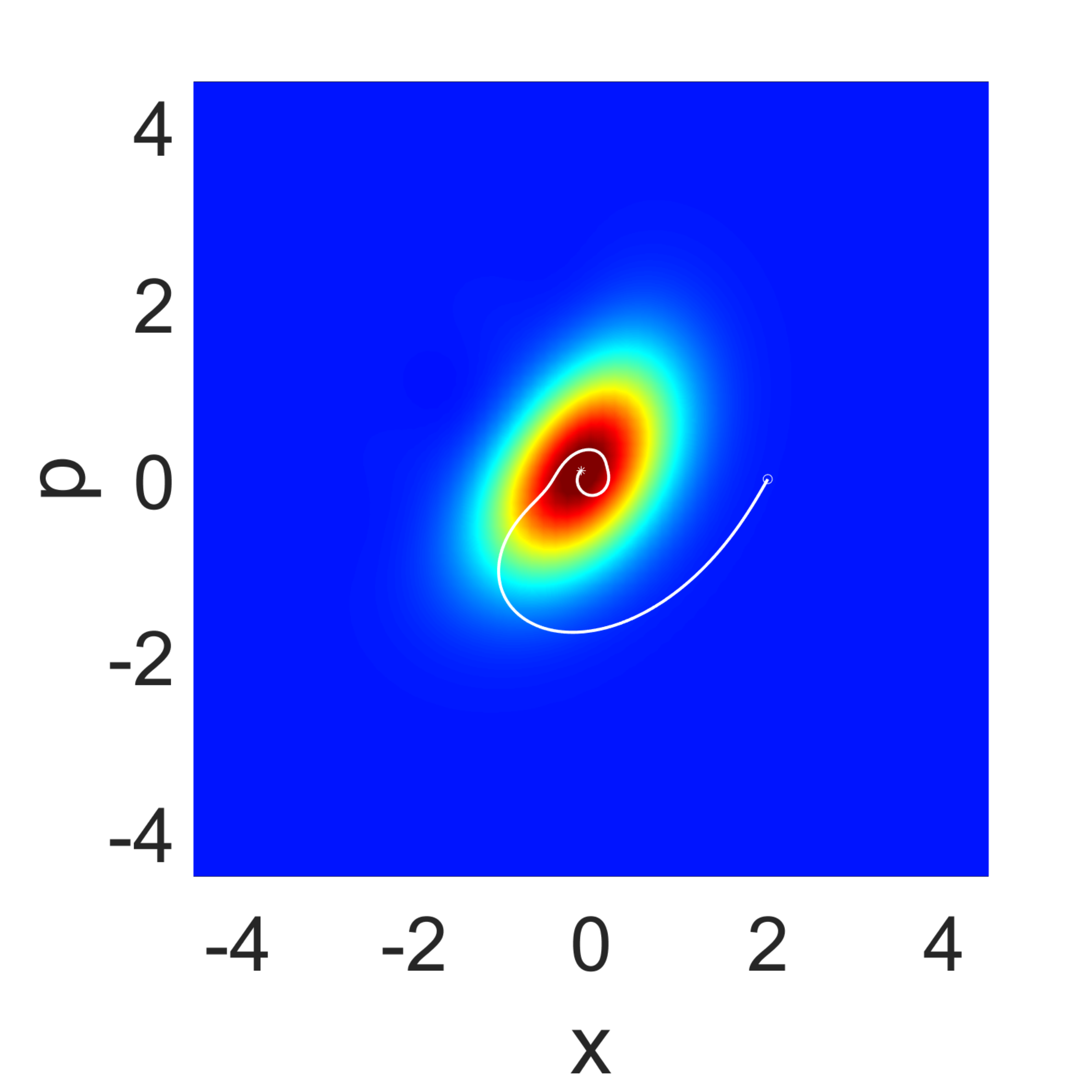} \\
\caption{Wigner function for the first four time evolved Hagedorn states $\hat{U}(t)\ket{n,a_0,z_0}$ as defined in \cref{eq:HagProp}, for the position measurement model \cref{eq:PosModel} (n from 0-4 from left to right) at times $t=0,5,10$ (top to bottom)}\label{fig:HagBasisProp}
\end{centering}
\end{figure}
These quantities determine the time evolution of the Hagedorn basis states $\ket{n,a_t,z_t}$, the first four of which are depicted for different times in \cref{fig:HagBasis}. In \cref{fig:HagBasis} we can see how the entire Hagedorn basis is squeezed and shifted uniformly with $a_t$  and $z_t$. In contrast in \cref{fig:HagBasisProp} we see how the propagated initial states $\hat U(t)\ket{n,a_0,z_0}$ behave differently with each state decaying towards a squeezed state at the origin, albeit at different rates. Note that the evolution of the Hagedorn basis states does not depend on the specific realisation of the quantum-jump trajectory. What differs between different quantum-jump realisations are the coefficients of the state in this basis. For a given initial state they remain constant between the jumps and are updated at each jump according to equation (\ref{eqn:Hagjumpcoeffmap}). The coefficients for our example just before and just after the first jump, and just after the second jump are depicted in the bottom row of figure \ref{fig:WigPlotsPos_QJ}.

\begin{figure}
\begin{centering}
	  \includegraphics[width=0.3\textwidth]{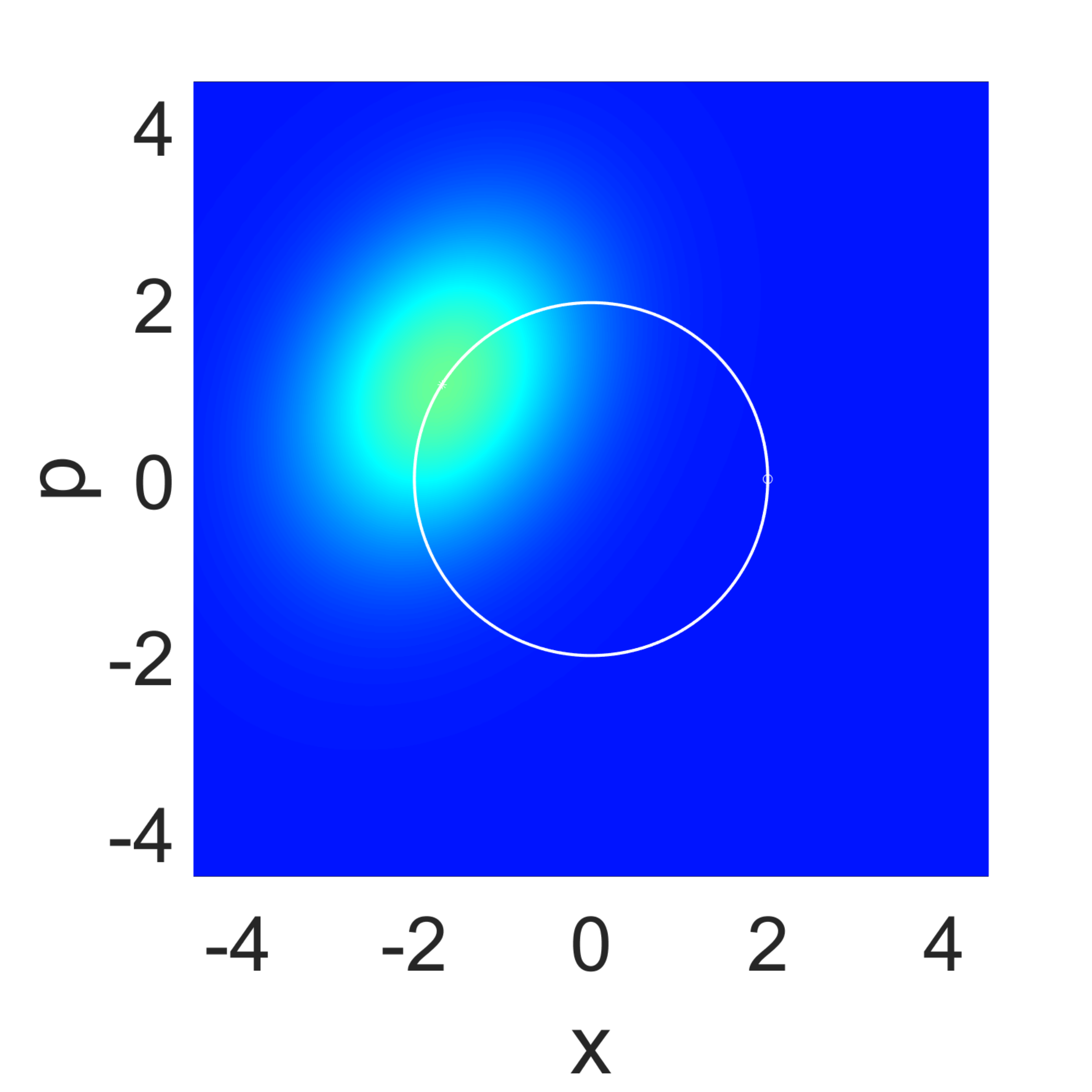}
	  \includegraphics[width=0.3\textwidth]{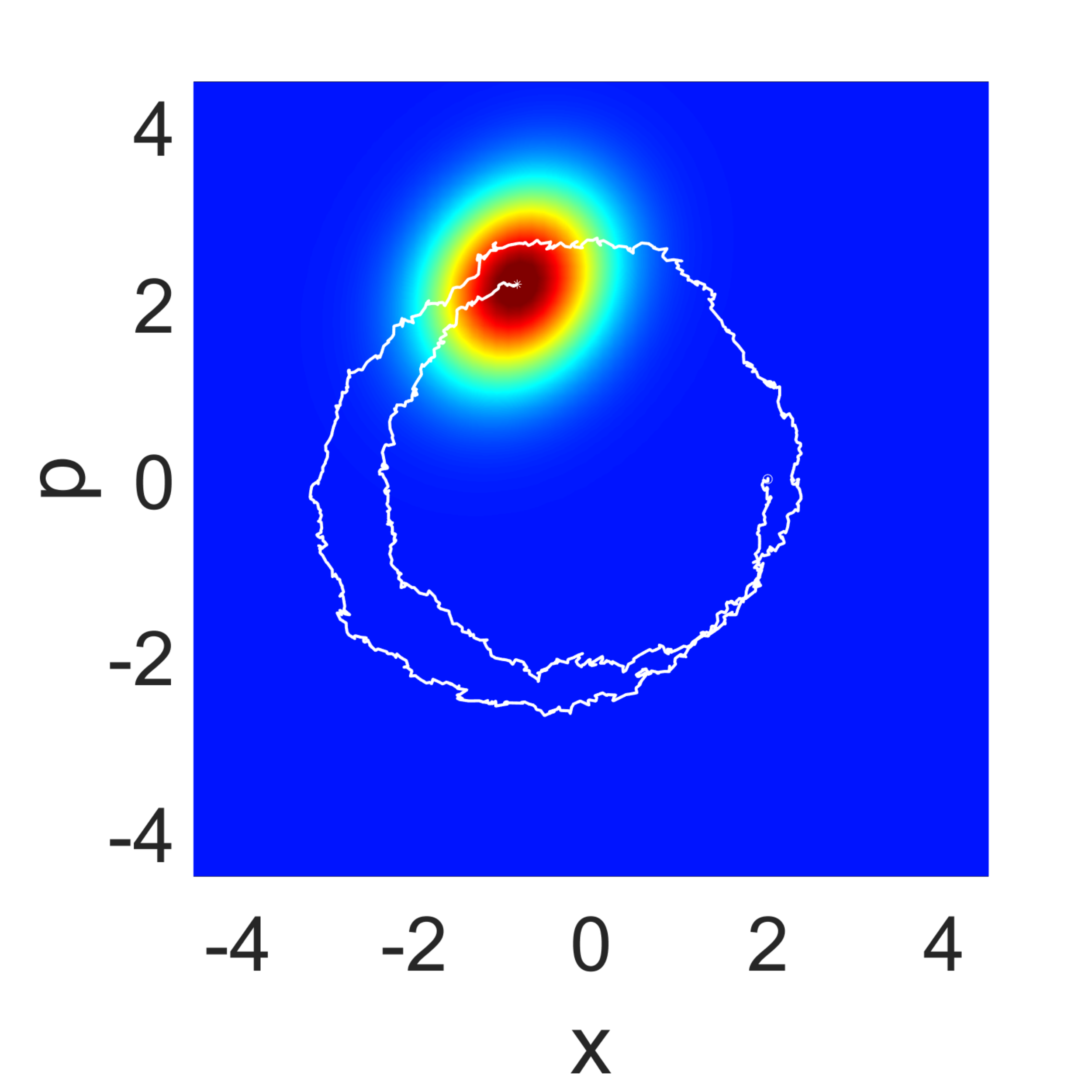}
	  \includegraphics[width=0.3\textwidth]{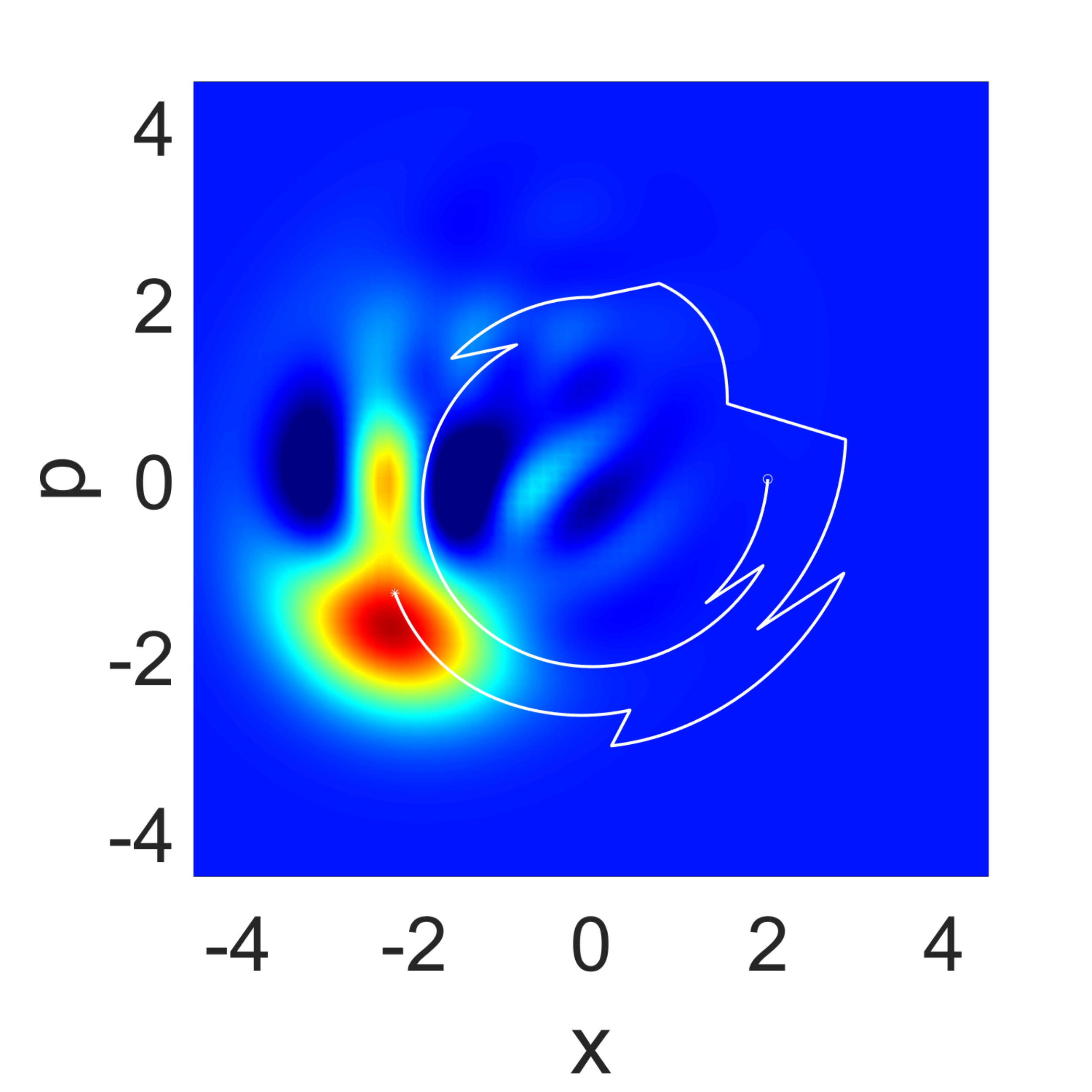}
\caption{Lindblad dynamics (left) compared with a single trajectory of the SSE (middle) and quantum jump (right) for the position measurement model \cref{eq:PosModel}, with parameters $\omega=1$ and $\gamma=0.2$. The initial Gaussian is a squeezed state (i.e. $a_0=(\tfrac{1}{\sqrt{2}}, i\sqrt{2})^T$ or $G=\left(\begin{smallmatrix}2&0\\0&\sfrac{1}{2}\end{smallmatrix}\right)$), centered at $z_0=\left(2,0\right)^T$. In each case, a snapshot of the Wigner function at $t=10$ is plotted in phase space, with a white line displaying the precedent central motion.} \label{fig:WigPlotsPos}
\end{centering}
\end{figure}

To summarise, we observe clear differences between the Lindblad, SSE and quantum-jump dynamics for the harmonic oscillator with position measurement, which can be understood to a large degree using the analytical treatment developed in the previous sections. Figure \ref{fig:WigPlotsPos} shows the Wigner functions of the state at $t=10$ for the three different realisations together with the central trajectory up until this time. In the right panel, corresponding to the Lindblad dynamics the state remains Gaussian, and its central motion follows the usual unitary harmonic oscillator trajectory. The increased uncertainties in position and momentum lead to the broadening of the Gaussian apparent here. The SSE dynamics in the central picture, on the other hand, also remains Gaussian in shape and stays well localised as predicted by the dynamical behaviour of $G$. The central trajectory performs a Brownian motion around the harmonic oscillator trajectory. Finally, the quantum-jump trajectory performs smooth stretches of damped harmonic oscillations interrupted by discrete jumps, and crucially, the state does not remain Gaussian.  

\FloatBarrier
\subsection{Example 2: Damped harmonic oscillator}
As a second example, we consider again a harmonic oscillator Hamiltonian and the non-Hermitian Lindbladian
\begin{equation} \label{eq:AnhModel}
\hat L=\sqrt{\frac{\gamma}{2}}(\hat x+i \hat p).
\end{equation}
For an initially coherent state, the resulting dynamics are rather trivial, as such a state is an eigenstate of both the Hamiltonian and the Lindbladian. As a result, for both the quantum-jump and SSE the stochastic terms vanish and the dynamics of all three descriptions are the same, simply transporting the initial state along the trajectories of the damped oscillator. Using a squeezed initial state instead, the motion becomes more interesting and Lindblad, SSE and quantum-jump trajectories differ.  

Explicitly the dynamical equations for the Lindblad evolution \cref{eq:LindSemi} become
\begin{equation} 
    \begin{aligned} \label{eq:LindAnh}
    \frac{d  z_t}{d t} &= (\omega\Omega -\frac{\gamma}{2 }) z_t, \\
  	\frac{dG}{dt} &=\omega(\Omega G-G\Omega) +\gamma( G - G^2).
    \end{aligned}
\end{equation}
The equation for the central dynamics is simply that of a damped oscillator. It can be written in terms of a second-order differential equation for $\tilde x$ taking the more familiar form
\begin{equation}
\frac{d^2 x_t}{d t^2}+\frac{\gamma}{2 }\frac{d x_t}{dt}+\omega x_t=0.
\end{equation}
Thus, the central dynamics are given by 
\begin{equation}
    \begin{pmatrix} x_t\\p_t\end{pmatrix}=e^{-\frac{\gamma t}{2 }}\begin{pmatrix}\cos(\omega t) &\sin(\omega t)\\-\sin(\omega t)&\cos(\omega t)\end{pmatrix}\begin{pmatrix}x_0\\p_0 \end{pmatrix}.
\end{equation}

\begin{figure}
  \centering
 	  \includegraphics[width=0.4\textwidth]{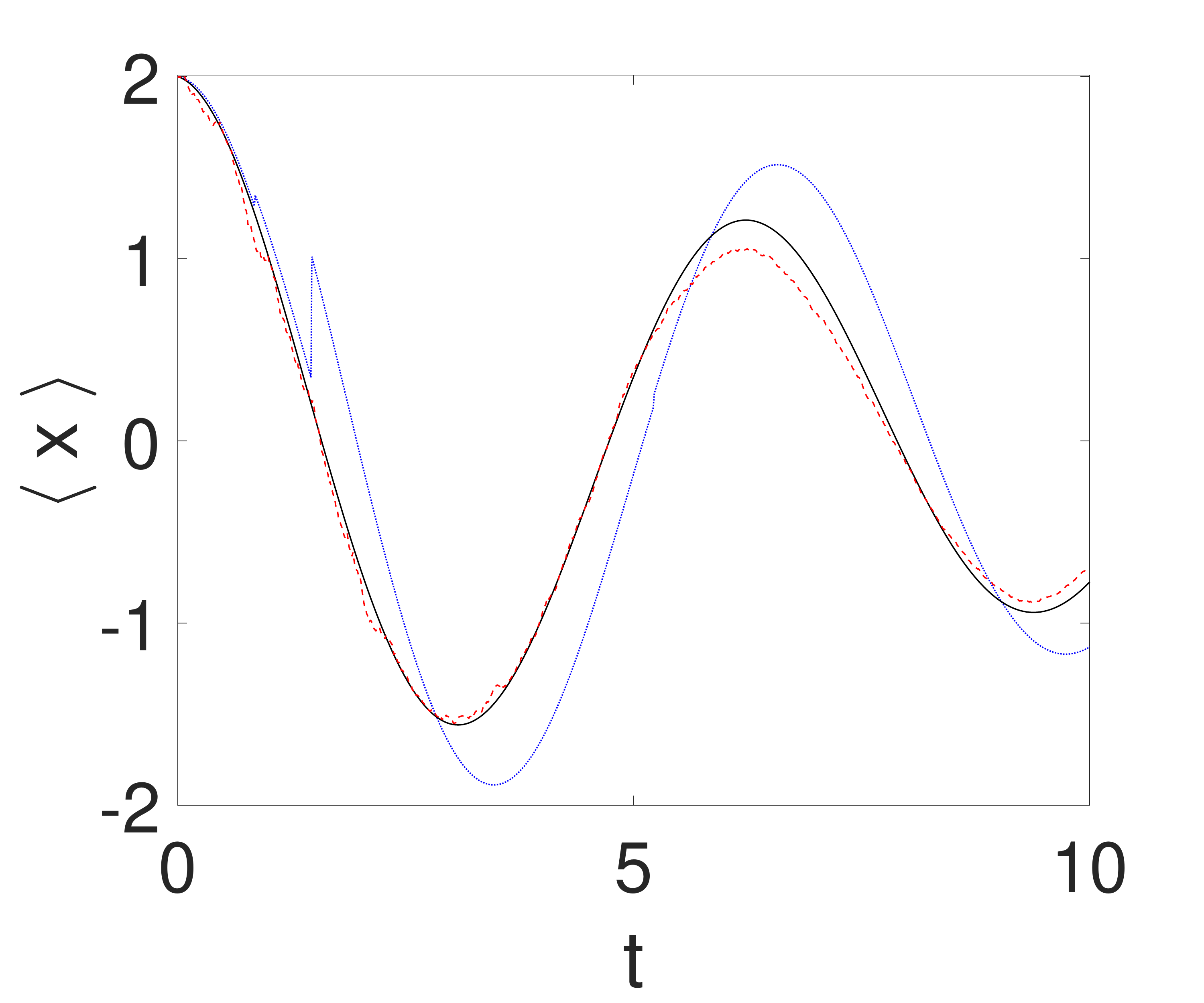}
	  \includegraphics[width=0.4\textwidth]{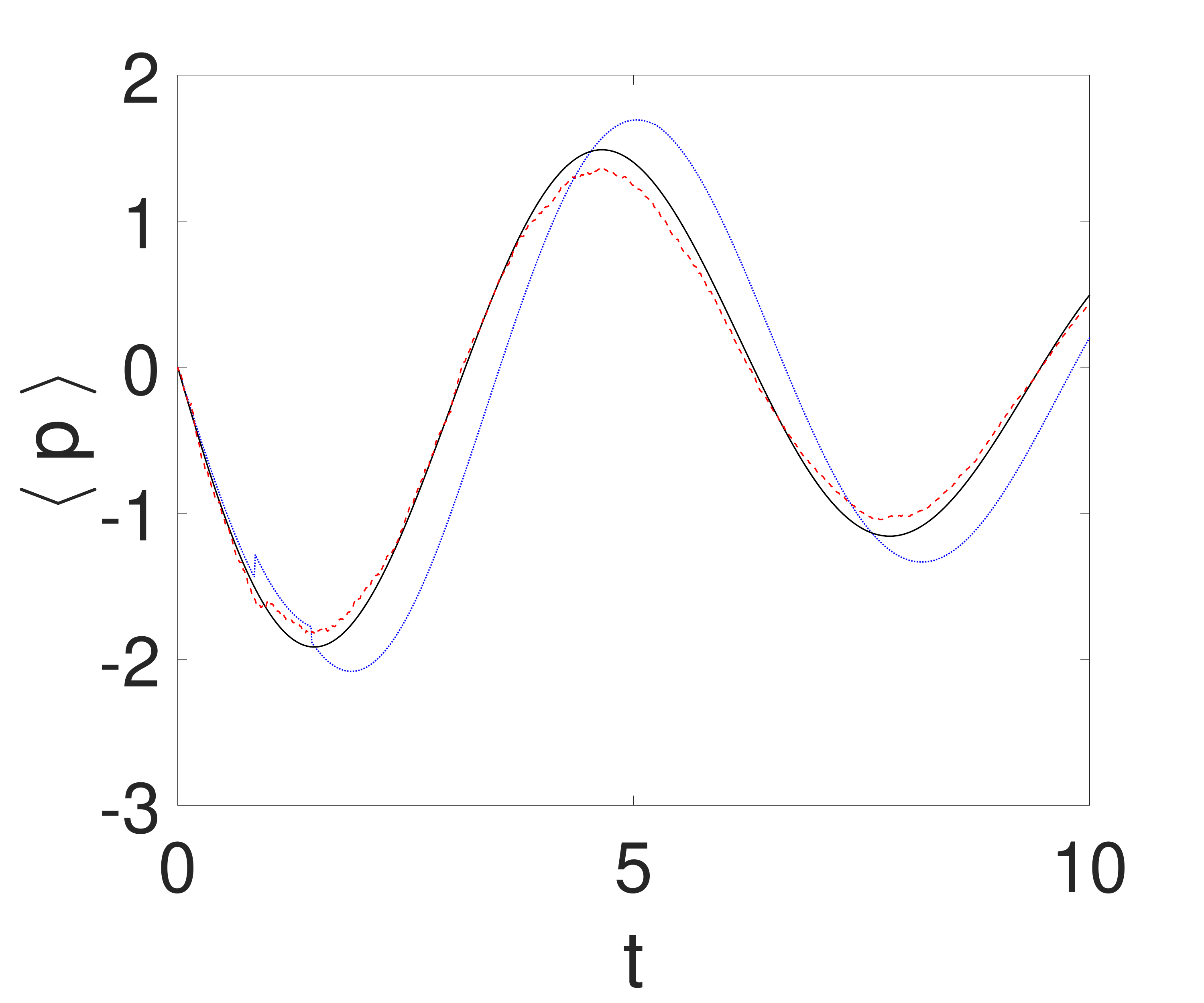}\\
	  \includegraphics[width=0.4\textwidth]{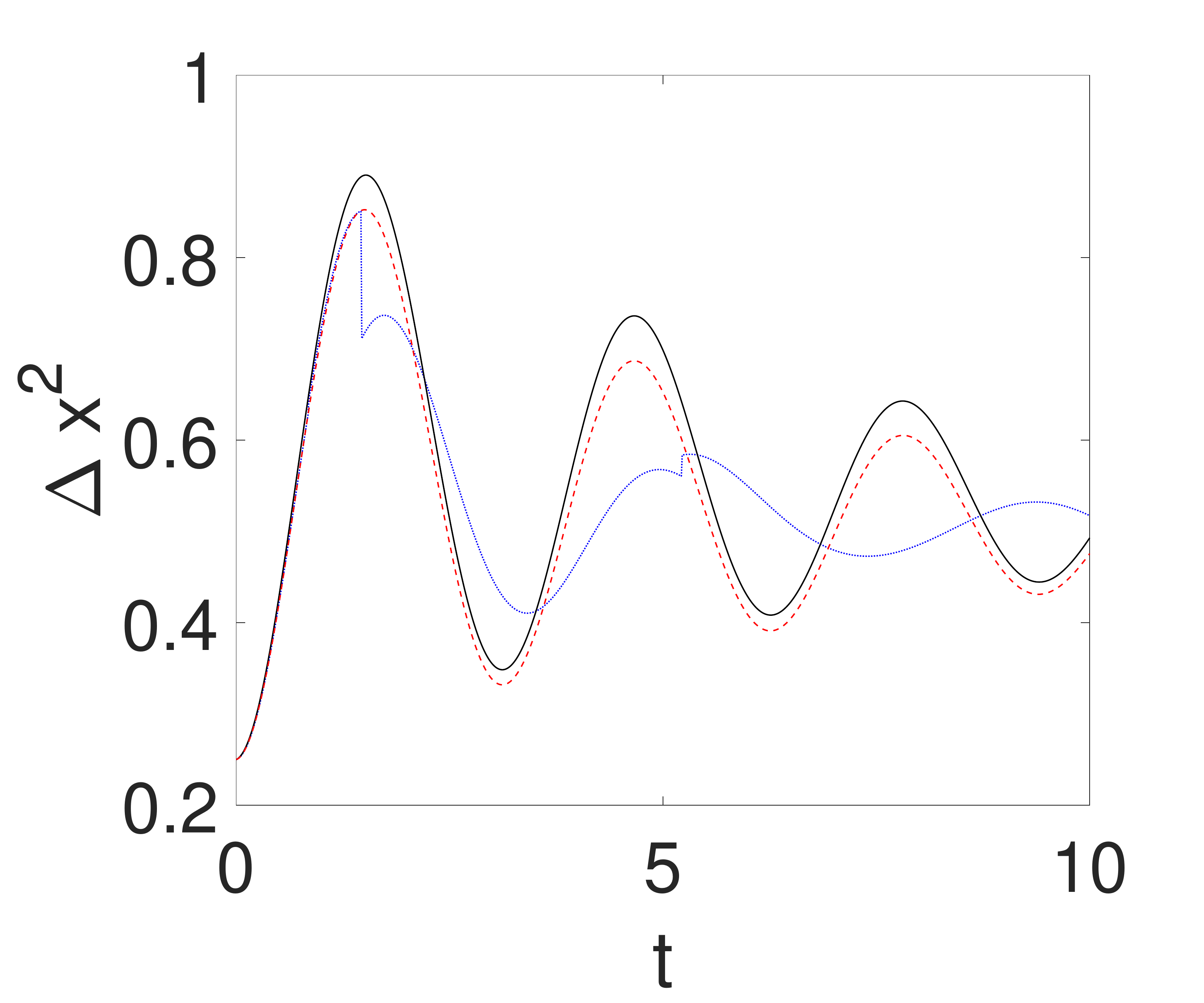}
	  \includegraphics[width=0.4\textwidth]{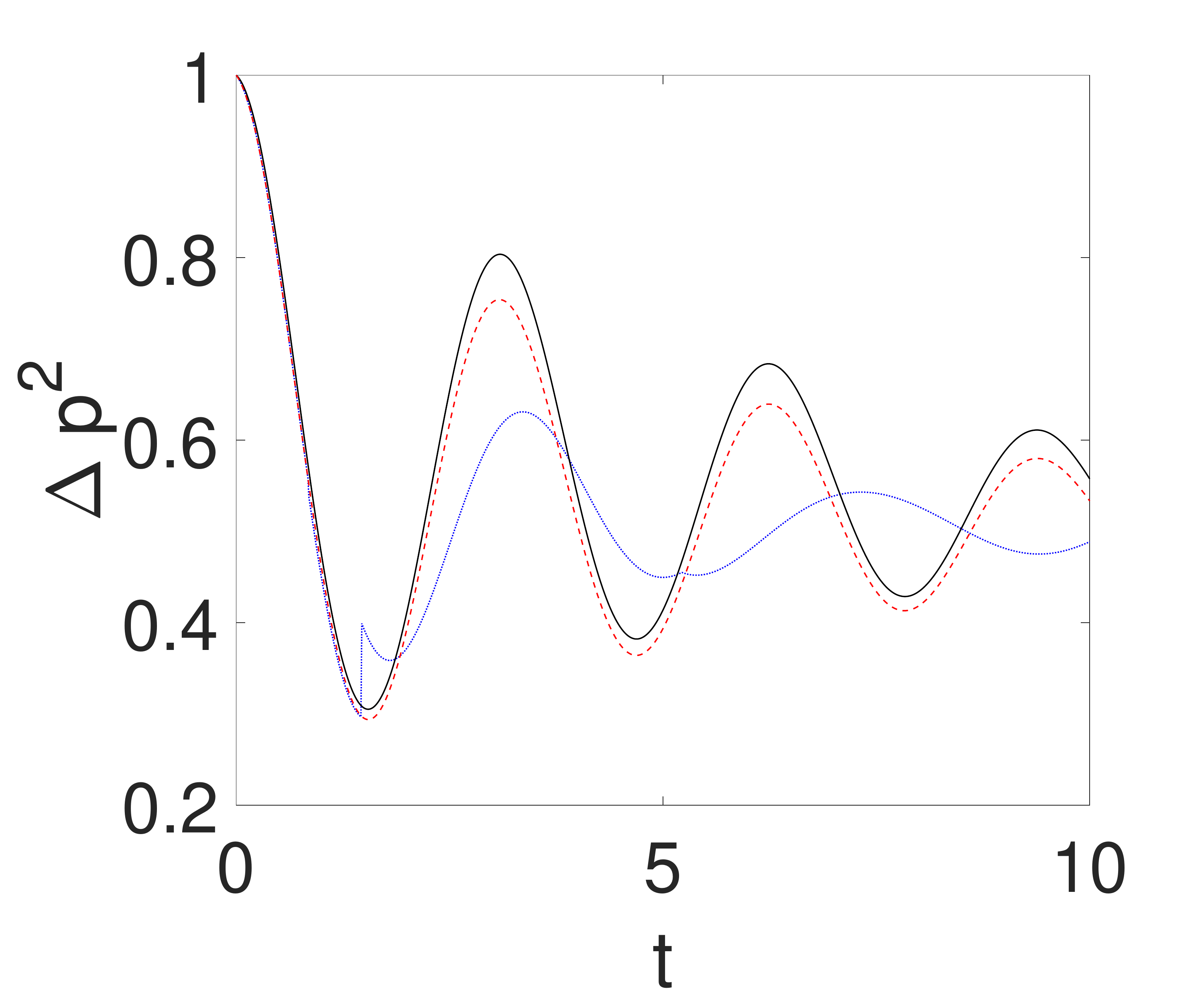}
\caption{  Lindblad dynamics (solid black line) compared with single trajectories of the SSE (dashed red line) and quantum-jump method (dotted blue line) for the damped oscillator model \cref{eq:AnhModel} with $\omega=1$ and $\gamma=0.2$. The initial Gaussian is a squeezed state (i.e. $a_0=(\tfrac{1}{\sqrt{2}}, i\sqrt{2})^T$ or $G=\left(\begin{smallmatrix}2&0\\0&\sfrac{1}{2}\end{smallmatrix}\right)$), centered at $ z_0=\left(2,0\right)^T$. We show the time dependence of the position expectation $\braket{\hat x}$ (top left), the momentum expectation $\braket{\hat p}$ (top right), the positional variance $\Delta x^2$ (bottom left) and the momentum variance $\Delta p ^2$ (bottom right).}\label{fig:SingleBehaviorAnh}
\end{figure}

In the equation for the covariance matrix $G$ we immediately observe the fixed point for $G=\mathbb{I}$, corresponding to a coherent state, which is indeed approached asymptotically by any initial Gaussian state. For an initially squeezed state with $G(0)=\big(\begin{smallmatrix}\zeta&&0\\0&&\sfrac{1}{\zeta}\end{smallmatrix} \big)$ we find the time-dependent covariances
\begin{equation}
    \Sigma(t)=\frac{\hbar}{2}
\begin{pmatrix}1&0\\0&1\end{pmatrix}+ \frac{\hbar e^{-\gamma t}}{4 \zeta}\begin{pmatrix}
 (\zeta -1)^2-\left(\zeta ^2-1\right) \cos (2 \omega t  ) &
   \left(\zeta ^2-1\right)  \sin (2 \omega t ) \\
  \left(\zeta ^2-1\right) \sin (2\omega t  ) &  (\zeta
   -1)^2+\left(\zeta ^2-1\right) \cos (2 \omega t  ). \\
\end{pmatrix}
\end{equation}
That is, we observe the usual oscillations with frequency $2\omega$ around the coherent state covariances, which are now damped, and $\Sigma(t)$ asymptotically approaches $\Sigma(t)\to\frac{\hbar}{2}\mathbb{I}$, while $\tilde z\to (0,0)^T$, resulting in a coherent state at the origin. This behaviour is clearly visible in figure \ref{fig:SingleBehaviorAnh}, in which the expectation values and uncertainties of position and momentum in the Lindblad case are depicted as black solid lines for an initial state with $\zeta=2$ centred at $\tilde z_0=(2,0)^T$. 

The dynamical \cref{eq:SSEParamz,eq:SSEParamG} in the SSE case become
\begin{equation} 
    \begin{aligned} \label{eq:SSEAnh}
    \frac{d  z_t}{d t} &= (\omega\Omega -\frac{\gamma}{2 }) z_t dt +\frac{\sqrt{\hbar \gamma}}{2}(G^{-1}-\mathbb{I})\begin{pmatrix}1 \\ 0\end{pmatrix}d\xi_R-\frac{\sqrt{\hbar \gamma}}{2}(G^{-1}-\mathbb{I})\begin{pmatrix}0 \\ 1\end{pmatrix}d\xi_I,\\
  	\frac{dG}{dt} &=\omega(\Omega G-G\Omega) +\frac{\gamma}{2 }( \mathbb{I} - G^2).
    \end{aligned}
\end{equation}
The deterministic part of the SSE and the Lindblad central dynamics agree as usual. The additional drift term is proportional to $G^{-1}-\mathbb{I}$, and vanishes for the coherent state $G=\mathbb{I}$. Although the evolution of $G$ differs from the Lindblad case, it too asymptotically approaches the fixed point $G=\mathbb{I}$, independent of the initial conditions. Thus, the stochastic contributions to the central trajectory become smaller over time, and the dynamics drive the state towards a coherent state at the origin, just as in the Lindblad case. 

The complexified linearised flow of the damped oscillator model \cref{eq:AnhModel} is explicitly given by
\begin{equation} \label{eq:SAnh}
    S(t)=\begin{pmatrix}
    \cosh \left(\frac{\gamma +2 i  \omega}{2 }t \right) & -i \sinh \left(\frac{\gamma +2 i  \omega}{2 }t \right) \\
 i \sinh \left(\frac{\gamma +2 i  \omega}{2}t \right) & \cosh \left(\frac{\gamma +2 i  \omega}{2}t \right) \\
    \end{pmatrix},
\end{equation}
For an initial squeezed state with $G(0)=\big(\begin{smallmatrix}\zeta&&0\\0&&\sfrac{1}{\zeta}\end{smallmatrix} \big)$ 
this yields the time dependent covariances in the SSE dynamics
\begin{equation}
    \Sigma(t)=\frac{\hbar}{2 f(t)}
\bigg(\begin{smallmatrix}
 (\zeta ^2+1) \cosh (\gamma  t)+2 \zeta  \sinh (\gamma  t)-(\zeta ^2-1) \cos (2  \omega t
   ) & (\zeta ^2-1) \sin (2 
   \omega t) \\
 (\zeta ^2-1) \sin (2  \omega t) &
   (\zeta ^2+1) \cosh (\gamma  t)+2 \zeta  \sinh (\gamma  t)+(\zeta ^2-1) \cos (2 \omega t
   ) \\
\end{smallmatrix}\bigg),
\end{equation}
with 
\begin{equation}
    f(t)=\left(\zeta ^2+1\right) \sinh (\gamma  t)+2 \zeta  \cosh (\gamma  t).
\end{equation}

For long times $t\to\infty$ this behaves as 
\begin{equation}
    \Sigma(t)=\frac{\hbar}{2}\begin{pmatrix}1&0\\0&1\end{pmatrix}+\frac{\hbar (\zeta -1)}{2(1+\zeta)} e^{-\gamma t}\begin{pmatrix}
 - \cos (2 \omega t) & \sin (2 \omega t) \\
  \sin (2  \omega t) & \cos (2 \omega t) \\
\end{pmatrix},
\end{equation}
with $\Sigma(t)$ approaching $\frac{\hbar}{2}\mathbb{I}$ just as in the Lindblad case. The example in figure \ref{fig:SingleBehaviorAnh} nicely demonstrates how the SSE and Lindblad dynamics approach the same limit in different ways. We also observe the reduced amplitude of the noise in the central dynamics, as the matrix $G$ approaches the identity.   

The quantum jump trajectory in this example again initially follows the Gaussian dynamics, here given by
\begin{equation}
  \frac{d  z_t}{d t}=\left(\omega \Omega  -\frac{\gamma}{\hbar} \Sigma(t)\right) z_t,
\end{equation} 
where $G(t)$ evolves dynamically as in the SSE case. This central dynamics is also a type of damped harmonic oscillator, however, it differs from the Lindblad case, both in the presence of a damping term in both position and momentum, and in the modulation of the damping induced by the time-dependent covariance matrix. In the example in figure \ref{fig:SingleBehaviorAnh}, however, we observe that this is a quantitative rather than qualitative difference in this case. 

\begin{figure}
\begin{centering}
	  \includegraphics[width=0.3\textwidth]{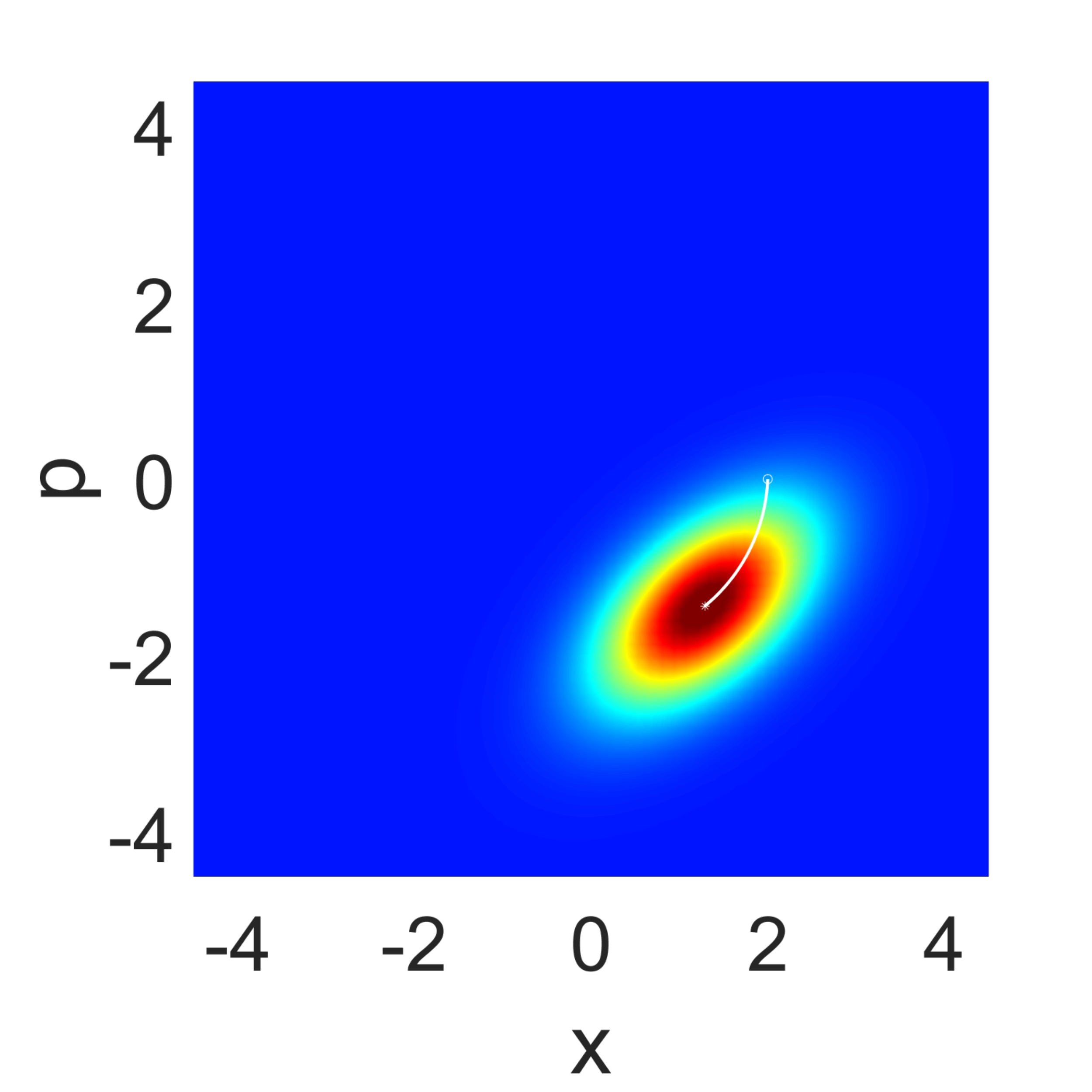}
	  \includegraphics[width=0.3\textwidth]{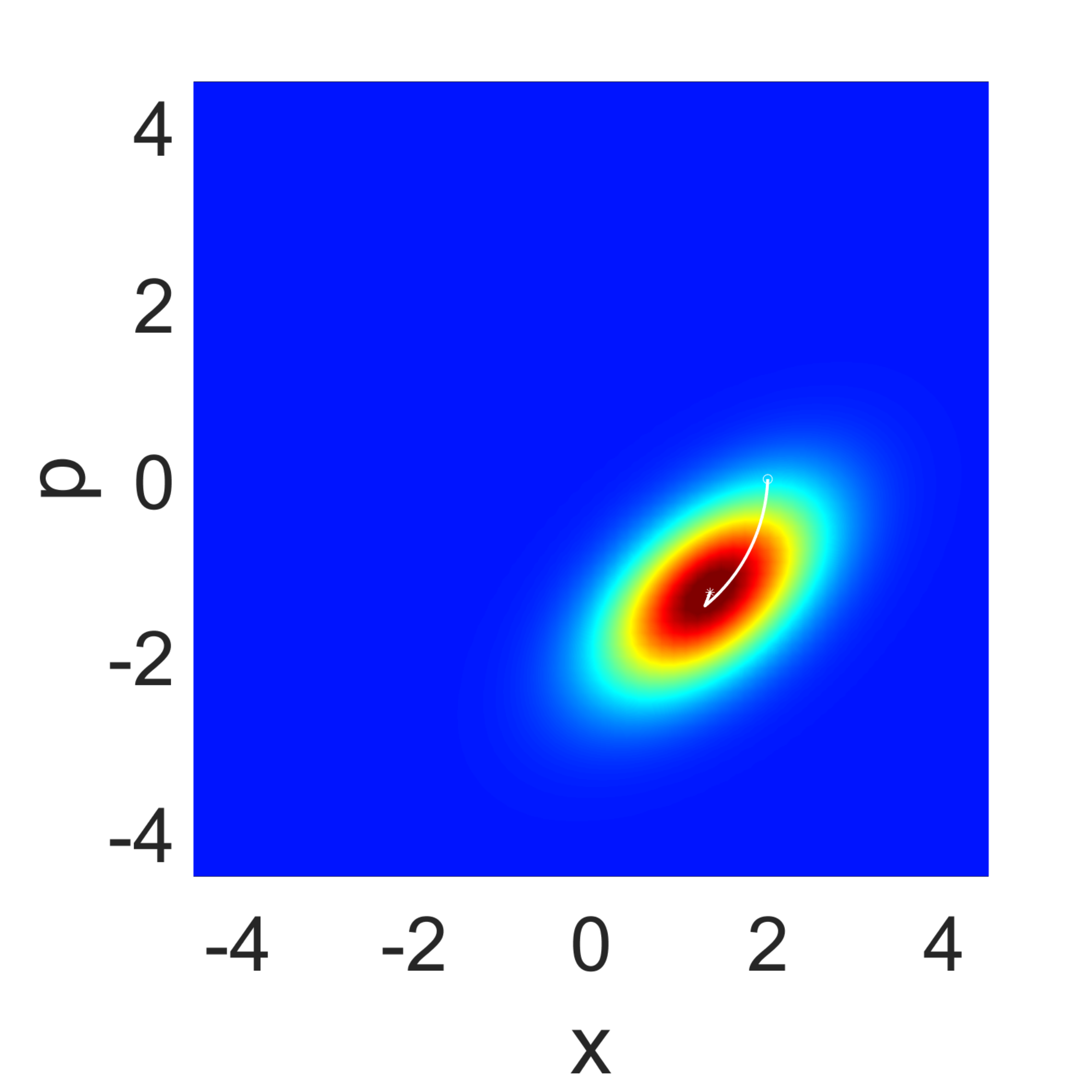}
	  \includegraphics[width=0.3\textwidth]{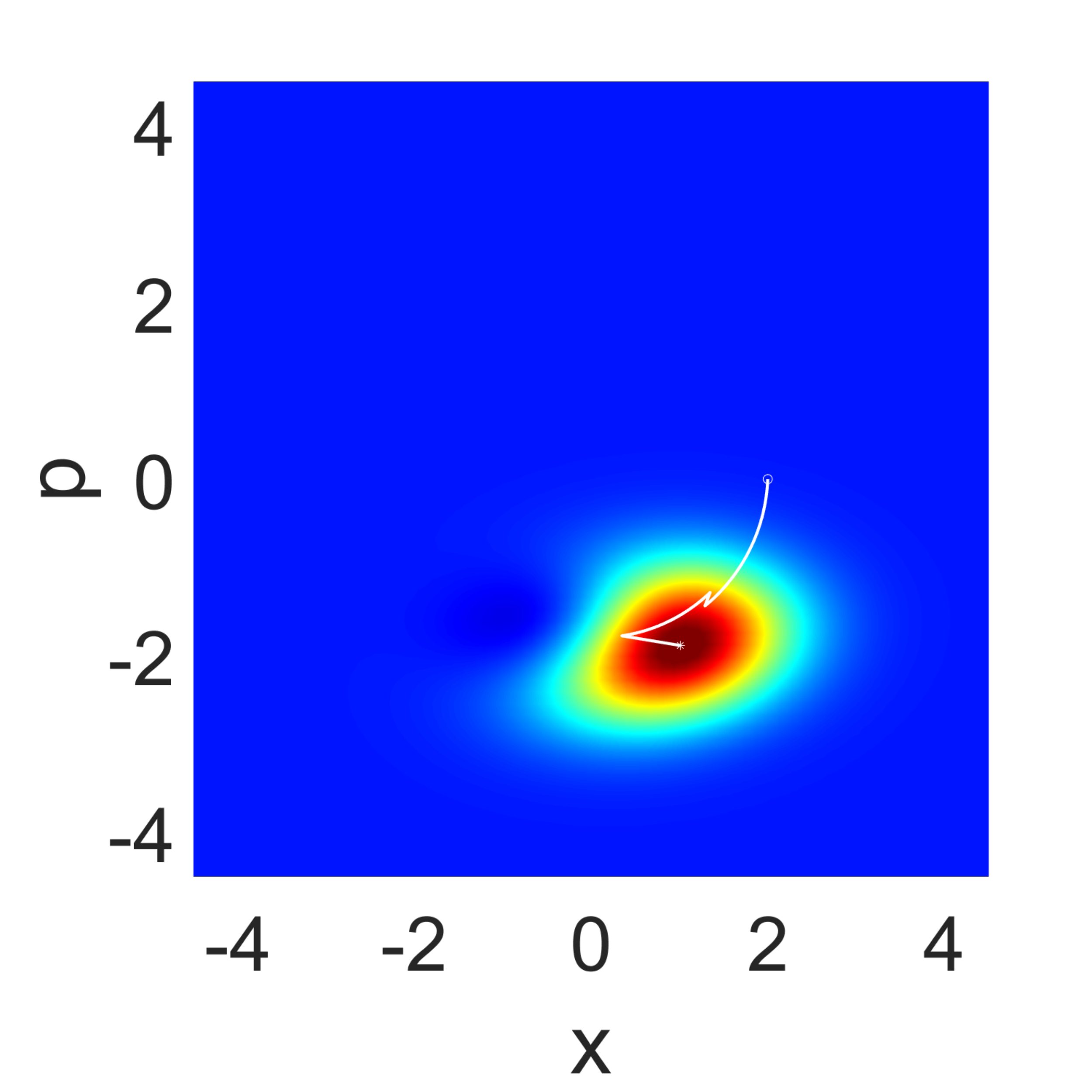} \\
	  \includegraphics[width=0.3\textwidth]{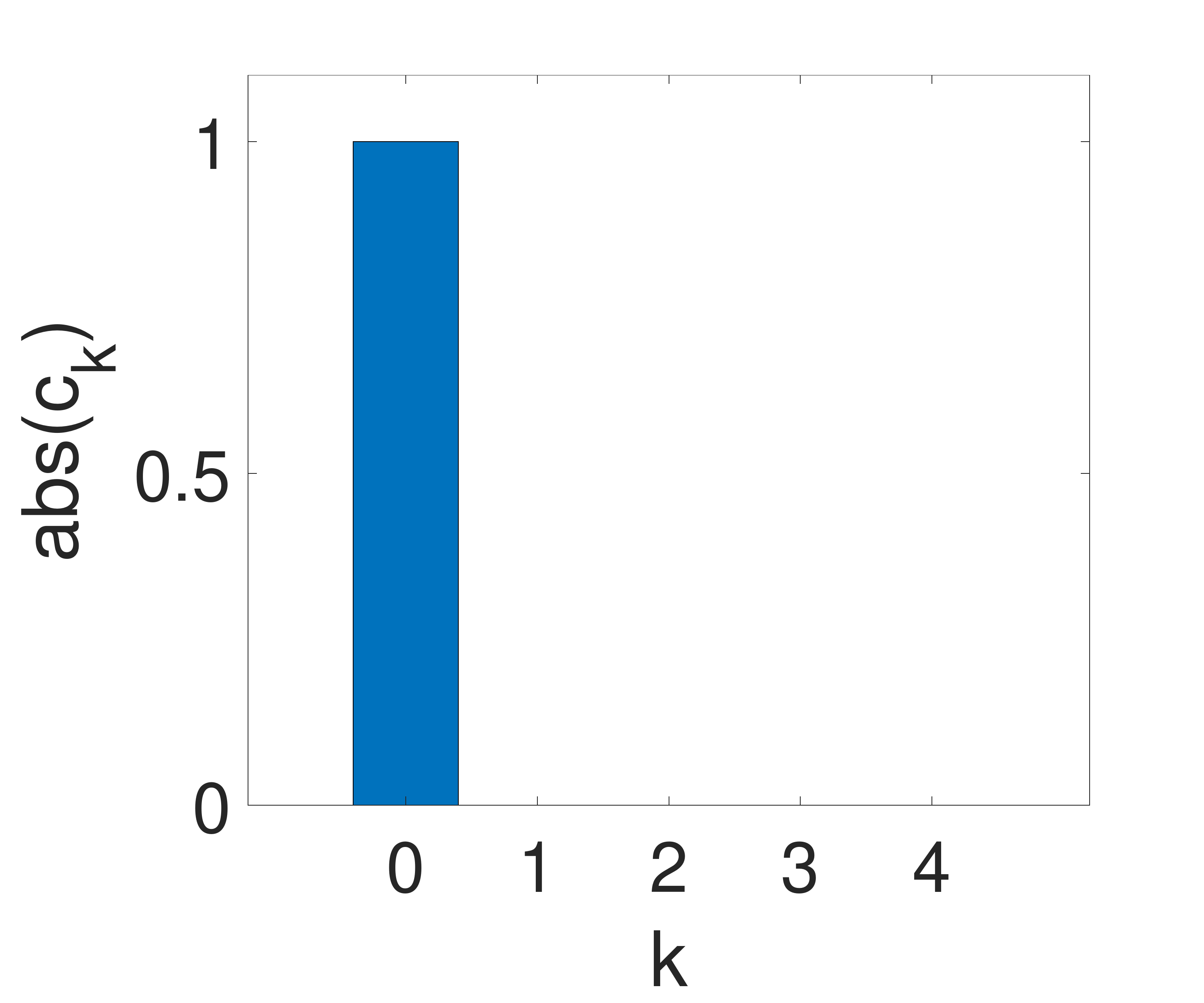}
	  \includegraphics[width=0.3\textwidth]{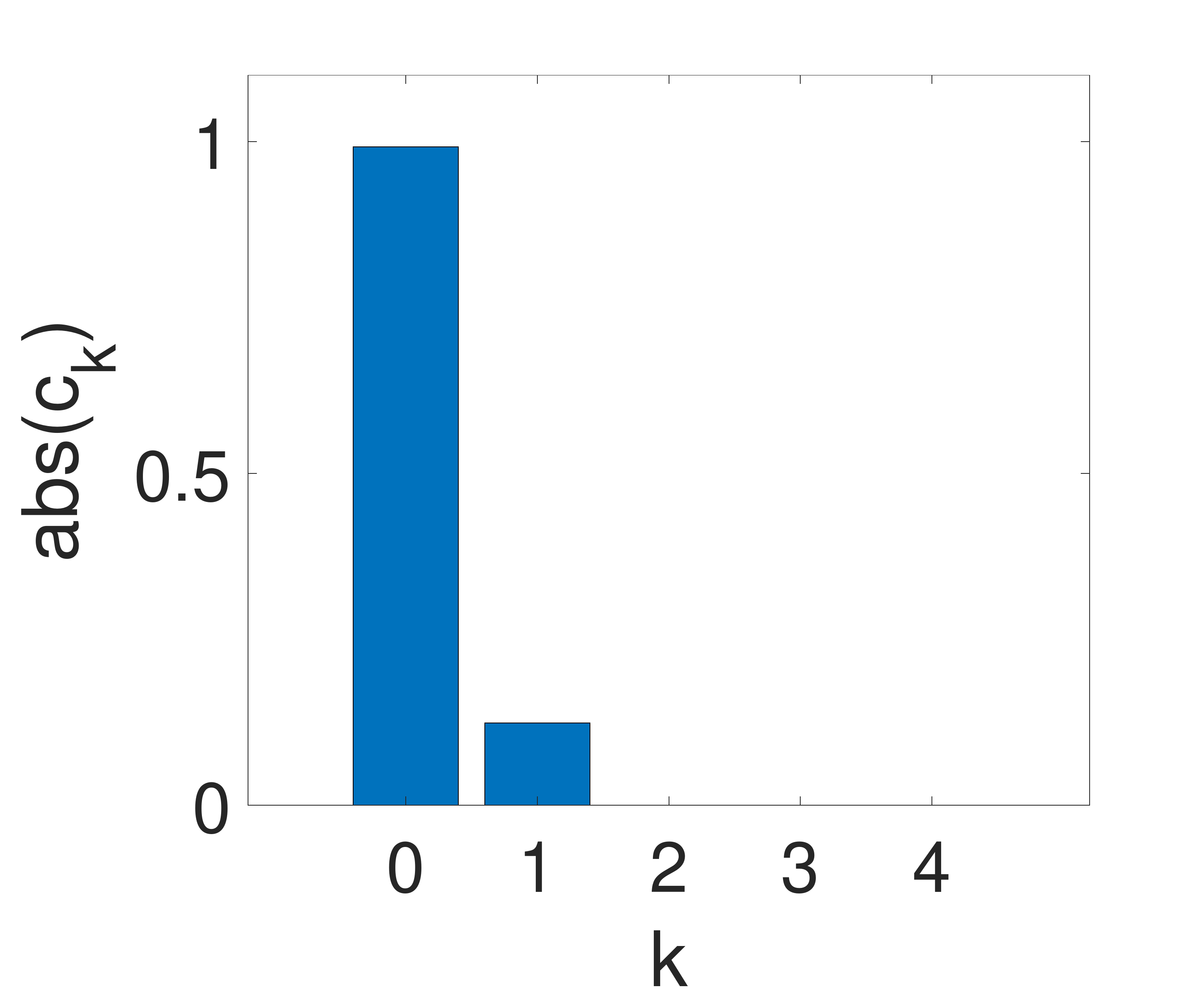}
	  \includegraphics[width=0.3\textwidth]{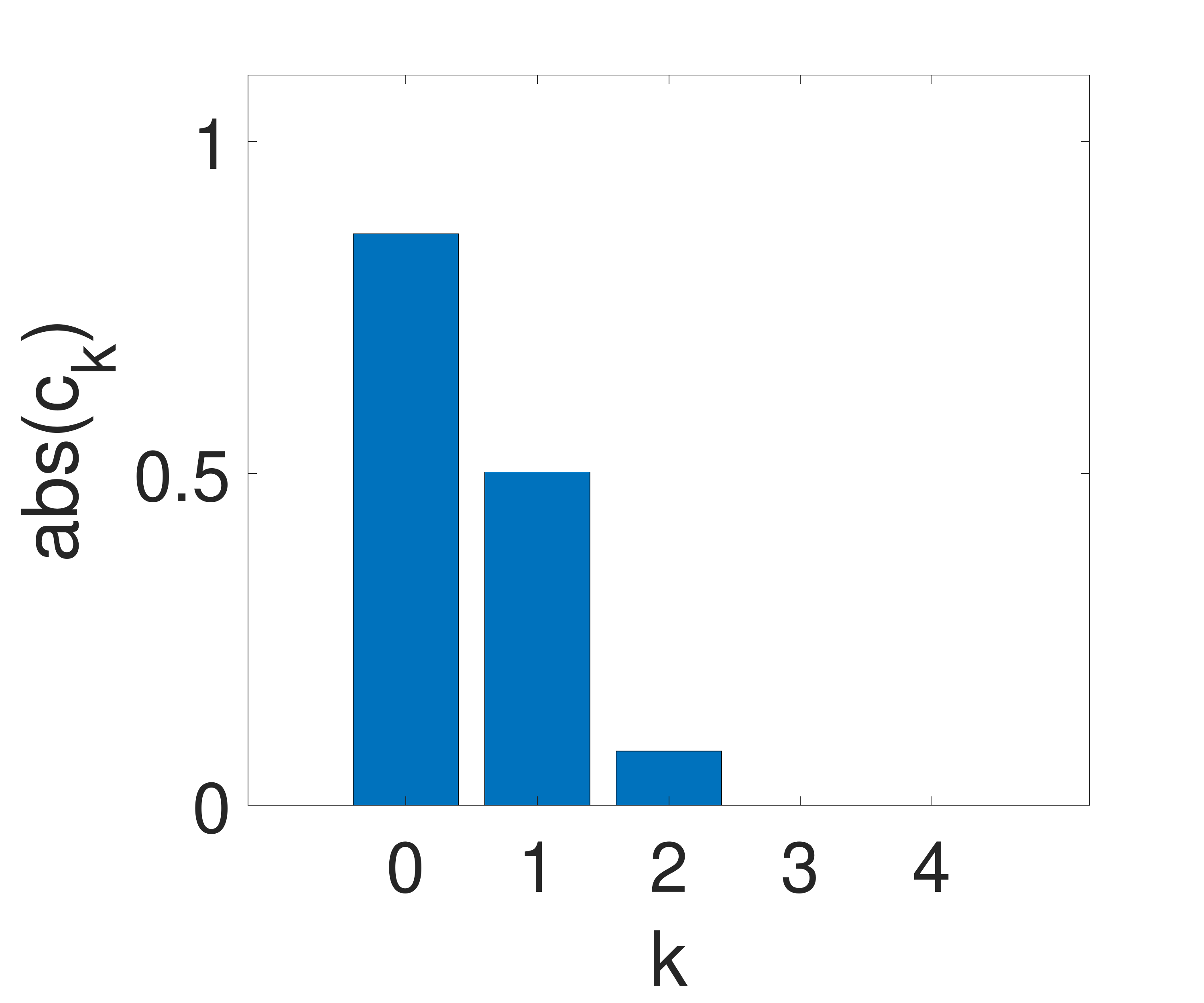}
\caption{Top row: Wigner functions of the single quantum jump trajectory in figure \ref{fig:SingleBehaviorAnh} at selected times (from left to right: t=0.84 (shortly before the first jump) and t=0.86 (just after the first jump) and t=1.48 (just after the second jump)). The white line traces the preceding central motion. Bottom row: Relative magnitudes of the coefficients of the state in the time evolved basis ($\hat U(t)\ket{n,a_0,z_0}$) at the same times as in the top row.} \label{fig:WigPlotsAnh_QJ}

\end{centering}
\end{figure}
\begin{figure}[tb] 
\begin{centering}
	  \includegraphics[width=0.4\textwidth]{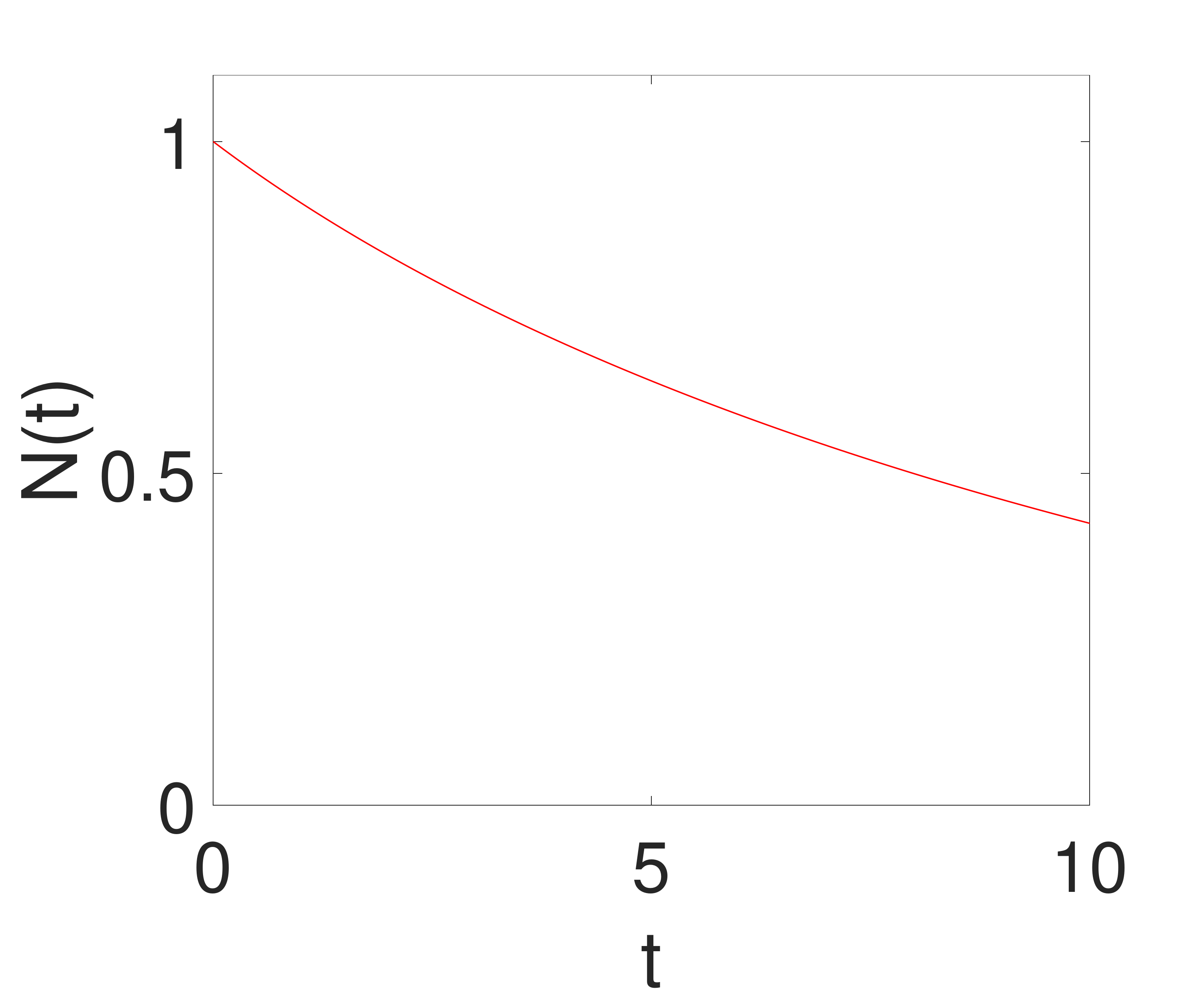}
	  \includegraphics[width=0.4\textwidth]{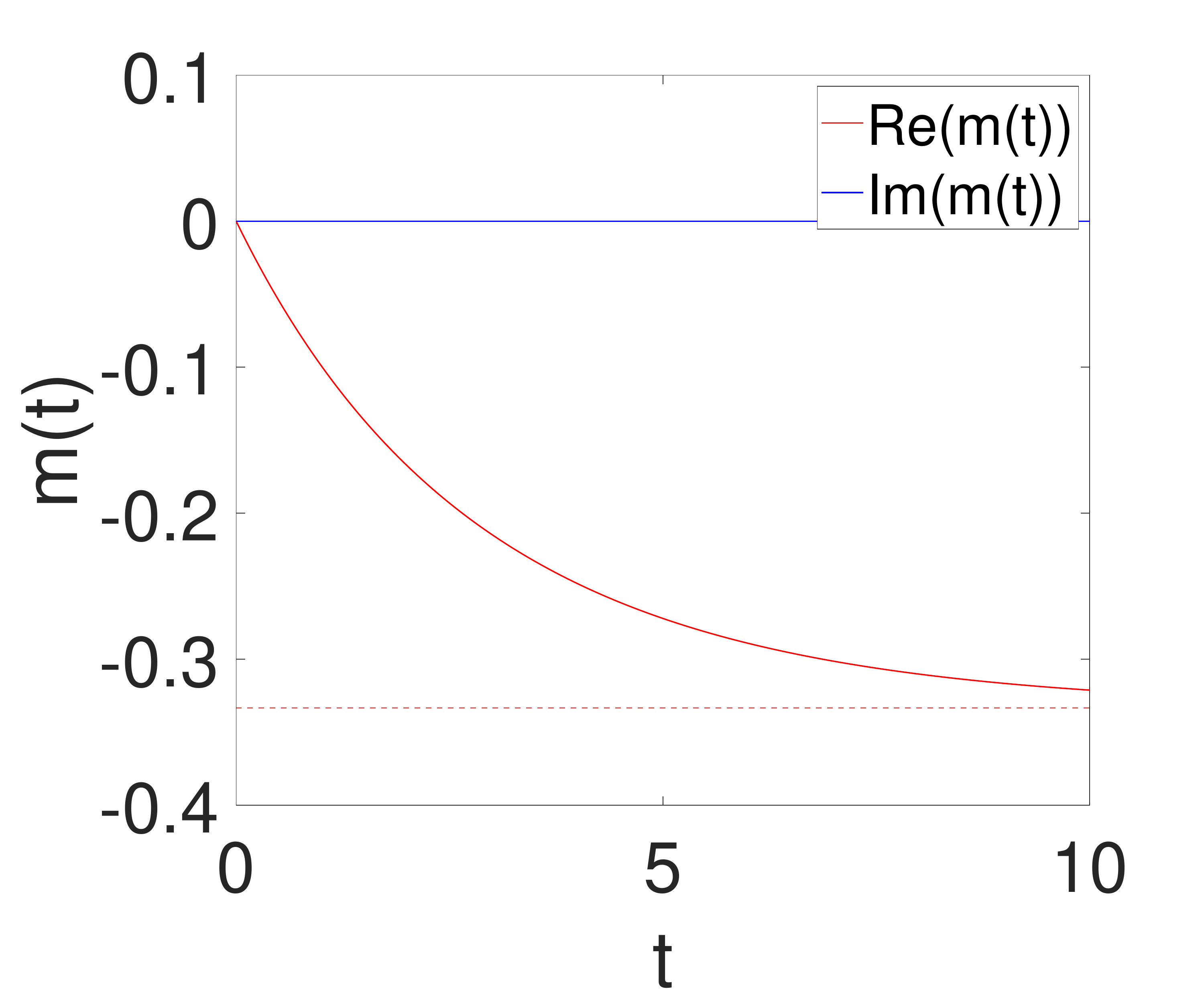}
\caption{Hagedorn basis parameter evolution for the damped oscillator model \cref{eq:AnhModel}. The figure on the left shows the norm $N(t)$ of the ground state, and the right figure the evolution of the parameter $M(t)$ with dotted lines depicting the asymptotic fixed values of $M(t)$. } \label{fig:HagedornVarsAnh}
\end{centering}
\end{figure}

Using \cref{eq:SAnh,eq:NDef,eq:HagM} we can derive the Hagedorn basis parameters 
\begin{equation}
\begin{aligned}
   N(t)&=\sqrt{\frac{2 \zeta}{\left(\zeta ^2+1\right) \sinh \left(\gamma  t\right)+2 \zeta \cosh \left(\gamma  t\right)}},\\
   M(t)&=-\frac{\left(\zeta ^2-1\right) \sinh (\gamma  t)}{\left(\zeta ^2+1\right) \sinh (\gamma  t)+2 \zeta  \cosh (\gamma  t)}.\\
\end{aligned}
\end{equation}
Much like the position example in the long time limit $N(t)$ tends to a simple exponential decay whilst $M(t)$ tends to a fixed value
\begin{equation}
\begin{aligned}
    N(t)&\to\frac{2 \sqrt{\xi}}{1+\xi}e^{-\frac{\gamma t}{2}}\\
    M(t)&\to\frac{1-\gamma}{1+\gamma}.\\
\end{aligned}
\end{equation}
Figure \ref{fig:WigPlotsAnh_QJ} illustrates the effect of the first two jumps in the particular realisation of figure \ref{fig:SingleBehaviorAnh}, showing the Wigner distributions shortly before the first jump and shortly after the first and second jumps, as well as the corresponding coefficients in the Hagedorn basis. 
\begin{figure}
\begin{centering}
	  \includegraphics[width=0.3\textwidth]{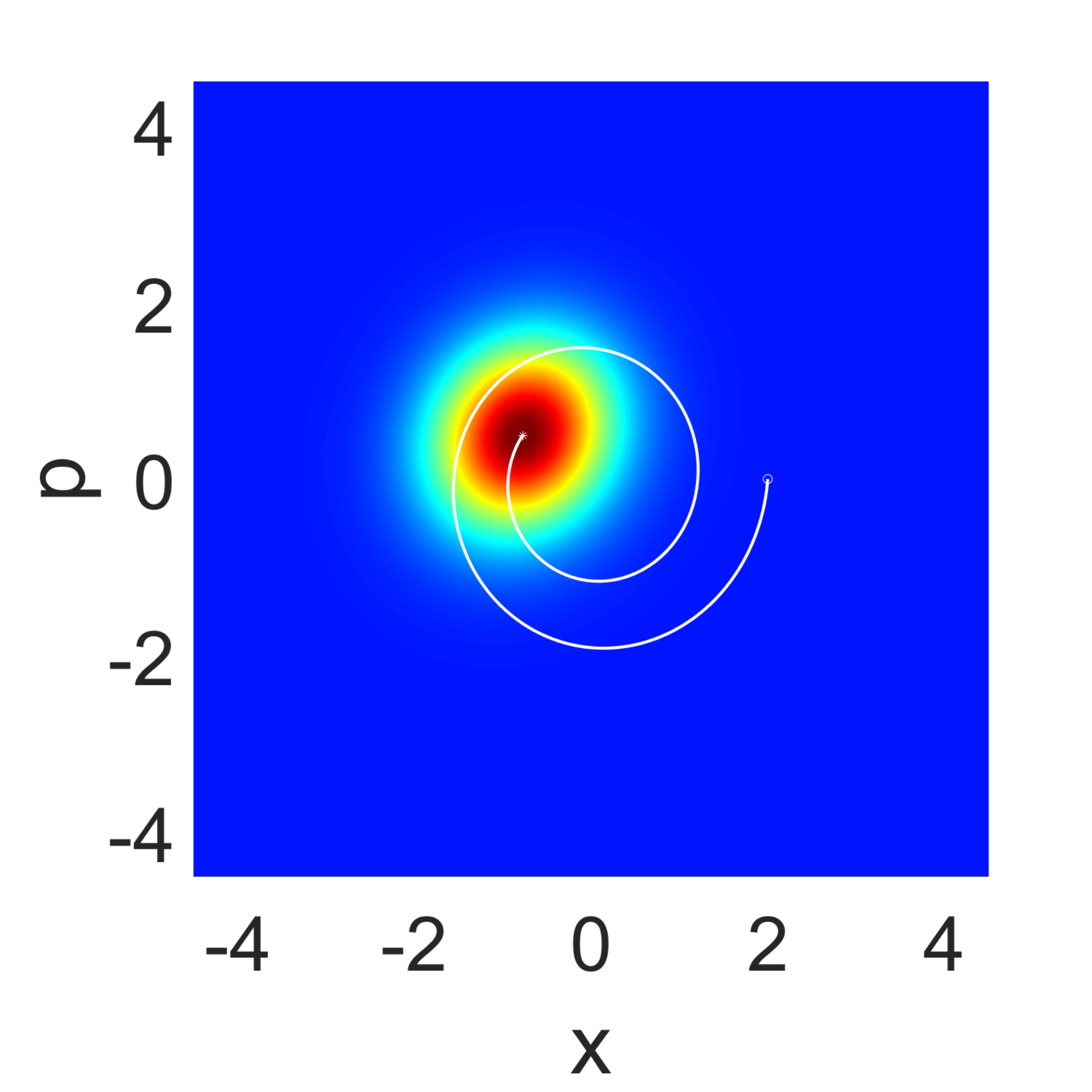}
	  \includegraphics[width=0.3\textwidth]{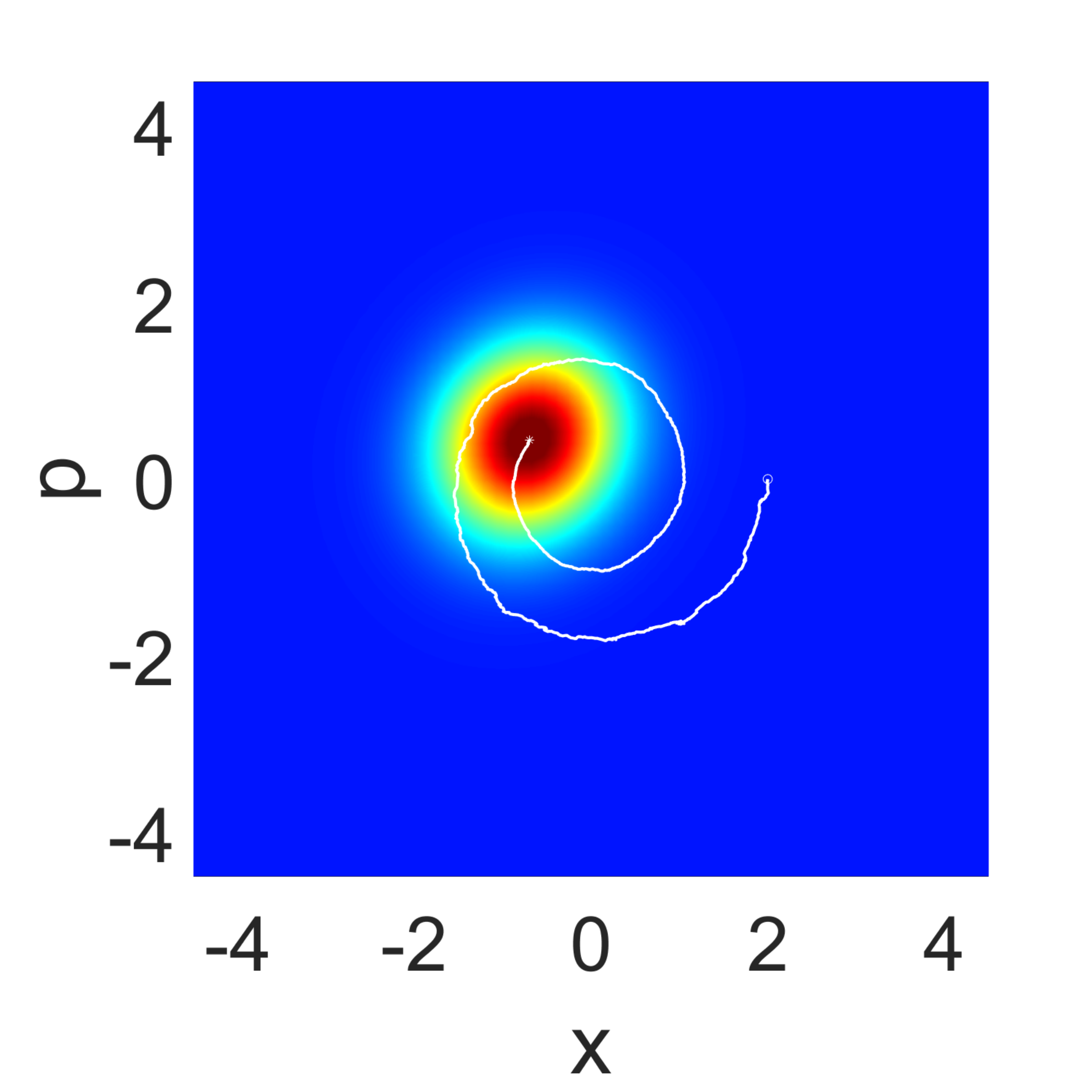}
	  \includegraphics[width=0.3\textwidth]{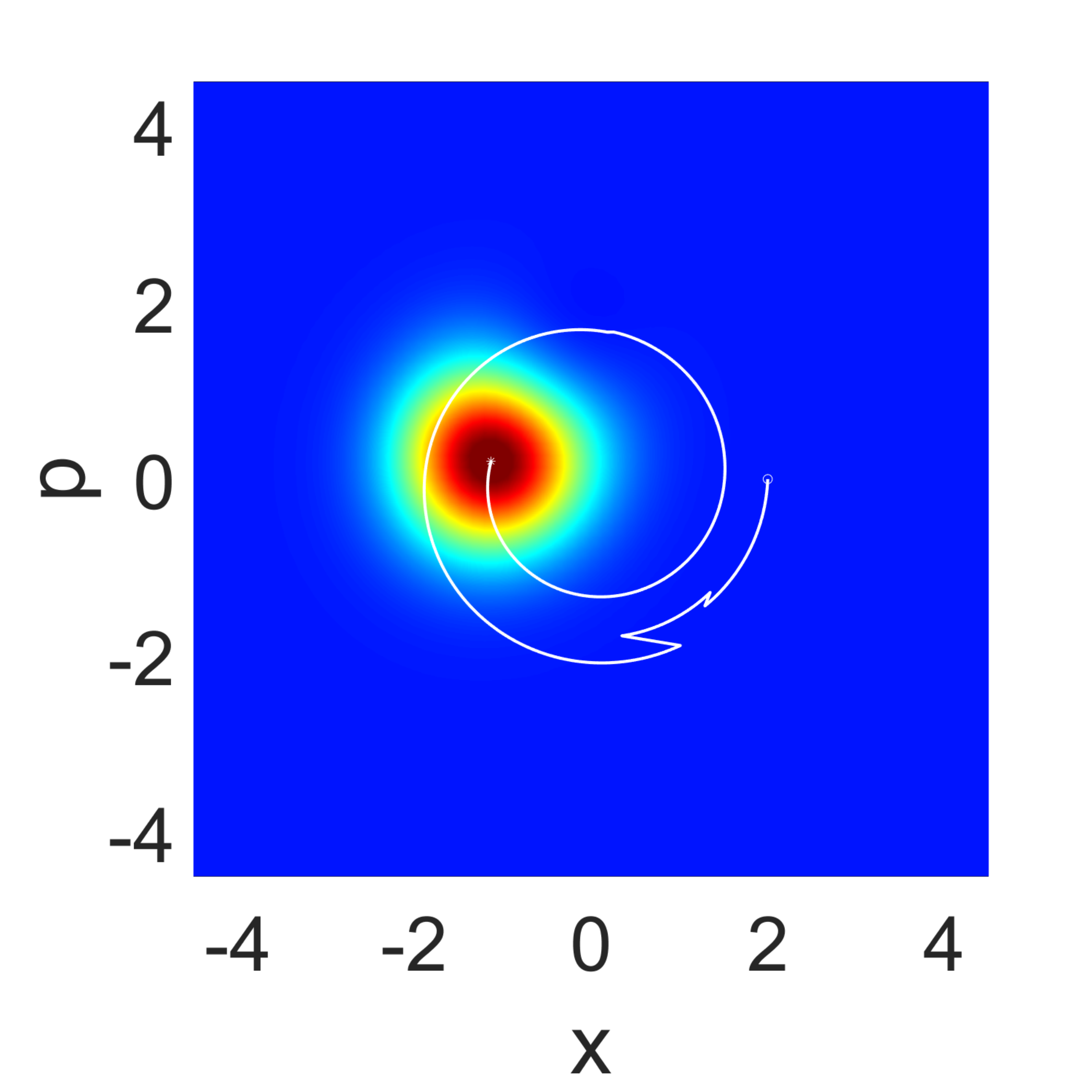}
\caption{Lindblad dynamics compared with a single quantum jump and SSE trajectory for the damped oscillator model \cref{eq:AnhModel}. In each case, a snapshot of the Wigner function at $t=10$ is plotted in phase space, with a white line displaying the precedent central motion. The left plot corresponds to the Lindblad dynamics, the middle one to the SSE and the right plot to the quantum jump dynamics.} \label{fig:WigPlotsAnh}
\end{centering}
\end{figure}
\Cref{fig:WigPlotsAnh} depicts the Wigner functions of the Lindblad, SSE and quantum-jump dynamics at time $t=10$, as well as the corresponding central trajectory up to this time. In comparison with the previous example, we now see dissipation for all three dynamics. The effect of the diminishing stochastic contribution in the SSE dynamics, specific to the Lindbladian considered here, is also visible in the figure.  While the final quantum-jump state is still non-Gaussian, every individual run asymptotically approaches a coherent state in the centre, in contrast to the dynamics resulting from the first example. 

\section{Summary and outlook} \label{sec:conclusion}
We have investigated the dynamics of initially Gaussian states in open quantum systems described in a Lindblad formalism with a quadratic Hamiltonian and linear Lindbladian in comparison to the two popular unravellings of the dynamics given by SSE and quantum-jump trajectories. In the SSE case, the state remains Gaussian for all times, where the central dynamics has a possibly damped deterministic contribution accompanied by a stochastic term that depends on the covariances of the state. The covariances themselves follow a deterministic time evolution, that is independent of the central trajectory, and that coincides with the dynamical equation found for quantum evolution generated by an effective non-Hermitian Hamiltonian. For the quantum-jump approach, initial Gaussian states do in general not stay Gaussian over time. Applying results from \cite{lasser2018non} we have formulated a method to describe quantum-jump trajectories utilising a family of solutions to the non-Hermitian Schr\"odinger equation that depends only on the dynamics of a $2\times2$ complex matrix known as the linearised flow. We have studied the similarities and differences of the dynamics resulting in the SSE, the quantum jump and the Lindblad descriptions for two important examples.

\section{Acknowledgements}
 We would like to thank Bradley Longstaff for his time checking calculations of the SSE Gaussian limit.

We acknowledge support  from  the Royal Society (Grants. No. URF\textbackslash R\textbackslash 201034 and RGF\textbackslash EA\textbackslash 180169) and from the European Research Council (ERC) under the European Union's Horizon 2020 research and innovation programme (grant agreement No 758453).

\appendix

\section{Quadratic Propagation}\label{app:A}
 
In this appendix we will discuss how to derive an expression for time evolved operators of the form $e^{-\frac{i}{\hbar} t \hat K}\hat A(b,\chi)\e^{\frac{i}{\hbar}t \hat{K}}$ where $\hat K$ is a possibly non-Hermitian operator of order at most two in $\hat p$ and $\hat q$. We will assume that the corresponding classical Hamiltonian is of the form 
\begin{equation}\label{eq:complex_Ham}
K(z)=\frac{1}{2} z\cdot K_2 z+k_1\cdot \Omega z+k_0
\end{equation}
where $K_2$ is a symmetric matrix, $k_1$ is a vector and $k_0$ a scalar. The solution  $z(t)$ to Hamilton's equations generated by $K(z)$ with initial conditions $z(0)=z$ is given by 
\begin{equation}
\Phi(t,z)=S(t)z+v(t)\,\, ,\quad\text{with}\quad v(t)=\int_0^t S(t-s)k_1\, d s\,\, ,
\end{equation}
where $S(t)=e^{t\Omega K_2}$ is symplectic, i.e., $S^T\Omega S=\Omega$. 

Now we consider a general operator $\hat P$ and $\hat P_t:=\e^{-\frac{i}{\hbar} t\hat K}\hat P\e^{\frac{i}{\hbar} t\hat K}$, which satisfies the differential equation 
\begin{equation}
i\hbar \partial_t \hat P_t=[\hat K, \hat P_t]\,\, .
\end{equation}
In the Weyl representation this equation takes the form 
\begin{equation}
    \partial_t P_t(z)=\{K(z), P_t(z)\}\,\, ,
\end{equation}
and here we used that $K(z)$ is at most quadratic, otherwise there are higher order  terms on the right hand side from the Moyal bracket. This equation is up to a sign equal to the Liouville equation in classical mechanics which is solved in terms of the classical flow, and hence we find 
\begin{equation}\label{eq:Egorov}
    P_t(z)=P(\Phi(-t,z))\,\, .
\end{equation}
This is a special case of a much more general result which is known as Egorov's Theorem, the only new aspect here is that we allowed a complex Hamilton function $K(z)$. 

If we apply this result to the case that $P=A(b,\chi)=\frac{i}{\sqrt{2\hbar}} b\cdot \Omega (z-\chi)$ we find with $\Phi(-t,z)-\chi=S(-t)(z-\Phi(t,\chi))$ and $\Omega S(-t)=S^T(t)\Omega$, which follows from $S^T\Omega S=\Omega $ and $S(-t)=S(t)^{-1}$, that
\begin{equation}
P_t(z)=A(S(t)b,\Phi(t,\chi))
\end{equation}
or 
\begin{equation}\label{eq:transported-annihilation}
    \hat U(t)\hat A(b,\chi)\hat U(-t)=\hat A\big(S(t)b,\Phi(t,\chi)\big)\,\, .
\end{equation}

 \section{A non-commutative Binomial expansion}\label{app:B}
 
 In this appendix we present the technical details of the expansion \eqref{eq:binom-exp}, we assume we have a pair of creation and annihilation operators $\hat A^{\dagger}, \hat A$, and a corresponding orthonormal basis $\ket{n}=\frac{1}{\sqrt{n!}}\, [\hat A^{\dagger}]^n\ket{0}$ such that $\hat A \ket{n}=\sqrt{n}\, \ket{n-1}$ and $\hat A^{\dagger}\ket{n}=\sqrt{n+1}\ket{n+1}$, and in particular $\hat A \ket{0}=0$. 
 
 We want to find the expansion coefficients $B_{mn}$ in 
 \begin{equation}
 \frac{1}{\sqrt{n!}}(h_{+}\hat A^{\dagger}+h_{-}\hat A+h_0)^n\ket{0}=\sum_{m=0}^n B_{mn} \ket{m}\, \, ,
 \end{equation}
where $h_{+},h_{-},h_0$ are complex constants. The idea is to use that $(h_{+}\hat A^{\dagger}+h_{-}\hat A+h_0)^n$ is proportional to the $n$th therm in the expansion of 
\begin{equation}
e^{s(h_{+}\hat A^{\dagger}+h_{-}\hat A+h_0)}\ket{0}=\sum_{n=0}^{\infty}\frac{s^n}{n!}\, (h_{+}\hat A^{\dagger}+h_{-}\hat A+h_0)^n\ket{0}\,\, ,
\end{equation}
and to use the Baker Campbell Haussdorf formula 
\begin{equation}
e^{s(h_{+}\hat A^{\dagger}+h_{-}\hat A+h_0)}=e^{sh_{+}\hat A^{\dagger}}e^{s(h_{-}\hat A+h_0)}  e^{-\frac{s^2}{2} [h_{+}\hat A^{\dagger}, h_{-}\hat A+h_0]}\,\, .    
\end{equation}
With $[\hat A^{\dagger}, \hat A]=-1$ we obtain for the commutator $[h_{+}\hat A^{\dagger}, h_{-}\hat A+h_0]=-h_{+}h_{-}$ and expanding the exponentials and equating the terms of order $s^n$ on both sides gives 
\begin{equation}
(h_{+}\hat A^{\dagger}+h_{-}\hat A+h_0)^n=n! \sum_{m+l+2k=n}\frac{1}{m!l!k!2^k} (h_{+}h_{-})^k (h_{+}\hat A^{\dagger})^m(h_{-}\hat A+h_0)^l\,\, .
\end{equation}
We can now use that $(h_{-}\hat A+h_0)^l\ket{0}=h_0^l\ket{0}$ and $(h_{+}A^{\dagger})^m\ket{0}=\sqrt{m!}\, h_{+}^m\ket{m}$ to find 
\begin{equation}
 \frac{1}{\sqrt{n!}}(h_{+}\hat A^{\dagger}+h_{-}\hat A+h_0)^n\ket{0}=
 \sum_{m=0}^n\sum_{l+2k=n-m}\frac{\sqrt{n!}(h_{+}h_{-})^kh_{+}^mh_0^l}{\sqrt{m!}l! k! 2^k}\,\, \ket{m}
\end{equation}
and the inner sum can be rewritten as
\begin{equation}
B_{mn}= h_{+}^m h_0^{n-m}\sum_{k=0}^{[\frac{n-m}{2}]}\frac{\sqrt{n!}}{\sqrt{m!}(n-m-2k)! k! 2^k} \bigg(\frac{h_{+}h_{-}}{h_0^2}\bigg)^k \,\, . 
\end{equation}

\section*{References}
\bibliographystyle{IEEEtran}
\bibliography{main}{}

\end{document}